\documentclass[10pt,a4paper]{article} 
\usepackage{graphicx}
\usepackage{style}
\usepackage{epstopdf}
\usepackage{epsfig}
\usepackage{amsmath,amssymb,bm,amsfonts,yfonts,bbm}
\newcommand{\be}{\begin{equation}}
\newcommand{\ee}{\end{equation}}
\newcommand{\bea}{\begin{eqnarray}}
\newcommand{\nn}{\nonumber}
\newcommand{\eea}{\end{eqnarray}}

\def\Hc{{\cal H}}

\newcommand{\drho}{\delta \rho}

\newcommand{\bp}{\mathbf{p}}
\newcommand{\bq}{\mathbf{q}}
\newcommand{\bk}{\mathbf{k}}
\newcommand{\br}{\mathbf{r}}
\newcommand{\bx}{\mathbf{x}}

\DeclareSymbolFont{mathscrUC}{U}{rsfs}{m}{n}  
\DeclareSymbolFont{mathscrLC}{OT1}{pzc}{m}{n} 

\makeatletter

\DeclareRobustCommand*{\mathscr}[1]{\gdef\F@ntPrefix{mathscr@char@}%
  \@EachCharacter #1\@EndEachCharacter}
\long\def\DoLongFutureLet #1#2#3#4{%
   \def\@FutureLetDecide{#1#2\@FutureLetToken
      \def\@FutureLetNext{#3}\else
      \def\@FutureLetNext{#4}\fi\@FutureLetNext}
   \futurelet\@FutureLetToken\@FutureLetDecide}
\def\DoFutureLet #1#2#3#4{\DoLongFutureLet{#1}{#2}{#3}{#4}}
\def\@EachCharacter{\DoFutureLet{\ifx}{\@EndEachCharacter}%
   {\@EachCharacterDone}{\@PickUpTheCharacter}}
\def\m@keCharacter#1{\csname\F@ntPrefix#1\endcsname}
\def\@PickUpTheCharacter#1{\m@keCharacter{#1}\@EachCharacter}
\def\@EachCharacterDone \@EndEachCharacter{}

\DeclareMathSymbol{\mathscr@char@A}{\mathord}{mathscrUC}{`A}
\DeclareMathSymbol{\mathscr@char@B}{\mathord}{mathscrUC}{`B}
\DeclareMathSymbol{\mathscr@char@C}{\mathord}{mathscrUC}{`C}
\DeclareMathSymbol{\mathscr@char@D}{\mathord}{mathscrUC}{`D}
\DeclareMathSymbol{\mathscr@char@E}{\mathord}{mathscrUC}{`E}
\DeclareMathSymbol{\mathscr@char@F}{\mathord}{mathscrUC}{`F}
\DeclareMathSymbol{\mathscr@char@G}{\mathord}{mathscrUC}{`G}
\DeclareMathSymbol{\mathscr@char@H}{\mathord}{mathscrUC}{`H}
\DeclareMathSymbol{\mathscr@char@I}{\mathord}{mathscrUC}{`I}
\DeclareMathSymbol{\mathscr@char@J}{\mathord}{mathscrUC}{`J}
\DeclareMathSymbol{\mathscr@char@K}{\mathord}{mathscrUC}{`K}
\DeclareMathSymbol{\mathscr@char@L}{\mathord}{mathscrUC}{`L}
\DeclareMathSymbol{\mathscr@char@M}{\mathord}{mathscrUC}{`M}
\DeclareMathSymbol{\mathscr@char@N}{\mathord}{mathscrUC}{`N}
\DeclareMathSymbol{\mathscr@char@O}{\mathord}{mathscrUC}{`O}
\DeclareMathSymbol{\mathscr@char@P}{\mathord}{mathscrUC}{`P}
\DeclareMathSymbol{\mathscr@char@Q}{\mathord}{mathscrUC}{`Q}
\DeclareMathSymbol{\mathscr@char@R}{\mathord}{mathscrUC}{`R}
\DeclareMathSymbol{\mathscr@char@S}{\mathord}{mathscrUC}{`S}
\DeclareMathSymbol{\mathscr@char@T}{\mathord}{mathscrUC}{`T}
\DeclareMathSymbol{\mathscr@char@U}{\mathord}{mathscrUC}{`U}
\DeclareMathSymbol{\mathscr@char@V}{\mathord}{mathscrUC}{`V}
\DeclareMathSymbol{\mathscr@char@W}{\mathord}{mathscrUC}{`W}
\DeclareMathSymbol{\mathscr@char@X}{\mathord}{mathscrUC}{`X}
\DeclareMathSymbol{\mathscr@char@Y}{\mathord}{mathscrUC}{`Y}
\DeclareMathSymbol{\mathscr@char@Z}{\mathord}{mathscrUC}{`Z}
\DeclareMathSymbol{\mathscr@char@a}{\mathord}{mathscrLC}{`a}
\DeclareMathSymbol{\mathscr@char@b}{\mathord}{mathscrLC}{`b}
\DeclareMathSymbol{\mathscr@char@c}{\mathord}{mathscrLC}{`c}
\DeclareMathSymbol{\mathscr@char@d}{\mathord}{mathscrLC}{`d}
\DeclareMathSymbol{\mathscr@char@e}{\mathord}{mathscrLC}{`e}
\DeclareMathSymbol{\mathscr@char@f}{\mathord}{mathscrLC}{`f}
\DeclareMathSymbol{\mathscr@char@g}{\mathord}{mathscrLC}{`g}
\DeclareMathSymbol{\mathscr@char@h}{\mathord}{mathscrLC}{`h}
\DeclareMathSymbol{\mathscr@char@i}{\mathord}{mathscrLC}{`i}
\DeclareMathSymbol{\mathscr@char@j}{\mathord}{mathscrLC}{`j}
\DeclareMathSymbol{\mathscr@char@k}{\mathord}{mathscrLC}{`k}
\DeclareMathSymbol{\mathscr@char@l}{\mathord}{mathscrLC}{`l}
\DeclareMathSymbol{\mathscr@char@m}{\mathord}{mathscrLC}{`m}
\DeclareMathSymbol{\mathscr@char@n}{\mathord}{mathscrLC}{`n}
\DeclareMathSymbol{\mathscr@char@o}{\mathord}{mathscrLC}{`o}
\DeclareMathSymbol{\mathscr@char@p}{\mathord}{mathscrLC}{`p}
\DeclareMathSymbol{\mathscr@char@q}{\mathord}{mathscrLC}{`q}
\DeclareMathSymbol{\mathscr@char@r}{\mathord}{mathscrLC}{`r}
\DeclareMathSymbol{\mathscr@char@s}{\mathord}{mathscrLC}{`s}
\DeclareMathSymbol{\mathscr@char@t}{\mathord}{mathscrLC}{`t}
\DeclareMathSymbol{\mathscr@char@u}{\mathord}{mathscrLC}{`u}
\DeclareMathSymbol{\mathscr@char@v}{\mathord}{mathscrLC}{`v}
\DeclareMathSymbol{\mathscr@char@w}{\mathord}{mathscrLC}{`w}
\DeclareMathSymbol{\mathscr@char@x}{\mathord}{mathscrLC}{`x}
\DeclareMathSymbol{\mathscr@char@y}{\mathord}{mathscrLC}{`y}
\DeclareMathSymbol{\mathscr@char@z}{\mathord}{mathscrLC}{`z}

\makeatother

\relax

\title{Renormalization-group flow of the effective action of cosmological large-scale structures}

\author[a]{Stefan Floerchinger,}
\author[b,c]{Mathias Garny,}
\author[b,d]{Nikolaos Tetradis,}
\author[b]{Urs Achim Wiedemann}

\affiliation[a]{Institut f\"{u}r Theoretische Physik, Universit\"{a}t Heidelberg, Philosophenweg 16, 69120 Heidelberg, Germany}
\affiliation[b]{Theoretical Physics Department, CERN, CH-1211 Gen\`eve 23, Switzerland}
\affiliation[c]{Centre de Physique Th{\'e}orique, Ecole Polytechnique, CNRS, 91128 Palaiseau Cedex, France}
\affiliation[d]{Department of Physics, University of Athens, Zographou 157 84, Greece}

\emailAdd{stefan.floerchinger@cern.ch}
\emailAdd{mathias.garny@cern.ch}
\emailAdd{ntetrad@phys.uoa.gr}
\emailAdd{urs.wiedemann@cern.ch}

\abstract{Following an approach of Matarrese and Pietroni, we derive the functional renormalization group (RG) flow of the effective action
of cosmological large-scale structures. Perturbative solutions of this RG flow equation are shown to be consistent with standard cosmological
perturbation theory. Non-perturbative approximate solutions can be obtained by truncating the a priori infinite set of possible effective actions to a
finite subspace. Using for the truncated effective action a form dictated by dissipative fluid dynamics, 
we derive RG flow equations for the scale dependence of the effective viscosity 
and sound velocity of non-interacting dark matter, and we solve them numerically. Physically, the effective viscosity and sound velocity account
for the interactions of long-wavelength fluctuations with the spectrum of smaller-scale perturbations. 
We find that the RG flow exhibits an attractor behaviour in the IR that
significantly reduces the dependence of the effective viscosity and sound velocity on the input values at the UV scale.
This allows for a self-contained computation of matter and velocity power spectra for which
the sensitivity to UV modes is under control. 
}

\begin{document}

\maketitle

\section{Introduction} \label{intro}

Understanding how density fluctuations in cold dark matter evolve under the influence of gravity is at the basis of analyzing data on large-scale 
structures and of constraining cosmological models from them. Technically, this task amounts to solving the collisionless Vlasov-Boltzmann 
equation for appropriate classes of initial conditions \cite{bernardeau0}. 
With recent progress in both computational power and coding techniques, $N$-body simulations describe by now
 structure evolution down to small scales where baryonic effects become increasingly important \cite{Kuhlen:2012ft, Vogelsberger:2014dza, Schneider:2015yka}. Even for pure cold dark matter simulations,
perturbative techniques cannot be expected to apply on scales where non-linearities become strong and the process of virialization
starts to dominate.
 Nevertheless, $N$-body simulations remain CPU-intensive, while suitable analytical techniques can help 
 to scan the relevant range of initial conditions, as well as to explore non-standard cosmological models.
 Also, there is the general desire to understand the result of complex simulations in simple analytical terms for certain limiting cases.
 Last but not least, future large-scale structure surveys, including EUCLID~\cite{Amendola:2012ys}, LSST~\cite{Abell:2009aa}, DESI~\cite{Levi:2013gra}
and eBOSS~\cite{Dawson:2015wdb},
will probe scales within the weakly non-linear regime with increasing precision. 
 For these and other reasons, much effort has gone recently into complementing $N$-body simulations with advanced analytical techniques. 

Cosmological perturbation theory is based on writing the Vlasov-Boltzmann equation as a hierarchy of equations for the
momentum moments of the phase-space distribution \cite{bernardeau0}. If truncated at the lowest moments, the resulting equations of motion
are of ideal fluid dynamic form. A linearized perturbative treatment of this `cosmological fluid'  works well for the small
fluctuations at sufficiently large scales
for redshift $z=0$. In addition, the leading non-linear corrections
computed within this framework capture the dominant effects on the density power spectrum at
scales $k \lesssim 0.075\, h/$Mpc \cite{Scoccimarro:1995if, bernardeau0, bernardeau, Carlson:2009it, Audren:2011ne, CrSc1, Taruya:2012ut}.
Moreover, unequal-time correlators exhibit a damping due to the stochastic background of large-scale bulk flows. This
effect can be rather accurately described within an ideal fluid dynamical framework by resumming certain classes of
perturbative contributions related to long-wavelength perturbations \cite{CrSc2, Bernardeau:2011vy}. While most of this effect cancels
out in equal-time correlators \cite{Jain:1995kx, Peloso:2013zw, Kehagias:2013yd, nonlinear}, its residual effect is important for the non-linear broadening and shift of the
baryon acoustic peak \cite{Crocce:2007dt, Smith:2007gi, Sherwin:2012nh, Baldauf:2015xfa, Senatore:2014via}. 
This influence of very long-wavelength modes on the scale of baryon acoustic oscillations (BAO)
can  be computed systematically by a suitable reformulation and resummation of perturbation
theory \cite{Baldauf:2015xfa, Zaldarriaga:2015jrj, Blas:2015qsi, Blas:2016sfa}.

However, beyond the linear approximation, fluctuations around the BAO scale are also
modified by interactions with the UV part of the spectrum of density fluctuations. Here, the standard 
cosmological perturbation theory fails, in the sense that higher-order `loop' corrections to the propagation of long-wavelength 
fluctuations become increasingly more sensitive to the UV part of the spectrum, see e.g.~\cite{Blas1}. 
Several properties are relevant for a discussion of this failure: First, the density contrast grows large in the UV. Second, at 
UV scales where the so-called single-stream approximation fails, higher moments of the phase-space distribution that are
not accounted for in standard perturbation theory become important. 
Third, $N$-body simulations~\cite{Bagla:1996zb,nishimichi,Pueblas:2008uv} as well as general considerations about the decoupling of virialized substructures \cite{Peebles:1980}
and the form of 
non-perturbative response functions \cite{Garny:2015oya}, indicate that the impact of UV modes on the BAO range is overestimated by standard perturbation theory. 
In other words, UV modes decouple more efficiently than expected perturbatively. 

To account for interactions with the UV part of density fluctuations, a consistent cosmological perturbation theory
needs to be based on effective dynamics, which is applicable only above some length scale and in which 
non-perturbative parameters absorb the effect of small-scale perturbations that are `integrated out'~\cite{Pietroni:2011iz,Manzotti:2014loa}. Since the lowest moments 
of the phase-space distribution that are dropped in the single-stream approximation define the stress tensor of an imperfect fluid, 
it is natural to absorb the integrated out UV physics in the viscous transport coefficients that parametrize the non-ideal stress tensor. 
This strategy is adopted for instance in recent works on the effective field theory of large-scale structure where the
viscous coefficients are not predictable within the effective theory, but are fixed by comparison of calculated correlations
functions with either $N$-body simulations or observations \cite{Baumann:2010tm, Carrasco:2012cv, Porto:2013qua, Foreman:2015lca, Baldauf:2015aha, Assassi:2015jqa, Abolhasani:2015mra}. 
For a discussion of alternative approaches leading to the same strategy, see \cite{Fuhrer:2015cia} and references therein.

The main point of the present paper is that the scale dependence of the viscous coefficients is calculable. 
Our main result is the formulation of a renormalization group for the coefficients that enter the effective
dynamics of large-scale cosmological perturbations, and an explicit calculation of the renormalization-group trajectories of the effective 
viscosity and sound velocity derived from it.  These coefficients account for momentum transfer with UV modes that are integrated out in the effective dynamics.

Technically, we start from
work of Matarrese and Pietroni~\cite{Max1} who formulated renormalization-group (RG) techniques for the mildly non-linear stages of large-scale structure evolution. For other applications of RG techniques to the problem of large scale structure, see Refs.~\cite{mcdonald,time}.
Based on the central elements of their proposal, which we recall in section~\ref{sec2}, we derive in section~\ref{sec3} how the effective 
action for long-wavelength cosmological perturbations flows with the coarse-graining scale. In section~\ref{sec4}, we demonstrate that, if solved
perturbatively, this exact functional RG flow equation for the effective action is consistent with results from standard cosmological perturbation
theory. In section~\ref{sec5}, we then employ the physical intuition that the effective action at mildly non-linear scales describes the
dynamics of an imperfect fluid to explore a truncated non-perturbative solution of the functional renormalization group. In this way, 
the dependence on the coarse-graining scale of the non-perturbative parameters entering the effective dynamics of large-scale structure
can be described explicitly by a set of coupled ordinary differential equations. In section~\ref{sec6}, we solve for this RG flow numerically and we discuss physical
implications. Furthermore, we compare the resulting power spectra of the density contrast and velocity scalar to those
measured in $N$-body simulations. We close with a short discussion and outlook.

\section{Generating functionals and effective action for the cosmological fluid}
\label{sec2}
In this preparatory section, we introduce basic concepts used throughout this work. We first recall the range of validity of
a description of dark matter as a pressureless ideal fluid. Following the approach of
Mattarese and Pietroni~\cite{Max1}, we discuss then how the resulting equations of motion for stochastic initial conditions 
can be obtained from a classical microscopic action after introduction of an auxiliary field. We give explicit expressions for 
generating functionals and the effective action, and we discuss the physical meaning of the resulting functional derivatives. 
Within this framework, we shall set up in subsequent sections a compact functional renormalization-group equation for
cosmological perturbations, which we shall solve for different situations of physical interest. 

\subsection{Range of applicability and equations of motion of the pressureless ideal cosmological fluid}
\label{sec2.1} 

At cosmologically late times when microscopic interaction rates are negligible, dark matter is described by the collisionless 
Vlasov-Boltzmann equation, which can be written as a coupled hierarchy of evolution equations for moments of
the dark matter phase-space distribution. The equations of fluid dynamics result from truncating this hierarchy at the second moment,
and the range of validity of this truncation determines the range of validity of fluid dynamics.
Assuming that at sufficiently early times,  non-interacting dark matter is close to
local equilibrium (an assumption that is justified by the presumed thermal origin of dark matter), and that local particle
velocities $v_p$ with respect to this equilibrium state are non-relativistically small, this truncation of the 
Vlasov-Boltzmann system can be justified for~\cite{Baumann:2010tm}
\begin{equation}
	k\, v_p\,  H^{-1} \ll 1\, .
	\label{eq2.1}
\end{equation}
Here, $k$ is the inverse wavelength of a perturbation in the dark matter system, and $H$ denotes the Hubble constant.
Within the lifetime $1/H$ of the Universe, a particle of velocity $v_p$ free-streams over a distance $v_p\,  H^{-1}$.  
According to equation~(\ref{eq2.1}), a system of non-interacting particles that is initially close to local 
equilibrium will remain close to local equilibrium for wavelengths $1/k$ that exceed significantly
the free-streaming distances $v_p\,  H^{-1}$. Therefore, on large scales, dark matter can be described as a cosmological 
fluid, not because microscopic interaction rates are sufficiently large to maintain local thermal equilibrium, but because the lifetime 
of the Universe is too short for dark matter 
to deviate strongly from local equilibrium. 

The simplest truncation of the Vlasov-Boltzmann hierarchy is the so-called single stream approximation in which the shear tensor is
set to zero. This system is typically considered for irrotational flows. 
The resulting fluid dynamic equations are then written in terms of two scalar fields only: the density 
perturbation $\delta\equiv\drho/\rho_m$ and the velocity divergence $\theta\equiv\vec{\nabla}\vec{v}$. 
We write the Fourier modes of these two scalar fields as a doublet, 
\begin{equation}
 \left(
\begin{array}{c}
\phi_{1}(\eta,\textbf{k})\\ \\ \phi_{2}(\eta,\textbf{k})
\end{array}
\right)
\equiv\left(
\begin{array}{c}
\delta_\bk(\tau)\\ \\-\dfrac{\theta_{\bk}( \tau)}{\mathcal{H}}
\end{array}
\right)\, .
\label{doublet}
\end{equation}
For the metric, we consider an ansatz of the form
\be
ds^2=a^2(\tau)\left[
-\left(1+2\Phi(\tau,\bx) \right)d\tau^2
+\left(1-2\Phi(\tau,\bx) \right) d\bx\, d\bx \right]\, ,
\label{metric} \ee
that accounts for the dominant scalar metric perturbations and 
includes only one Newtonian potential. This is sufficient for the 
treatment of subhorizon perturbations. Distances in time and space are measured with respect to the background metric which corresponds to \eqref{metric} with $\Phi = 0$. 
For the present work we concentrate on the dynamics of perturbations and assume that the background expansion follows the usual dynamics with negligible backreaction effects.
In an essentially Newtonian approximation, discussed and motivated e.g. in \cite{CrSc1,viscousdm}, the dynamic equations for the Fourier modes of the scalar fields evolve according to
\be
\partial_\eta \phi_a (\bk)= -\Omega_{ab}(\bk,\eta) \phi_b (\bk)+
 \int d^3 p \, d^3 q \, \delta^{(3)}(\bk-\bp-\bq)
\gamma_{abc}(\bp,\bq,\eta)\, \phi_b (\bp)\, \phi_c(\bq)\, .
\label{eom}
\ee
Here, $\Hc=\dot{a}/a$, $\eta=\ln a(\tau)$, 
\be
\Omega(\bk,\eta)=\left(
\begin{array}{cc}
~~~0 &~~~-1
\\
-\frac{3}{2} \Omega_m &~~~ 1 +\frac{\Hc'}{\Hc}
\end{array} \right)\, ,
\label{ome} \ee
and 
the prime denotes a derivative with respect to $\eta$. 
The non-zero elements of $\gamma_{abc}$ are 
\bea
\gamma_{121}(\bq,\bp,\eta)=\gamma_{112}(\bp,\bq,\eta)&=&\frac{(\bp+\bq)\bq}{2q^2} \label{gamma1}, \\
\gamma_{222}(\bp,\bq,\eta)&=&\dfrac{(\bp+\bq)^{2} \bp\cdot\bq}{2 p^{2} q^{2}}\, .\label{gamma3} 
\eea 

The evolution is particularly simple in an Einstein-de Sitter (EdS) Universe with 
$\Omega_m=1$ and $\Hc'/\Hc=-1/2$, and we concentrate on this case. 
Any $\Lambda$CDM cosmology can be mapped,
through an appropriate change of variables, to one with $\Omega_m=1$ to a very
good approximation \cite{bernardeau0}.
The retarded linear propagator $g_{ab}$ satisfies
\be
\left( \delta_{ac} \, \partial_\eta + \Omega_{ac}(\bk,\eta) \right) g_{cb}^\text{R} (\bk, \eta,\eta^\prime) = \delta_{ab} \delta\left(\eta-\eta^\prime\right)\, 
\label{linpropa}
\ee
for $\eta >\eta'$. It takes the form
\be
g_{ab}^\text{R}(\eta-\eta^\prime) = \frac{e^{\eta-\eta^\prime}}{5} \begin{pmatrix} 3 && 2 \\ 3 && 2 \end{pmatrix} \Theta\left(\eta-\eta^\prime\right)
 - \frac{e^{-3(\eta - \eta^\prime) / 2}}{5} \begin{pmatrix} -2 && ~~2 \\ ~~3 && -3 \end{pmatrix} \Theta\left(\eta-\eta^\prime\right)\, ,
\label{eq2.9}
\ee
where the decaying mode in the second term becomes unimportant in the limit of large $\eta-\eta^\prime$.

\subsection{Classical evolution for stochastic initial conditions and representation as a generating functional}
\label{sec2.2}

The task is to solve eq. (\ref{eom}) for stochastic initial conditions corresponding to a given primordial spectrum. To this
end, we recall now basic elements of a formulation~\cite{Max1} of the problem in terms of generating functionals.
One starts by introducing an auxiliary field $\chi$, in order to 
construct an action whose extremum gives eq.\ (\ref{eom}),
\be
\begin{split}
S[\phi,\chi]= \int d\eta \Bigg[ 
&\int d^3k \,\chi_a(-\bk,\eta)\left(\delta_{ab} 
\partial_\eta+\Omega_{ab}\right) \phi_b( {\bf k},\eta) 
\Bigg.
\label{action} \\
&- \Bigg.  
 \int d^3k\, d^3 p \, d^3 q \, \delta^{(3)}(\bk-\bp-\bq)
\gamma_{abc}(\bk, \bp,\bq)
\chi_a(-\bk,\eta)\phi_b(\bp,\eta)
\phi_c( \bq ,\eta) \Bigg] \,.
\end{split} \ee
The evolution of $\phi$ is classical, so that the probability 
of its evolution to $\phi(\eta_f)$ starting 
from a given initial condition $\phi(0)$ corresponds to a delta
functional. Here and in what follows, we set for notational simplicity the initial time to $\eta_0 = 0$. This probability can be 
given a functional-integral representation as
\be
P[ \phi(\eta_f);\, \phi(0)] = {\cal N} \int {\cal D}''\phi\,{\cal D} 
\chi \,e^{i S[\phi,\chi]}\,,
\label{funct} \ee
with the double prime denoting that 
the field $\phi$ is kept fixed at the initial and final times.
Supplementing this expression by sources and integrating over the final fields $\phi_a(\eta_f)$, 
one obtains a generating functional for fixed initial conditions $\phi_a(0)$, 
\be
	Z[J_a, K_b; \phi(0)] = \int {\cal D}\phi_a(\eta_f) \int {\cal D}''\phi_a\, {\cal D} \chi\,  \exp\left[ i\,S[\phi,\chi] + J_a\, \phi_a + K_b\, \phi_b  \right]\, .
\ee
Stochastic initial conditions are characterized via a probability distribution $w[\phi_a(0),C]$. In the simplest case, $w$ is specified in terms of two-point correlations only, 
that is in terms of  the initial power spectrum $P^0$,
\be
	w[\phi_a(0),C] = \exp\left[-\frac{1}{2} \int d{\bf k}\, \phi_a({\bf k},0)\, C_{ab}(k)\, \phi_b(-{\bf k},0) \, \right] \, ,\qquad C^{-1}_{ab}(k) = P^0_{ab}(k)\, .
\ee
The corresponding generating functional for stochastic initial conditions is
\be
	Z[J_a, K_b; P^0] = \int {\cal D}\phi_a(0)\,  w[\phi_a(0),C_{ab}] \,   Z[J_a, K_b; \phi(0)] \, .
	\label{eq2.14}
\ee
This is the central object of our discussion. All physical information about cosmological perturbations $\delta_\bk$, $\theta_{\bk}$, their correlation
and their dependence on time or redshift can be obtained from suitable functional derivatives of (\ref{eq2.14}). 

For the linear theory ($\gamma_{abc}\to 0$), one finds \cite{Max1}
\be
\begin{split}
Z_0&[J,\, K;\,P^0]
=\exp\Bigg\{
-  \int d\eta d\eta^\prime\, d^3 k \Bigg.
\label{z0} \\
&\Bigg. \left[\frac{1}{2} J_a(-\bk,\eta) 
P^L_{ab}(\bk;\eta, \eta^\prime) J_b(\bk,\eta^\prime)
+i J_a(-\bk,\eta) 
g_{ab}(\bk,\eta,\eta^\prime) 
K_b(\bk,\eta^\prime)\right]\Bigg\}\,,
\end{split}
\ee
where $ P^L_{ab}$ is the linear power spectrum
\be
P^L_{ab}(\bk;\eta,\eta^\prime) = g_{ac}(\bk,\eta,0) g_{bd}(-\bk,\eta^\prime,0) 
P^0_{cd}(\bk)\, ,
\label{linearps}
\ee
and $g_{ab}(\bk,\eta,\eta^\prime)$ the retarded linear propagator of eq. (\ref{linpropa}).
Formally, the generating functional of the interacting theory can then be written as
\be
Z[J,\, K;\,P^0]=
 \exp\left\{-i\int d\eta\, \gamma_{abc}\left( 
\frac{-i\delta}{\delta K_a}\frac{-i\delta}{\delta J_b}
 \frac{-i\delta}{\delta J_c}
\right)\right\} \, Z_0[J,\, K;\,P^0]\,,
\label{interz} \ee
where we have suppressed the momentum dependence and the
corresponding integrations. 

More directly, one obtains from \eqref{eq2.14}
\be
\begin{split}
Z[J, \, K;\,P^0] 
=\int {\cal D} & \phi {\cal D} \chi 
\exp \bigg\{ -\frac{1}{2}   \,  \phi_a(0) C_{ab} 
\phi_b(0) 
\label{genf2} \\
&+ 
i \int  d\eta \left[ \chi_a ( \delta_{ab}\partial_\eta+\Omega_{ab}) \phi_b  -  
\gamma_{abc} \chi_a \phi_b \phi_c + 
J_a \phi_a +  K_b \chi_b\right]\bigg\} \,.
\end{split}
 \ee
 To all orders in perturbation theory, this expression is equivalent to 
\be
\begin{split}
Z[J, \, K;\,P^0] 
=\int {\cal D} & \phi {\cal D} \chi 
\exp \bigg\{ -\frac{1}{2}  \,  \chi_a(0) P^0_{ab} \chi_b(0) 
\label{genf} \\
&+ 
 i \int  d\eta \left[ \chi_a ( \delta_{ab} {\partial}_\eta+\Omega_{ab}) \phi_b  -  
\gamma_{abc} \chi_a \phi_b \phi_c + 
J_a \phi_a +  K_b \chi_b\right]\bigg\} \,, 
\end{split}
 \ee
Eq.(\ref{genf}) illustrates that the field $\chi$ is more than an auxiliary book-keeping device. As the spectrum couples to
$\chi_b$ only, this field carries information about the statistics of initial conditions.

\subsection{The generating functional W}
\label{sec2.3}

One can now define the generator of connected Green's functions 
\be
W[J,K; P^0]=-i \log Z[J,K; P^0].
\label{eq:DefWFunctional}
\ee
The full power spectrum
$P_{ab}$ and the propagator $G_{ab}$ 
can be obtained through second functional derivatives of 
$W$,
\begin{eqnarray}
\left. \frac{\delta^2 W}{\delta J_a(-\bk,\eta)\, \delta J_b(\bk',\eta^\prime)}\right|_{J,\,K=0} 
&=& i \delta({\bk}-{\bk}') \, P_{ab}(\bk,\eta,\eta^\prime)\,,\nonumber\\
\left. \frac{\delta^2 W}{\delta J_a(-\bk,\eta)\, \delta K_b(\bk',\eta^\prime)}\right|_{J,\,K=0} 
&=& - \delta({\bk}-{\bk}') G_{ab}^R(\bk,\eta,\eta^\prime) \, , \nonumber\\
\left. \frac{\delta^2 W}{\delta K_a(-\bk,\eta)\, \delta J_b(\bk',\eta^\prime)}\right|_{J,\,K=0} 
&=& - \delta({\bk}-{\bk}') G_{ab}^A(\bk,\eta,\eta^\prime) \, , \nonumber\\
\left. \frac{\delta^2 W}{\delta K_a(-\bk,\eta)\, \delta K_b(\bk',\eta^\prime)}\right|_{J,\,K=0} 
&=& 0\, .
\label{funder} \end{eqnarray}
The retarded propagator is subject to the boundary condition\footnote{This boundary condition does not follow directly from the continuum version of the formalism discussed above. It can be added as an additional constraint or implemented in a formalism with a discretized time coordinate.} $G^R_{ab}(\bk,\eta,\eta^\prime)=0$ for $\eta<\eta^\prime$ and one has
\be
G^A_{ab}(\bk, \eta,\eta^\prime) = G^R_{ba}(-\bk,\eta^\prime,\eta).
\label{eq:relationGRGA}
\ee
Higher order functional derivatives of $W$ with respect to $J$ yield connected correlation functions (or cumulants) of the field $\phi$ while mixed derivatives with respect to $J$ and $K$ yield connected correlations of $\phi$ and $\chi$.

\subsection{The effective action $\Gamma$ and its physical interpretation}
\label{sec2.4}

The effective action is the Legendre transform with respect to both source fields,
\be
\Gamma[\phi,\chi; P^0] = \int d\eta d^3 \bk \left\{ J_a \phi_a + K_b \chi_b \right\} - W[J,K;P^0], \label{eq:GammaLegendre}
\ee
where
\be
\phi_a(\bk,\eta) = \frac{\delta}{\delta J_a(\bk, \eta)} W, \quad\quad\quad \chi_b(\bk,\eta) = \frac{\delta}{\delta K_b(\bk, \eta)} W,
\ee
are now expectation values. We use the same notation as for the integration variables in \eqref{genf} for simplicity.

Interestingly, for vanishing initial spectrum,  the functional integral over $\chi$ in \eqref{genf} can be performed and leads to a functional $\delta$-function which constrains $\phi$ to be a solution of the equation of motion (with boundary condition $\phi_a(0)=0$)
\be
(\delta_{ab}\partial_\eta+\Omega_{ab}) \phi_b - \gamma_{abc} \phi_b \phi_c = - K_a\, ,\qquad \hbox{for $P^0 \to 0$.}
\label{eom2}
\ee
Using this in \eqref{eq:DefWFunctional} together with \eqref{eq:GammaLegendre} yields then (up to an irrelevant additive constant)
\be
\Gamma[\phi,\chi; 0] = - S[\phi,\chi]\, ,\qquad \hbox{for $P^0 \to 0$.}
\ee
For the more general (and more realistic) case of $P^0\neq 0$, the effective action is modified compared to the microscopic action by the effect of initial state fluctuations. 

The definition of $\Gamma$ as a Legendre transform implies that its second functional derivative is inverse to the second functional derivative of $\Gamma$ in \eqref{funder},\footnote{We display these relations for vanishing sources but they hold in similar form also for general $J$ and $K$.}
\begin{eqnarray}
\left. \frac{\delta^2 \Gamma}{\delta \phi_a(-\bk,\eta)\, \delta \phi_b(\bk',\eta^\prime)}\right|_{J,\,K=0} 
&=& 0 \; , \nonumber \\
\left. \frac{\delta^2 \Gamma}{\delta \phi_a(-\bk,\eta)\, \delta \chi_b(\bk',\eta^\prime)}\right|_{J,\,K=0} 
&=& - \delta({\bk}-{\bk}') D^A_{ab}(\bk,\eta,\eta^\prime) \, , \nonumber\\
\left. \frac{\delta^2 \Gamma}{\delta \chi_a(-\bk,\eta)\, \delta \phi_b(\bk',\eta^\prime)}\right|_{J,\,K=0} 
&=& - \delta({\bk}-{\bk}') D^R_{ab}(\bk,\eta,\eta^\prime) \, , \nonumber\\
\left. \frac{\delta^2 \Gamma}{\delta \chi_a(-\bk,\eta)\, \delta \chi_b(\bk',\eta^\prime)}\right|_{J,\,K=0} 
&=& - i \delta({\bk}-{\bk}')  H_{ab}(\bk,\eta,\eta^\prime) \, .
\label{funder2} 
\end{eqnarray}
We have introduced here the derivative operator $D^R_{ab}$ that is inverse to the retarded correlation function such that
\be 
\int d\eta^\prime \; D^R_{ab}(\bk,\eta,\eta^\prime) G^R_{bc}(\bk,\eta^\prime,\eta^{\prime\prime}) = \delta_{ac} \, \delta(\eta-\eta^{\prime\prime})\,  ,
\label{3.15}
\ee
and similarly for $D^A_{ab}$. The inverse retarded propagator $D^R_{ab}(\bk,\eta,\eta^\prime)$ is causal, such that $D^R_{ab}(\bk,\eta,\eta^\prime)=0$ for $\eta < \eta^\prime$. According to \eqref{eq:relationGRGA} one has
\be
D^A_{ab}(\bk, \eta,\eta^\prime) = D^R_{ba}(-\bk,\eta^\prime,\eta).
\label{eq:relationDRDA}
\ee
The object $H_{ab}$ is defined by
\begin{eqnarray}
H_{ab}(\bk,\eta,\eta^{\prime}) & = & \int d \eta^{\prime\prime} d \eta^{\prime\prime\prime} \; D^R_{ac}(\bk,\eta,\eta^{\prime\prime}) P_{cd}(\bk,\eta^{\prime\prime},\eta^{\prime\prime\prime}) D^A_{db}(\bk,\eta^{\prime\prime\prime},\eta^{\prime}) \nonumber \\ 
& = & \int d \eta^{\prime\prime} d \eta^{\prime\prime\prime} \; D^R_{ac}(\bk,\eta,\eta^{\prime\prime}) D^R_{bd}(-\bk,\eta^{\prime},\eta^{\prime\prime\prime}) P_{cd}(\bk,\eta^{\prime\prime},\eta^{\prime\prime\prime})  .
\label{eq:defh}
\end{eqnarray}
For some purposes it is useful to decompose
\be
H_{ab}(\bk,\eta,\eta^\prime) = P^0_{ab}(\bk) \delta(\eta) \delta(\eta^\prime) + \Phi_{ab}(\bk,\eta,\eta^\prime) \, .
\label{3.19}
\ee
For vanishing non-linear terms $\gamma_{abc}\to 0$, one has formally $\Phi_{ab} \to 0$. 

\subsection{Equations of motion for expectation values}

The effective action $\Gamma[\chi,\phi]$ is subject to ``renormalized field equations''
\begin{eqnarray}
\frac{\delta}{\delta \phi_a(\bx,\eta)} \Gamma[\phi,\chi] & = & J_a(\bx,\eta), \label{eq:fieldEqnGammaGeneral1} \\
\frac{\delta}{\delta \chi_a(\bx,\eta)} \Gamma[\phi,\chi] & = & K_a(\bx,\eta), \label{eq:fieldEqnGammaGeneral2}
\end{eqnarray}
that follow directly from the definition of the effective action as a Legendre transform \eqref{eq:GammaLegendre}. In practice, these equations are most relevant for vanishing source fields $J=K=0$. In the case of $K=0$ it follows from the partition function \eqref{genf} that the expectation value of the auxiliary field vanishes, $\chi=0$. This is therefore always the solution of the field equation \eqref{eq:fieldEqnGammaGeneral1}. However, the second equation in \eqref{eq:fieldEqnGammaGeneral2} yields a non-trivial evolution equation for the expectation value $\phi(\bk,\eta)$ which reads at vanishing source
\be
\frac{\delta}{\delta \chi_a(\bx,\eta)} \Gamma[\phi,\chi]{\Big |}_{\chi=0} = 0.
\ee
While the variation of the microscopic action $S[\phi,\chi]$ yields the microscopic equations of motion, the effective actions of motion for field expectation values that follow from the variation of $\Gamma[\phi,\chi]$ are similar in form but can include additional terms not present in the microscopic equations. These additional terms are due to the initial state fluctuations. We will discuss this in more detail below. 

From this discussion it follows that terms in $\Gamma[\phi,\chi]$ that are linear in $\chi$ but contain one or more powers of $\phi$ parametrize the effective field equation for $\phi$. On the other side, terms involving only the field $\chi$ parametrize statistical information. For example, the power spectrum can be obtained from inverting eq. \eqref{eq:defh},
\be
P_{ab}(\bk,\eta,\eta^\prime) \delta(\bk-\bk^\prime) = \int d\eta^{\prime\prime} d\eta^{\prime\prime\prime} G^R_{ac}(\bk,\eta,\eta^{\prime\prime}) G^R_{bd}(-\bk^\prime,\eta^\prime,\eta^{\prime\prime\prime}) i \frac{\delta^2 \Gamma[\phi, \chi]}{\delta \chi_c(-\bk,\eta^{\prime\prime}) \delta \chi_d(\bk^\prime,\eta^{\prime\prime\prime})},
\label{eq3.20}
\ee
where all objects on the right hand side have to be evaluated on the solution of the field equations \eqref{eq:fieldEqnGammaGeneral1},
\eqref{eq:fieldEqnGammaGeneral2} for vanishing sources.
Measurable higher-order correlation functions  (such as the bispectrum) can be obtained analogously from higher (third) functional derivatives of 
$W$ expressed as the (inverse) Legendre transform of $\Gamma$. In appendix~\ref{appa}, we collect simple facts about functional
derivatives of $W$ and $\Gamma$ in a compact notation, and we provide an explicit expression for the bispectrum in (\ref{eqa10}).

\section{Coarse grained effective action and functional RG equation}
\label{sec3}

In this section, we construct a coarse-grained effective action $\Gamma_k$ that defines the effective dynamics of long-wavelength perturbations (i.e. perturbations
of wavenumber $|\bq| < k$), by resumming all effects of stochastic initial fluctuations of wavenumber $|\bq| > k$. Our aim is then to understand how the effective 
coarse-grained field equations change with the coarse-graining scale $k$.

There are several motivations for such a program: First, while field equations derived from the effective action $\Gamma$ differ from those derived from the microscopic action $S[\phi,\chi]$ by the resummation of initial state fluctuations on {\it all} scales $\bq$, one may be interested in physical applications that
follow explicitly the dynamics of all modes $|\bq| < k$ and that therefore resum only the effects from the UV part of the initial spectrum. 
Second, as explained in section~\ref{sec2.1}, the validity of a fluid dynamical description~(\ref{eom}) is limited to sufficiently small wavenumbers, say $|\bq| < k_m$. However, because of the non-linear terms in ~(\ref{eom}), long-wavelength fluid dynamic modes interact also with 
initial fluctuations of wavenumber $|\bq| > k_m$. As this UV part of the initial fluctuations lies outside the validity of a fluid dynamic description, it is a priori unclear whether the effective dynamics of long-wavelength modes up to scale $k_m$ is of fluid dynamic form, or whether
it will be dominated by contributions from the UV sector of initial fluctuations. An explicit equation for the exact effective action $\Gamma_k$ for the IR dynamics of modes $|\bq| < k$ may clarify this point. Third, we have argued in a recent work that, at least for 
some range of scales $k_m$, the interactions of fluid dynamic modes with UV fluctuations can be expected to be accounted for by a description of the coarse-grained system as a non-ideal viscous fluid in terms of a cutoff dependent effective viscosity and effective sound velocity that depends on the initial spectrum. 
The construction of the exact effective action $\Gamma_k$ given here will clarify the theoretical basis for this phenomenologically successful proposal. Fourth, 
 control over the flow of $\Gamma_k$ with coarse-graining scale $k$  will inform us about how to match best the effective coarse-grained dynamical description to some more fundamental dynamics at the scale $k_m$. 

We note that, in principle, the exact resummation of UV physics may be very complicated, so that $\Gamma_k$ may contain a large number of interaction terms that are not contained in the microscopic action $S[\phi,\chi]$.

\subsection{Construction of a $k$-dependent effective action and its exact flow equation}

We construct a coarse-grained version of the effective action for which only statistical fluctuations at small scales are taken into account, while those at large scales are suppressed. This is done by modifying the initial power spectrum so that it includes only modes with wavevector $|\bq|$ larger than 
the coarse-graining scale $k$,
\be
P^0_k(\bq)=P^0(\bq)\, \Theta(|\bq|-k).
\label{thet} 
\ee
Here, we use a sharp cutoff at the scale $k$, but more smooth regulator functions could be used as well. 
While we started in section~\ref{sec2} from the same set-up as ref. \cite{Max1}, it is at this point that our analyses start
to differ. First, we coarse grain over UV degrees of freedom, while ref.~\cite{Max1} focused on IR modes. Second, 
the formulation of ref. \cite{Max1} employed mainly the generating function of connected Green's functions,
while we shall find several technical advantages in formulating the problem as a RG flow of the effective action.

Because the modified spectrum \eqref{thet} vanishes for $|\bq| < k$, the corresponding modes $\phi_a(\bq, \eta)$ are kept fixed in the initial conditions at $\eta=0$. Only the modes with $|\bq| > k$ are effectively fluctuating. When the regulator scale $k$ is lowered, the effect of fluctuations with smaller and smaller initial wavevectors is gradually taken into account. 

The modified spectrum (\ref{thet}) defines a coarse-grained generating functional
\be
 W_k[J,K]\equiv W[J,K;P^0_k]\, .
 \ee
This functional satisfies an exact equation that determines how it is modified when new modes are 
eliminated or incorporated in the initial spectrum,
\be
\partial_k W_k[J,K] =  \frac{i}{2} \int d^3 \bq \; \partial_k \left(P_k^0\right)_{ab}({\bq}) \left\{ \chi_a(\bq,0) \chi_b(-\bq,0) -i \frac{\delta^2 W_k[J,K]}{\delta K_a(\bq,0) \delta K_b(-\bq,0)} \right\}.
\label{eq:PolchinskiEqn}
\ee 
Here $\chi_a(\bq,0)=\frac{\delta}{\delta K_a(\bq,0)} W_k[J,K]$ is the expectation value for arbitrary sources. Eq.\ \eqref{eq:PolchinskiEqn} is the analog of the Polchinski equation \cite{Polchinski:1983gv} in the present context.

We now define the flowing action by adding to the Legendre transform of $W_k[J,K]$
\be
\tilde \Gamma_k[\phi,\chi] = \int d\eta d^3 \bk \left\{ J_a \phi_a + K_b \chi_b \right\} - W_k[J,K], \label{eq:GammaLegendreScale}
\ee
the difference between the full and the modified initial spectrum,
\be
\Gamma_k[\phi,\chi] = \tilde \Gamma_k[\phi,\chi] - \frac{i}{2} \int d^3\bq \; \chi_a(\bq,0) \left( P^0_{ab}(\bq) - (P^0_k)_{ab}(\bq) \right) \chi_b(-\bq,0).
\label{eq3.24}
\ee
Let us comment on this step in more detail: in general, for fixed expectation values $\phi$, $\chi$ the sources $J$, $K$  can depend on the coarse-graining scale $k$, i.e. $\partial_k J$, $\partial_k K$ do not need to vanish. 
Nevertheless, because of properties of the Legendre transform, the $k$-dependence of $\tilde \Gamma_k[\phi,\chi]$ is set by the $k$-dependence of the generating functional for connected Green's functions, $\partial_k W_k[J,K] = - \partial_k \tilde\Gamma_k[\phi,\chi]$. 
The first term on the right hand side of (\ref{eq:PolchinskiEqn}) is the disconnected part of the two-point Green's function. This is a trivial
contribution to the derivative $\partial_k W_k[J,K]$ that arises simply from changing the spectrum at initial time $\eta = 0$. 
Subtracting from $\tilde \Gamma_k$ by hand the second term in (\ref{eq3.24}) is convenient, since it removes this trivial part of the scale dependence. 
Using $\partial_k W_k[J,K] = - \partial_k \tilde\Gamma_k[\phi,\chi]$ one finds then from \eqref{eq:PolchinskiEqn}
\be
\partial_k \Gamma_k[\phi,\chi] =  \frac{1}{2} \, \text{Tr} \left\{ \left( \Gamma_k^{(2)}[\phi,\chi] - i \left(P^0_k - P^0\right) \right)^{-1}  \partial_k P^0_k \right\} 
\label{eq:WetterichEqn}
\ee
We used here a symbolic notation where $P^0_k$ and $P^0$ are matrices with non-zero entries in the $\chi$-$\chi$ block and at the initial time $\eta=0$, only. Eq.\ \eqref{eq:WetterichEqn} is the analog of the Wetterich equation in the present context \cite{Wetterich:1992yh}. It describes the flow of $\Gamma_k[\phi,\chi]$ as a functional under the process of integrating out more and more initial state fluctuations.

Analogs of equation \eqref{eq:WetterichEqn} are frequently used in other contexts of quantum and statistical field theory. Applications of the functional renormalization-group formalism range from critical phenomena to quantum gravity, for reviews see \cite{Aoki:2000wm, Bagnuls:2000ae, Berges:2000ew, Polonyi:2001se, Pawlowski:2005xe, Gies:2006wv, Delamotte:2007pf, Rosten:2010vm, Litim:2011cp, Reuter:2012id}.

\subsection{Flow equations for correlation functions}
In complete analogy to other applications of the functional renormalization 
group, flow equations for (inverse) propagators, effective vertices and noise terms can be obtained from $\Gamma_k[\phi_a,\chi_b]$ by taking functional derivatives. For instance, under the process of
integrating out more and more initial state fluctuations, the flow of the inverse propagators $D^R_{ab,k}(\bq,\eta,\eta^\prime)$ is obtained by combining eqs.~ (\ref{eq:WetterichEqn}) and (\ref{funder2}),
\begin{eqnarray}
- \delta({\bq}-{\bq}') \partial_k D^R_{ab,k}(\bq,\eta,\eta^\prime) 
&=&
\left. \frac{\delta^2 \partial_k \Gamma_k}{\delta \chi_a(-\bq,\eta)\, \delta \phi_b(\bq',\eta^\prime)}\right|_{J,\,K=0} 
 \nonumber\\
&=&  \frac{1}{2} \, \text{Tr} \left\{ W_k^{(2)}  \frac{\delta \Gamma^{(2)}_k}{\delta \chi_a(-\bq,\eta)}   W_k^{(2)} 
\frac{\delta \Gamma^{(2)}_k}{\delta \phi_b(\bq',\eta^\prime)}  W_k^{(2)}  \partial_k P^0_k \right\} 
 \nonumber\\
&& +  \frac{1}{2} \, \text{Tr} \left\{ W_k^{(2)} \frac{\delta \Gamma^{(2)}_k}{\delta \phi_b(\bq',\eta^\prime)} W_k^{(2)} \frac{\delta \Gamma^{(2)}_k}{\delta \chi_a(-\bq,\eta)}   
 W_k^{(2)}  \partial_k P^0_k \right\} 
 \nonumber\\
&& - \frac{1}{2} \, \text{Tr} \left\{ W_k^{(2)}    \frac{\delta^2  \Gamma^{(2)}_k}{\delta \chi_a(-\bq,\eta)\, \delta \phi_b(\bq',\eta^\prime)}
 W_k^{(2)}  \partial_k P^0_k \right\} \, .
\label{flowow} 
\end{eqnarray}
Similar expression can be written for the inverse advanced propagator $D^A_{ab,k}(\bq,\eta,\eta^\prime)$, and the flow of the evolved spectrum 
$H_{ab,k}(\bq,\eta,\eta^\prime)$. Here, we made use of the relation between the second order functional derivatives of the effective action and $W_k$, 
\begin{equation}
W_k^{(2)}[J,K] = \left( \tilde \Gamma_k^{(2)}[\phi,\chi] \right)^{-1} 
= \left( \Gamma_k^{(2)}[\phi,\chi] - i \left(P^0_k - P^0\right) \right)^{-1} \, .
\end{equation}
The right-hand side of \eqref{flowow} is evaluated for vanishing sources.  $W_k^{(2)}$ defines the full spectrum and the full propagator of the coarse-grained theory, for which initial state 
fluctuations are integrated out above the scale $k$, see eq.~\eqref{funder}. The functional derivatives $ \delta \Gamma^{(2)}_k/\delta \chi_a$, $ \delta \Gamma^{(2)}_k/\delta \phi_b$ in equation 
\eqref{flowow} are exact three-point vertices of this coarse-grained theory and the second functional derivative of $\Gamma^{(2)}_k$ in the last line of \eqref{flowow} is an exact four-point vertex. 

\begin{figure}
\centering
\includegraphics[width=0.6\textwidth]{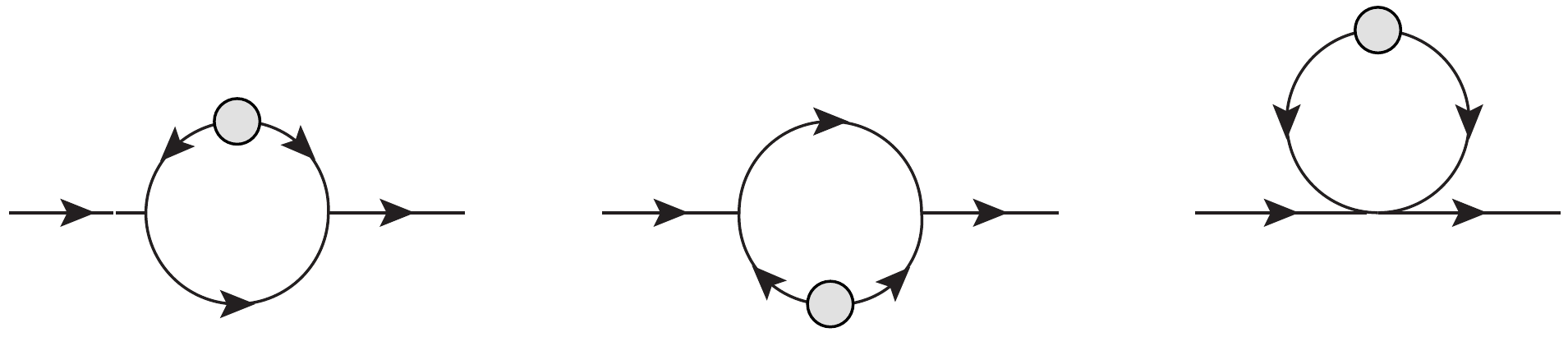}
\includegraphics[width=0.6\textwidth]{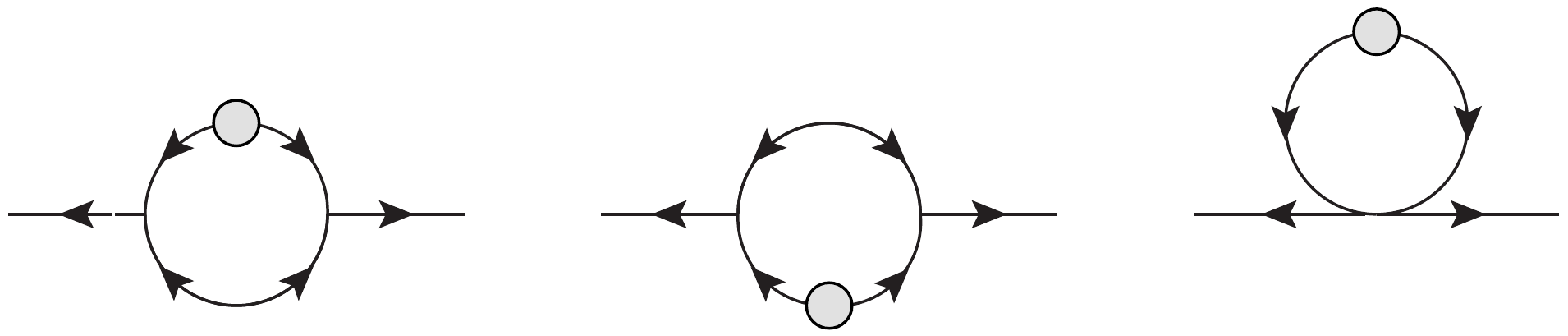}
\caption{Graphical representation of the loop expressions in eq.\ \eqref{flowow} that describes the renormalization-group modifications of the inverse retarded propagator $D_{ab,k}^R(\bq,\eta,\eta^\prime)$ according to \eqref{flowow} (upper diagrams). The filled circles denote scale derivatives of the modified initial density spectrum $\partial_k P_k^0$. The lower diagrams represent analogous loop expressions for the flow of the correlation function $H_{ab,k}(\bq,\eta,\eta^\prime)$ as defined in \eqref{eq:defh} (flow equation not shown).}
\label{fig:FlowDR}
\end{figure}

As we explain now, the three terms in \eqref{flowow} have a simple representation in terms of one-loop expressions. Since $P^0_k$ is a matrix with non-zero entries in the $\chi$-$\chi$ block
and at initial times only, and since $W_k^{(2)}$ vanishes in the $K$-$K$ block for vanishing sources, the factors $W_k^{(2)}  \partial_k P^0_k  W_k^{(2)} $ in ~\eqref{flowow} correspond to
a term  $\partial_k P^0_k$ evolved from initial time with exact propagators up to times $\eta$, $\eta'$ respectively. In the last term of ~\eqref{flowow} this factor runs in a tad-pole loop, in the
other two terms it connects to two exact three-point vertices and the loop is closed by insertion of another factor $W_k^{(2)} $. For a modified power spectrum of the form \eqref{thet},
one has $\partial_k P^0_k(\bar\bq) =  P^0(\bar\bq) \, \delta\left(|\bar\bq|-k \right)$. Therefore, the flow of the exact inverse propagator $D^R_{ab,k}$ at scale $k$ is determined by the contribution of the 
momentum shell $|\bar\bq| = k$ of the internal loop integral. A graphical representation of eq.\ \eqref{flowow} is shown in figure \ref{fig:FlowDR}.

\section{Connecting the functional RG to standard perturbation theory}
\label{sec4}
The solution of the flow equation \eqref{eq:WetterichEqn} is equivalent to an infinite series of perturbative contributions to the effective action 
$\Gamma[\phi_a,\chi_b]$. It is typically difficult to find exact solutions of this flow equation, since such solutions would correspond to a complete 
summation of perturbation theory to all orders. Instead, one has to take recourse to constructing approximate solutions. This can be done for instance 
by summing perturbative expressions to a certain loop order. Alternatively, one can construct approximate solutions 
based on truncations. In this section, we follow the first approach. We present explicit applications and discuss their physical relevance.  
An application following the latter approach will be given in section \ref{sec5}.

For very large coarse-graining scale $k$, the modified initial spectrum $P^0_k(\bq)$ vanishes in the relevant regime of wavevectors $\bq$ and one has the limiting behaviour
\be
\lim_{k\to \infty} \Gamma_k[\phi,\chi] = - S[\phi,\chi] - \frac{i}{2} \int d^3 \bq \; \chi_a(\bq,0) P^0_{ab}(\bq) \, \chi_b(-\bq,0)\, .
\label{pertinitdata}
\ee
In the following, we shall often introduce a very large, fundamental scale $\Lambda$, assuming that $\Gamma_\Lambda$ is given 
by the limit (\ref{pertinitdata}). We can then exploit that at scale $\Lambda$, the RG evolution is given by the microscopic action on the right
hand side of (\ref{pertinitdata}), i.e., by standard cosmological perturbation theory.

\subsection{Iterative solution of RG flow}

To solve the RG flow  \eqref{eq:WetterichEqn} with \eqref{pertinitdata} as initial value, it is useful to introduce a formal $k$-derivative $\tilde\partial_k$
that acts only on the modified initial spectrum $P_k^0$, such that
\be
\partial_k \Gamma_k[\phi,\chi] = \frac{i}{2} \tilde \partial_k \, \text{Tr} \left\{ \ln \left( \Gamma_k^{(2)}[\phi_a,\chi_b] - i \left(P^0_k - P^0\right) \right) \right\} ,
\label{eq:WetterichEqnFormalScaleDerivative}
\ee
In the evolution from a very large scale $\Lambda$ to $k$,
the effective action will then receive a correction  $\Delta \Gamma_{k,\Lambda}[\phi,\chi]$, 
\begin{equation}
\Gamma_k[\phi,\chi] = \Gamma_\Lambda[\phi,\chi] + \Delta \Gamma_{k,\Lambda}[\phi,\chi]\, .
\end{equation} 
The iterative solution is based on the assumption that $\Delta \Gamma_{k,\Lambda}[\phi,\chi]$ is small in a formal sense. This is in particular the case when $k$ and $\Lambda$ are 
close to each other. To lowest approximation, one can then drop the term $\Delta \Gamma_{k,\Lambda}[\phi,\chi]$ on the right hand side of \eqref{eq:WetterichEqnFormalScaleDerivative} and obtain
\begin{equation}
\partial_k \Gamma_k[\phi,\chi] = \frac{i}{2} \tilde \partial_k \, \text{Tr} \left\{ \ln \left( \Gamma_\Lambda^{(2)}[\phi,\chi] - i \left(P^0_k - P^0\right) \right) \right\}\, .
\label{eq:WetterichEqnFormalScaleDerivative2}
\end{equation}
The formal scale derivative $\tilde \partial_k$ can now be exchanged for a conventional derivative $\partial_k$, and the resulting expression can be integrated directly with respect to $k$,
\begin{equation}
\Gamma_k[\phi,\chi] = \Gamma_\Lambda[\phi,\chi] + \frac{i}{2} \, \text{Tr} \left\{ \ln \left( \Gamma_\Lambda^{(2)}[\phi,\chi] - i \left(P^0_k - P^0\right) \right) - \ln \left( \Gamma_\Lambda^{(2)}[\phi,\chi] - i \left(P^0_\Lambda - P^0\right) \right) \right\}\, .
\label{eq:WetterichEqnIterativeSolution}
\end{equation}
The term on the right hand side can be seen as the effective action $\Gamma_k[\phi,\chi]$ in an approximation where it differs from $\Gamma_\Lambda[\phi,\chi]$ by a one-loop term. Using this term as the first-order approximation for $\Delta \Gamma_{k,\Lambda}[\phi,\chi]$ leads to the second-order solution (the two-loop term) and so on.

\subsection{Retarded self-energy and propagator at one-loop}
\label{sec4.2}

As a first application of the RG flow of the effective action $\Gamma_\Lambda$ for cosmological perturbations, we show here how the scale dependence of the retarded propagator can be derived in this framework. We start from the inverse $D^R_{k,ab}(\bq,\eta,\eta^\prime)$ of the full retarded propagator  at coarse-graining scale $k$  which is given by the second functional derivative 
\begin{eqnarray}
\left. \frac{\delta^2 \Gamma_k}{\delta \chi_a(-\bq,\eta)\, \delta \phi_b(\bq',\eta^\prime)}\right|_{J,\,K=0} 
&=& - \delta({\bq}-{\bq}') D^R_{k,ab}(\bq,\eta,\eta^\prime)  \nonumber\\
&=& -\, \delta({\bq}-{\bq}') \left( \delta_{ab} \, \partial_\eta + \Omega_{ab}(\bq,\eta) \right)\, \delta\left(\eta - \eta^\prime \right)\nonumber \\
&&  -\, \Sigma_{k, ab}(\bq,\eta;\bq',\eta')
\, .
\label{LambdaInverseProp} 
\end{eqnarray}
The scale-dependent `self-energy' correction $\Sigma_{ab,k}(\bq,\eta;\bq',\eta')$ to $D^R_{ab,k}(\bq,\eta,\eta^\prime)$ takes then the form 
\begin{eqnarray}
	&&\Sigma_{k, ab}(\bq,\eta;\bq',\eta') = 
	\frac{-i}{2} \left( 
 \frac{\delta^2  \text{Tr} \left\{ \ln \left( \Gamma_\Lambda^{(2)}[\phi,\chi] - i \left(P^0_k - P^0\right) \right)  \right\}   
}{\delta \chi_a(-\bq,\eta)\, \delta \phi_b(\bq',\eta^\prime)}    - \frac{\delta^2  \text{Tr} \left\{ \ln \left( \Gamma_\Lambda^{(2)}[\phi,\chi]  - i \left(P^0_\Lambda - P^0\right)\right)  \right\}   
}{\delta \chi_a(-\bq,\eta)\, \delta \phi_b(\bq',\eta^\prime)} \right)\, ,\nonumber\\
\label{eq4.9}
\end{eqnarray}
where the functional derivatives on the right hand side give rise to two different types of terms, 
\begin{eqnarray}
\frac{i}{2}
 \frac{\delta^2  \text{Tr} \left\{ \ln \left( \Gamma_\Lambda^{(2)}[\phi,\chi] - i \left(P^0_k - P^0\right) \right)  \right\}   
}{\delta \chi_a(-\bq,\eta)\, \delta \phi_b(\bq',\eta^\prime)}    
&=& \frac{-i}{2} \text{Tr} \left\{ W_{\Lambda,k}^{(2)}  \frac{\delta \Gamma^{(2)}_\Lambda}{\delta \chi_a(-\bq,\eta)}   W_{\Lambda,k}^{(2)} 
\frac{\delta \Gamma^{(2)}_\Lambda}{\delta \phi_b(\bq',\eta^\prime)}  \right\} 
 \nonumber\\
 && + \frac{i}{2} \text{Tr} \left\{ W_{\Lambda,k}^{(2)}  \frac{\delta \Gamma^{(2)}_\Lambda}{\delta \chi_a(-\bq,\eta)\, \delta \phi_b(\bq',\eta^\prime)}    \right\} \, .
 \label{eq4.10}
\end{eqnarray}
We recall that we choose $\Lambda$ so large that $\Gamma_\Lambda$ can be approximated by the right hand side of (\ref{pertinitdata}).
All terms entering the right hand side of \eqref{eq4.10} can therefore be expressed perturbatively. In particular, since  
the second functional derivatives of $\Gamma_\Lambda[\phi,\chi]$ are the bare leading-order expressions of standard perturbation theory, given 
by writing \eqref{funder2} with the free inverse propagators and with the free spectrum $H_{ab}(\bq,\eta,\eta') \to \left(P^0\right)_{ab}(\bq) \delta(\eta)\, \delta(\eta')$ on the right and side, one finds
\begin{eqnarray}
	&&W_{\Lambda,k}^{(2)}\left(a,\bq,\eta;b,\bq',\eta'\right) \equiv \left( \Gamma_\Lambda^{(2)}[\phi,\chi] - i \left(P^0_k - P^0\right) \right)^{-1}\, .
	\nonumber \\
	&& \qquad \qquad = \delta\left(\bq+\bq' \right)
	 \left(
 	\begin{array}{cc}  i g^R_{aa'}(\eta,0) \left(P_k^0\right)_{a'b'}(\bq)\, g^A_{b'b}(0,\eta') \ -g_{ab}^R(\eta,\eta') \\
 				-g_{ab}^A(\eta,\eta') \ \qquad \qquad \qquad0 
 		\end{array}\right)\, .
 		\label{eq4.11}
\end{eqnarray}
Also, the first functional derivatives of $\Gamma_{\Lambda}^{(2)}$ in (\ref{eq4.10}) correspond
to perturbative three-point vertices $\gamma_{abc}$ at the fundamental UV scale $\Lambda$. For instance, for the $\phi$-$\phi$-block of the first functional derivative in (\ref{eq4.10}), one finds
\begin{eqnarray}
&& \left.  \frac{\delta \Gamma^{(2)}_\Lambda \left(c,\bp,\xi;c',\bp',\xi' \right)}{\delta \chi_a(\bq,\eta)} \right|_{\phi\phi} 
\equiv -  \frac{\delta S[\phi,\chi]}{\delta \chi_a(\bq,\eta)\, \delta \phi_c(\bp,\xi)\,  \delta \phi_{c'}(\bp',\xi')}
\nonumber\\
&& \qquad \qquad  = 2 \delta\left(\bq+\bp+\bp' \right)\, \delta\left(\eta-\xi \right)\, \delta\left(\eta-\xi' \right)\, \gamma_{acc'}\left(-\bq,\bp,\bp'\right)\, .
\label{eq4.12}
\end{eqnarray}

The second contribution on the right hand side of (\ref{eq4.10}) is of tad-pole type. It vanishes at the fundamental, microscopic scale $\Lambda$, since
there is no fundamental four-point vertex. (This would be different had we started the evolution from another scale, since effective four-point vertices can be generated in the evolution). Therefore, we have to consider only the first term on the right hand side of (\ref{eq4.10}). Using the perturbative expressions
\eqref{eq4.11} and  \eqref{eq4.12}, one finds then by explicit calculation
\begin{eqnarray}
	&&\Sigma_{k, ab}(\bq,\eta;\bq',\eta') 
= 4 \delta\left(\bq-\bq'\right) \int d{\bf r}\, 
g_{c'f}^R(\eta,0) \left[P_k^0({\bf r}) - P^0({\bf r}) \right]_{fh}\, g^A_{hd'}(0,\eta')\nonumber \\
&& \qquad \qquad  \qquad \qquad \qquad \qquad \qquad \times \gamma_{acc'}(\bq,{\bf r}+\bq,-{\bf r}) g_{cd}^R(\eta,\eta') \gamma_{dd'b}({\bf r}+\bq,{\bf r},\bq)\, .
\label{eq4.13}
\end{eqnarray}
The integrand in the loop integral of (\ref{eq4.13}) takes the form
\begin{equation}
 - P_k^0({\bf r}) + P^0({\bf r}) = P^0({\bf r})\, \Theta\left(|{\bf r}|-k\right)\, ,
 \label{eq4.16}
\end{equation}
and the integration is therefore limited to the UV modes $|{\bf r}| > k$. 

Note, that the one-loop correction to the inverse propagator \eqref{eq4.13} corresponds formally already to a resummation of infinitely many terms that contribute to the propagator (in the sense of a generalized geometric series). The resummation is done by virtue of the one-particle irreducible scheme according to the definition of the effective action $\Gamma_k$. Assuming that the self-energy correction $\Sigma_{k, ab}$ is small, one can approximate the corresponding scale-dependent correction 
$\delta G_{k,ab}$ to the full retarded propagator by
\begin{equation}
	G_{k,ab}^R(\bq,\eta,\eta') \equiv g_{ab}^R(\eta,\eta') + \delta G_{k,ab}^R(\bq,\eta,\eta')\, ,
	\label{eq4.14}
\end{equation}
where the correction to the linear propagator is of the form 
\begin{equation}
\delta G^R_{k,ab}(\bq,\eta;\bq',\eta') = - \int_0^{\eta} d\bar\eta \int_0^{\bar\eta} d\bar\eta' \,  g^R_{a\bar a}(\eta,\bar\eta) \Sigma_{k, \bar a\bar b}(\bq,\bar\eta;\bq',\bar\eta') g^{R}_{\bar b b}(\bar\eta',\eta')\, .
\label{eq4.15}
\end{equation}
The results (\ref{eq4.13}) and (\ref{eq4.15}) have several properties, on which we comment now:
\begin{enumerate}
\item
In the limit $k\to \infty$, when one does not coarse grain over any scale, $\Sigma_{k, ab}$ vanishes. This follows directly from the defining relation (\ref{eq4.9}),
and is the expected result. At the fundamental, microscopic scale, one recovers the perturbative data (\ref{pertinitdata}). Therefore, 
equation~\eqref{LambdaInverseProp} returns the free propagator for $k\to \infty$. 
\item
In the opposite limit, $k\to 0$, one has coarse grained over all scales.
In this limit, $\lim_{k\to 0} \delta G_{k,ab}^R$ is exactly the one-loop correction of the free propagator obtained in standard perturbation theory, see eq. (22) of Ref.~\cite{CrSc2}.
This agreement can be regarded as an important consistency check for our formulation. 
In general, equation (\ref{eq4.9}) for $\Sigma_{k,ab}^R$ takes the form of a one-loop expression written in 
terms of effective propagators and effective vertices at coarse-graining scale $\Lambda$. Since we chose $\Lambda$ 
to be the fundamental scale, where propagators and vertices are perturbative, we had to recover the standard one-loop result in the limit $k\to 0$.
\item
For finite values of $k$, $ \delta G_{k,ab}^R$ denotes the correction to the free propagator in the effective theory obtained from integrating over all modes larger than $k$. 
In a previous paper~\cite{viscousdm}, we matched a perturbative expression for $\delta G_{k,ab}^R$ to viscous fluid dynamics at a matching scale $k_m$ below which a fluid dynamic 
description of cosmological perturbations applies. To this end, we motivated by heuristic arguments a form of $\delta G_{k,ab}^R$, obtained from the standard
one-loop result by restricting the integration over internal loop momenta to $|\bq| > k_m$.
We then used viscous fluid dynamics as the effective theory for cosmological
perturbations at scales $|\bq| < k_m$. We obtained an improved description of large-scale structure in the range of baryon acoustic oscillations~\cite{viscousdm}. 
With the derivation of (\ref{eq4.13}), we have shown in this section how this originally
heuristic procedure emerges from the functional renormalization group. 
\item
To first approximation, the results for $\Sigma_{k, ab}$ and for $\delta G_{k,ab}^R$ are equivalent and easily derived from each other, see (\ref{eq4.15}). We note, however, 
that $\Sigma_{k, ab}$ is simpler because it does not contain any time integration. We shall profit from this simplicity in subsequent calculations. This is one of
the reasons that prompted us to formulate the RG flow for $\Gamma_k$ and not for  $\delta G_{k,ab}^R$. 
\end{enumerate}

We close this section with a more general remark:
At a sufficiently large, fundamental scale $\Lambda$,  the dynamics of cosmological perturbations is not characterized fully by the equations of motion \eqref{eom} of the two-component field 
$\phi = \left(\delta , -\theta/ \mathcal{H}\right)$, since non-linear effects such as shell crossing are not captured in this framework.  One may therefore wonder whether a solution \eqref{eq:WetterichEqnIterativeSolution} for the effective action $\Gamma_k$ at a lower scale $k$ can really be based on initial data \eqref{pertinitdata} at a very large scale $\Lambda$ that do 
not account for all known non-linear effects. 
However, the impact of shell-crossing is known to be rather small at BAO scales \cite{Pueblas:2008uv, nishimichi}, and therefore one expects that the precise form of the
microscopic action is actually not very relevant. We will explicitly confirm this expectation later on. 
Note that, if this were not the case, one would have to identify a sufficiently low starting scale $\Lambda$ at which the characterization of cosmological perturbations in terms of the 
perturbative dynamics of the two-component field $\phi = \left(\delta , -\theta/ \mathcal{H}\right)$ is justified. The effective propagators and vertices at that lower scale $\Lambda$
would contain all relevant information about the physics at scales larger than $\Lambda$. In this case, the result for (\ref{eq4.9}) would differ from that of a standard perturbative treatment, 
but our formulation in terms of the functional renormalization-group flow for $\Gamma_k$ would still apply. 

\subsection{Renormalized spectrum at one-loop}
\label{sec4.1.3}
Paralleling the calculation of the retarded propagator in section~\ref{sec4.2}, we can derive under the same assumptions from 
\begin{equation}
\left. \frac{\delta^2 \Gamma_k}{\delta \chi_a(-\bq,\eta)\, \delta \chi_b(\bq',\eta^\prime)}\right|_{J,\,K=0} 
= - i\, \delta({\bq}-{\bq}') \{ P^0_{ab}(\bq)\, \delta(\eta)\, \delta(\eta^\prime) + \Phi_{k,ab}(\bq,\eta,\eta') \} 
\label{1loopspectrum} 
\end{equation}
an explicit expression for the one-loop correction to the spectrum,
\begin{eqnarray}
	\Phi_{k,ab}(\bq,\eta,\eta') &=& 2\int d{\bf r}\, \left[ 1-\Theta(|{\bf r}|-k)\, \Theta(|{\bf r}-{\bf q}|-k)  \right]
	\nonumber \\
	&& \times\,  g_{df}^R(\eta',0) \left(P^0\right)_{fh}({\bf r})\, g_{hc}^A(0,\eta)\, \gamma_{acc'}(\bq,{\bf r},-{\bf r}+\bq)
		\nonumber \\
	&& \times\,  g_{c'\bar{f}}^R(\eta,0) \left(P^0\right)_{\bar{f}\bar{h}}({\bf r}-\bq)\, g_{\bar{h}d'}^A(0,\eta')\, \gamma_{bd'd}(-\bq,{\bf r}-\bq, {\bf r}) .
\end{eqnarray}
Writing the initial spectrum in terms of the growing mode, $\left(P^0\right)_{fh}({\bf r}) = P^0_{fh}({\bf r})\, u_f\, u_h$, we find in the long-wavelength limit $|\bq| \ll k$
\begin{eqnarray}\label{Phi1L}
	\Phi_{k}(\bq,\eta,\eta') \Bigg\vert_{|\bq| \ll k} &=& \frac{2\pi}{15}
	 \begin{pmatrix} 7 && -5 \\ -5 && 15 \end{pmatrix}
	e^{2\eta + 2 \eta'}\int_k^{\infty} r^2\, dr\, \frac{q^4}{r^4} P^0({\bf r})\, P^0({\bf r})\, .
\end{eqnarray}
%

\section{Viscous fluid dynamics as a truncation to solve the functional RG}
\label{sec5}

The differential flow \eqref{eq:WetterichEqn} of $\Gamma_k[\phi,\chi]$ is an exact equation that captures all orders of perturbation theory as well as non-perturbative effects.
The iterative perturbative solution discussed in section~\ref{sec4} does not fully exploit the information contained in eq.~\eqref{eq:WetterichEqn}. 
While one cannot expect to find exact, analytic solutions for $\Gamma_k[\phi,\chi]$,  one can find approximate solutions in terms of truncations. The idea here is to use additional 
physics insight and intuition to truncate the {\it a priori} infinite space of possible actions $\Gamma_k[\phi,\chi]$ (the configuration space of the renormalization-group flow) to a finite subspace. Formally, one writes the flowing action in the form
\be
\Gamma_k[\phi,\chi] = \int d\eta \, d^3 \bx \sum_j \; \alpha_j(k) {\cal O}_j[\phi,\chi]\, ,
\label{eq:operatorExp}
\ee
where ${\cal O}_j[\phi,\chi]$ are a set of conveniently chosen ($k$-independent) operators and the $\alpha_j(k)$ are $k$-dependent coefficients. In the non-equilibrium situation relevant for cosmology, the operators ${\cal O}_j[\phi,\chi]$ can also depend on time. 

By projecting the flow equation \eqref{eq:WetterichEqn} to the space of functionals spanned by \eqref{eq:operatorExp}, one can derive renormalization-group equations for the coefficients $\alpha_j(k)$. If the index $j$ sums over a discrete, finite set, these are ordinary, coupled differential equations that can be solved analytically or numerically.

In general one expects that the dynamics of long-wavelength modes is dominated by the conservation laws that follow from the symmetries of the problem \cite{KadanoffMartin, Hohenberg:1977ym}, and that can
be formulated in terms of (viscous) fluid dynamics. To solve the RG flow equation \eqref{eq:WetterichEqn} approximately,
we therefore aim in this section at identifying a truncation of the flowing action in terms of a viscous fluid. To this end, we introduce first in section~\ref{sec5.1} the classical action whose variation leads to 
viscous fluid dynamics and which defines the form of the truncated action \eqref{eq:operatorExp}. We then define in section~\ref{sec5.2} the procedure via which we project  the RG flow on
this truncated action. This will allow us to derive differential equations for the coefficients  $\alpha_k(k)$ that enter the truncation of $\Gamma_k$, and to solve them numerically.

\subsection{The action of viscous fluid dynamics as a specific truncation of $\Gamma_k$}
\label{sec5.1}

Fluid dynamics is usually not formulated in terms of an action but in terms of equations of motion. For the ideal fluid discussed in section \ \ref{sec2.1},
we have seen in section~\ref{sec4} how the auxiliary field formalism provides a relation between equations of motion and the functionals $S[\phi,\chi]$ or $\Gamma_k[\phi, \chi]$. Here, we extend this formulation to first order in derivatives, i.e., to first-order viscous fluid dynamics, assuming that the fluid flow velocity is rotation free and can therefore be characterized completely in terms of its divergence.\footnote{The reader may wonder how dissipative terms can arise from the variation of an (effective) action: Our formalism using the auxiliary field $\chi$ is analogous to the classical limit of the Schwinger-Keldysh formalism and it is well known that the latter provides a way to describe dissipation, see e.\ g.\ \cite{Calzetta:1986cq}. The standard QFT formalism using the Feynman or time-ordered effective action does not allow this but a properly analytically continued effective action is another possibility \cite{Floerchinger:2016gtl}.}
Following Ref.~\cite{viscousdm}, the equations of motion of viscous fluid dynamics are of the form \eqref{eom}, but with a modified matrix $\hat{\Omega}(\bq, \eta)$ in which two additional coefficients enter,
\be
\hat \Omega(\bq,\eta)=\left(
\begin{array}{cc}
~~~0 &~~~-1
\\
-\frac{3}{2} \Omega_m + \gamma_s(\eta) \, \bq^2 &~~~ 1 +\frac{\Hc'}{\Hc} + \gamma_\nu(\eta) \, \bq^2
\end{array} \right).
\label{ome2} \ee
Here, the coefficient $\gamma_s$ is related to an effective sound velocity $c_s$\, ,
\be
\gamma_s=\frac{c_s^2}{{\cal H}^2}=\frac{d p/ d\rho}{{\cal H}^2}.
\ee
Similarly, $\gamma_\nu$ is related to a combination of effective shear viscosity $\eta$ and bulk viscosity $\zeta$,
\be
\gamma_\nu = \frac{4\eta/3 + \zeta}{(\rho+p){\cal H} a}\, .
\ee
In addition to the linear term described by $\hat \Omega(\bq,\eta)$, the effective sound velocity and viscosity appear also in the non-linear (vertex) terms $\gamma_{abc}$. However, as explained in Ref.~\cite{viscousdm}, dissipative modifications of these vertices are not unique and they are subleading by one power of $\bq^2$. We have checked explicitly that the following discussion and in particular the calculations in section \ref{sec5.3} do not depend on their inclusion. We therefore neglect vertex corrections for simplicity. 

A truncation of the effective action $\Gamma_k[\phi,\chi]$ that corresponds to these equations of motion is of the form
\begin{equation}
\begin{split}
\Gamma_k[\phi,\chi] = \int d\eta \Bigg[ 
&\int d^3q \,\chi_a(-\bq,\eta)\left(\delta_{ab} 
\partial_\eta+\hat \Omega_{ab}(\bq, \eta)\right) \phi_b( {\bf q},\eta) \\
&-  \int d^3k\, d^3 p \, d^3 q \, \delta^{(3)}(\bk-\bp-\bq)
\gamma_{abc}(\bk, \bp,\bq)
\chi_a(-\bk,\eta)\phi_b(\bp,\eta)
\phi_c( \bq ,\eta) \\
& -\frac{i}{2} \int d^3q\, \chi_a(\bq,\eta) H_{ab,k}(\bq,\eta,\eta^\prime) \chi_b(\bq,\eta^\prime) +\ldots \Bigg]\, ,
\end{split} 
\label{eq:truncationGamma}
\end{equation}
where $\hat \Omega(\bq,\eta)$ is given by eq.~\eqref{ome2} and the coefficients $\gamma_s(\eta)$ and $\gamma_\nu(\eta)$ depend now on the renormalization-group scale $k$. The object $H_{ab,k}(\bq,\eta,\eta^\prime)$ also depends on $k$. It parametrizes the statistical information of the full power spectrum, see eqs. \eqref{eq:defh} and \eqref{eq3.20}. The ellipses in \eqref{eq:truncationGamma} stand for terms of higher orders in $\chi$, which parametrize, for example, information about the bispectrum. 

Albeit truncated, the effective action \eqref{eq:truncationGamma} shares
all symmetries of the full problem. Galilean invariance of the effective action gives rise to Ward identities that relate
in particular two-point and three-point functions in the effective action~\cite{Peloso:2013zw, Kehagias:2013yd, Creminelli:2013mca}. One checks easily that these Ward identities are satisfied for the effective action \eqref{eq:truncationGamma},
irrespective of whether viscous corrections proportional to $\gamma_s(\eta)$ and $\gamma_\nu(\eta)$ appear only in the two-point functions, or whether they are added also
to the three-point vertices.

In the specific application of the renormalization-group formalism that we will use below, we will make the additional assumption that the time dependence of $\gamma_s(\eta)$ and $\gamma_\nu(\eta)$ can be written in the form of a power law,
\begin{equation}
\gamma_{s,k}(\eta) = \lambda_s(k) \, e^{\kappa(k) \eta}, \quad\quad\quad\quad \gamma_{\nu,k}(\eta) = \lambda_\nu(k) \, e^{\kappa(k) \eta},
\label{eq:truncgammasnupower}
\end{equation}
with three RG-scale dependent but time-independent coefficients $\lambda_s(k)$, $\lambda_\nu(k)$ and $\kappa(k)$. This ansatz  is
motivated by perturbation theory and will be shown to be self-consistent.

\subsection{Projection prescription by Laplace transforms}
\label{sec5.2}

To project the flow equation \eqref{eq:WetterichEqn} on a truncated effective action of viscous-fluid form \eqref{eq:truncationGamma}, we focus on the modes with small wavevectors.
We start from the following observations:

The effective viscosity and sound velocity terms affect, according to \eqref{ome2} and \eqref{eq:truncationGamma}, the terms in $\Gamma_k[\phi,\chi]$ that are proportional to $\chi$ and $\phi$. In the notation of \eqref{funder2}, this corresponds to the inverse propagators $D^R_{ab}(\bq,\eta,\eta^\prime)$ and $D^A_{ab}(\bq,\eta,\eta^\prime)$, respectively. A flow equation for the kernel $D^R_{ab}(\bq,\eta,\eta^\prime)$ can be obtained as a second functional derivative of the flow equation for $\Gamma_k[\phi,\chi]$, cf.\ eq.\ \eqref{flowow}. Therefore, we shall determine the coefficients $\gamma_s$ and $\gamma_\nu$
 (or, alternatively, the functions $\lambda_s(k)$, $\lambda_\nu(k)$ and the exponent $\kappa(k)$ used to parametrize them) by analyzing the retarded inverse propagator $D^R_{ab}(\bq,\eta,\eta^\prime)$. 
 As we work to lowest order in small wavevectors $\bq$, we need in particular a projection prescription from the flowing action that determines the time-dependence of the coefficients 
 $\gamma_s$ and $\gamma_\nu$. To this end, we take recourse to the Laplace transform of the inverse propagator $D^R_{ab}(\bq,\eta,\eta^\prime)$, as we explain now.

\subsubsection{Determining growth factors via Laplace transforms (case of local dynamics)}
We start from the equation which relates the retarded Green's function to its inverse, 
\begin{equation}
\int_{\eta_0}^\eta d\eta^\prime \; D^R_{ab}(\bq, \eta,\eta^\prime) G^R_{bc}(\bq, \eta^\prime,\eta_0) = \delta_{ac} \delta (\eta-\eta_0)\, .
\label{eq:SigmaGRelaNonlocal2}
\end{equation}
In the following, we write the inverse as a function of $\eta$ and $\Delta\eta = \eta - \eta'$. The
Laplace transform of $D^R_{ab}$ with respect to $\Delta\eta$ takes naturally into account that the retarded inverse propagator has support for $\Delta \eta > 0$ only, 
\begin{equation}
D^R_{ab}(\bq,\eta ; s) = \int_0^\infty d\Delta \eta \; e^{-s\Delta\eta} D^R_{ab}(\bq,\eta,\eta-\Delta\eta)\, .
\end{equation}
 One can then rewrite \eqref{eq:SigmaGRelaNonlocal2} in terms of this Laplace transform 
 \begin{equation}
 	\int d\eta' \frac{1}{2\pi i} \int_{\gamma - iS}^{\gamma + iS} ds\, D^R_{ab}(\bq,\eta ; s)\, e^{s(\eta - \eta')}\, G_{bc}^R(\bq, \eta^\prime,\eta_0) = \delta_{ac} \delta (\eta-\eta_0)\, .
 	\label{eq5.9}
 \end{equation}
  Here, the inverse Laplace transform is written in the standard way by an integration over the variable $s$ along a contour in the complex plane that goes parallel to the imaginary
  axis ($S \to \infty$) at a fixed real value $\gamma$ chosen to lie to the right of the rightmost pole of $ D^R_{ab}(\bq,\eta ; s)$. In general, $ D^R_{ab}(\bq,\eta ; s)$ 
 is a non-symmetric matrix that can be written in terms of its left- and right- eigenvectors $w_a^j(\bq,\eta ; s)$ and
 $v_a^j(\bq,\eta ; s)$ with eigenvalues $\lambda_j(\bq,\eta ; s)$, such that
 \begin{eqnarray}
 	D^R_{ab}(\bq,\eta ; s) &=& \sum_{j=1}^2   \lambda_j(\bq,\eta ; s)\, v_a^j(\bq,\eta ; s)\, w_b^j(\bq,\eta ; s)\, ,
 	\label{hi1}\\
 	w_a^j(\bq,\eta ; s)\, v_a^i(\bq,\eta ; s) &=& \delta^{ij}\,, \quad\quad\quad\quad w_a^j(\bq,\eta ; s)\, v_b^j(\bq,\eta ; s) = \delta_{ab} \, .
 \end{eqnarray}
 We now make the ansatz that the $s$-dependence of $ D^R_{ab}(\bq,\eta ; s)$ can be approximated by linearizing around the zeros of the two eigenvalues 
 $\lambda_j(\bq,\eta ; s) = \bar\lambda_j(\bq,\eta) \left(s - s_j(\bq,\eta)\right)$,
 \begin{eqnarray}
 	D^R_{ab}(\bq,\eta ; s) &=& \sum_{j=1}^2  \bar\lambda_j(\bq,\eta) \, \left(s - s_j(\bq, \eta)\right) \, v_a^j(\bq,\eta ; s_j(\bq,\eta))\, w_b^j(\bq,\eta ; s_j(\bq,\eta))  \nonumber \\
 	                                    &=&  s\, M_{ab}(\bq,\eta) - N_{ab}(\bq,\eta) \, .
 	\label{ansatzLaplace}
 \end{eqnarray}
The last line defines the matrices
 \begin{eqnarray}
 	M_{ab}(\bq,\eta) &=& \sum_{j=1}^2  \bar\lambda_j(\bq,\eta) \, v_a^j(\bq,\eta ; s_j(\bq,\eta))\, w_b^j(\bq,\eta ; s_j(\bq,\eta)) \, , \nonumber \\
 	N_{ab}(\bq,\eta) &=&  \sum_{j=1}^2  \bar\lambda_j(\bq,\eta) \, s_j(\bq, \eta) \, v_a^j(\bq,\eta ; s_j(\bq,\eta))\, w_b^j(\bq,\eta ; s_j(\bq,\eta)) \, .
 	\label{ansatzLaplaceMatrices}
 \end{eqnarray}
The linearized ansatz for $D^R_{ab}(\bq,\eta ; s)$ as described above can represent a wide class of situations where the effective linear equations of motion are first-order differential equations in time. Indeed, inserting \eqref{ansatzLaplace} into \eqref{eq5.9} leads to
 \begin{equation}
 	\int d\eta' \frac{1}{2\pi i} \int_{\gamma - iS}^{\gamma + iS} ds\, \left[M_{ab}(\bq,\eta) \, \partial_\eta - N_{ab}(\bq,\eta)  \right]\, e^{s(\eta - \eta')}\, G_{bc}^R(\bq, \eta^\prime,\eta_0) = \delta_{ac} \delta (\eta-\eta_0)\, .
 	\label{eq5.13}
 \end{equation}
If one uses that the Laplace transform of unity is a delta function, the $s$-integration returns $\delta(\eta -\eta')$, and the $\eta'$-integration is trivial. Eq.~\eqref{eq5.13} reduces then to
 \begin{equation}
 	 {\Big (} M_{ab}(\bq,\eta) \partial_\eta - N_{ab}(\bq,\eta) {\Big )}\, G_{bc}^R(\bq, \eta,\eta_0) = \delta_{ac} \delta (\eta-\eta_0)\, .
 	\label{eq5.14}
 \end{equation}
Moreover, assuming that $M_{ab}(\bq,\eta)$ can be inverted and that at the initial time $\eta_0$ one has $M_{ab}(\bq,\eta_0) = \delta_{ab}$ leads to
 \begin{equation}
 	 {\Big (} \delta_{ab} \partial_\eta - \Omega_{ab}(\bq,\eta) {\Big )}\, G_{bc}^R(\bq, \eta,\eta_0) = \delta_{ac} \delta (\eta-\eta_0)\, ,
 \end{equation}
with $\Omega_{ab}(\bq,\eta) = M^{-1}_{ad}(\bq,\eta)N_{db}(\bq,\eta)$. By construction, the eigenvalues of the matrix $\Omega_{ab}(\bq,\eta)$ are the zero-crossings $s_j(\bq,\eta)$ with $j=1,2$. 

These arguments show that all situations where the inverse propagator is of the form 
\begin{equation}
D^R_{ab}(\bq,\eta, \Delta\eta) = \left(\delta_{ab}\partial_\eta - \Omega_{ab}(\bq,\eta)\right)\, 
\delta(\eta-\eta^\prime)\,
\label{eq5.15}
\end{equation}
can be captured by the above described linear ansatz. This includes in particular the bare inverse propagator for a $\Lambda$CDM cosmology \eqref{linpropa}, the closely related inverse propagator of an Einstein-de Sitter Universe for which $\Omega$ is time-independent, and the case \eqref{ome2} that encodes viscous corrections to these inverse propagators. In all these cases,
for $\bq \to 0$ the growth factors $s_j(\bq, \eta)$, $j=1,2$, 
for the growing and decaying 
modes at time $\eta$, can be obtained from the zero-crossings of the determinant $ \text{det} \, D^R_{ab}(\bq,\eta ; s) $, i.e.
\begin{equation}	
	\text{det} \, D^R_{ab}\left(\bq,\eta ; s_j(\bq,\eta)\right) = 0 \quad\quad\quad \hbox{defines}\quad\quad\quad s_j(\bq,\eta)\, , \quad j=1,2\, .
\label{eq:definitionsjqn}
\end{equation}
For example, in an Einstein-de Sitter universe, one finds from eq.\ \eqref{ome2} for viscous fluid dynamics to lowest order in $\bq^2$
\begin{equation}
\begin{split}
s_1 = & 1 - \left(\frac{2}{5}\gamma_s(\eta) + \frac{2}{5}\gamma_\nu(\eta) \right) \bq^2 , \\
s_2 = & -\frac{3}{2} + \left(\frac{2}{5}\gamma_s(\eta) - \frac{3}{5} \gamma_\nu(\eta) \right) \bq^2 .
\end{split}
\label{eq:sjViscousDM}
\end{equation}
Here, $s_1(0,\eta)=1$ and $s_2(0,\eta)=-3/2$ are the standard linear growing mode $\sim e^{\eta}$ and decaying mode $\sim e^{-3\eta/2}$, respectively (see eq.~\eqref{eq2.9}).
The leading dissipative corrections to these growth factors for large but finite wavelength are given by the $\bq^2$-dependent terms in \eqref{eq:sjViscousDM}. In particular, one 
finds, as expected, that finite effective viscosity and sound velocity tame the exponentially growing mode. 

For the case of time-dependent $\Omega_{ab}(\bq, \eta) $, there is no systematic procedure that leads from $s_j(\bq, \eta)$ to 
an explicit form for the retarded propagator. However, the partial information about the time evolution obtained from characterizing 
$s_j(\bq,\eta)$ to first subleading order in $\bq^2$ will be sufficient for our purposes. 
 
\subsubsection{Growth factors and projection prescription: the case with memory integrals} 

In general, the full inverse retarded propagator for cosmological perturbations is not of the form \eqref{eq5.15}, but it reads 
\begin{equation}
D^R_{ab}(\bq,\eta,\eta^\prime) = \delta_{ab} \delta^\prime(\eta-\eta^\prime) + \Omega_{ab}(\bq, \eta) \delta(\eta-\eta^\prime) + \Sigma^R_{ab}(\bq,\eta,\eta^\prime)\, .
\label{eq:fullKernelDR}
\end{equation}
Here, the `self-energy' term $\Sigma^R_{ab}(\bq,\eta,\eta^\prime)$ represents corrections arising from the averaging over initial state fluctuations. It includes memory effects
and it is thus non-local in time. The perturbative one-loop expression \eqref{eq4.13} for $\Sigma^R_{ab}(\bq,\eta,\eta^\prime)$ illustrates this general property. As a 
consequence, $D^R_{ab}(\bq,\eta,\eta-\Delta\eta)$ (or $D^R_{ab}(\bq,\eta; s)$) can be a complicated function of $\Delta\eta=\eta-\eta^\prime$ (or of $s$), and the
determinant $\text{det} \, D^R_{ab}\left(\bq,\eta ; s\right) $ can have more than two zero crossings. 

However, the memory integral $\Sigma^R_{ab}(\bq,\eta,\eta^\prime)$ vanishes in the limit $\bq \to 0$ of infinite wavelength. Therefore, amongst all the zero crossings of 
$\text{det} \, D^R_{ab}\left(\bq,\eta ; s\right)$ that we expand around $\bq^2 = 0$, 
there are exactly two that connect smoothly to the known growth factors of the growing and decaying long-wavelength fluid dynamic modes. 
Based on these observations, we are now in a position to state our \\
{\it Projection prescription:} For any RG-evolved expression of the full inverse propagator $D^R_{ab}\left(\bq,\eta,\Delta\eta \right)$, 
\begin{enumerate}
\item Determine the zero crossings $s_j(\bq, \eta)$ of the determinant of the Laplace transform $\text{det} \, D^R_{ab}\left(\bq,\eta ; s\right)$.
\item Expand the zero crossings $s_j(\bq, \eta)$ for small $\bq^2$,
\begin{equation}
s_j(\bq^2,\eta) = s_j(0,\eta) + \bq^2 s_j^\prime(0,\eta) + \ldots\, ,
\label{zerocrossingpara}
\end{equation}
and identify the two zero crossings that smoothly connect to the background values (i.e. $s_1(0,\eta)=1$ and $s_2(0,\eta)=-3/2$ for EdS) for small $\bq$.
\item Project on the truncated effective action of viscous fluid dynamic form by requiring that the $O(\bq^2)$  corrections to \eqref{eq:sjViscousDM}
match the $O(\bq^2)$ corrections obtained from the expansion \eqref{zerocrossingpara} for the RG-evolved full inverse propagator,
\begin{equation}
\begin{split}
s_1^\prime(0,\eta)  \equiv &  - \left(\frac{2}{5}\gamma_s(\eta) + \frac{2}{5}\gamma_\nu(\eta) \right)\,  , \\
s_2^\prime(0,\eta) \equiv & \left(\frac{2}{5}\gamma_s(\eta) - \frac{3}{5} \gamma_\nu(\eta) \right)\, .
\end{split}
\label{eq:projectionprescription}
\end{equation}
This leads to unique expressions for the effective fluid properties in the truncated effective action \eqref{eq:truncationGamma}.
\end{enumerate}
In summary, this projection prescription maps a non-local self-energy that involves memory integrals onto the effective action for local viscous fluid dynamics in such a way that the long-wavelength behaviour of the growth factors $s_j(\bq^2,\eta)$ is preserved to $O(\bq^2)$. 

As a first illustration of this projection prescription, we consider the perturbative one-loop result $\Sigma^R_{ab,k}(\bq,\eta,\eta^\prime)$ calculated in \eqref{eq4.13}.
This is a non-local memory integral. One can Laplace transform this expression explicitly and determine the zeros of the corresponding determinant 
$\text{det} \, D^R_{ab}\left(\bq,\eta ; s_j(\bq,\eta)\right)$. Since $\Sigma^R_{ab,k}(\bq,\eta,\eta^\prime) \propto e^{2\eta}\, \Theta(\Delta\eta)$, and since the Laplace 
transform of the $\Theta$-function is $1/s$, the determinant is a fourth-order polynomial in $s$ with zeros at
\begin{equation}
\begin{split}
s_1 = & 1 - \frac{187}{175} \bq^2 \sigma_{v,k}^2 e^{2\eta} + O(\bq^4) \, , \\
s_2 = & -\frac{3}{2} - \frac{29}{25} \bq^2 \sigma_{v,k}^2 e^{2\eta} + O(\bq^4) \,  ,\\
s_3 = & \frac{29}{25} \bq^2 \sigma_{v,k}^2 e^{2\eta} + O(\bq^4) \,  ,\\
s_4 = & -\frac{5}{2} + \frac{187}{175} \bq^2 \sigma_{v,k}^2 e^{2\eta} + O(\bq^4) \,  .\\
\end{split}\label{eq:sj1loopEdS}
\end{equation}
Here, we have introduced the shorthand
\be
\sigma_{v,k}^2 = \frac{4\pi}{3} \int_k^\infty dq P^0(q)\, .
\ee
In the limit $\bq \to 0$, the growth factors $s_1$ and $s_2$ in eq.~\eqref{eq:sj1loopEdS} smoothly connect to the ones of the standard EdS growing and decaying mode. This identifies
the two growth factors that govern the propagation of fluid dynamic modes. Identifying their $\bq^2$-dependence with that of viscous fluid dynamics, we find from eq.~\eqref{eq:projectionprescription}
\begin{equation}
\begin{split}
\gamma_s(\eta) = & \frac{31}{70} \sigma_{v,k}^2 e^{2\eta}, \\ 
\gamma_\nu(\eta) = & \frac{78}{35} \sigma_{v,k}^2 e^{2\eta}\, .
\end{split}
\end{equation}
In the parametrization of eq.~\eqref{eq:truncgammasnupower}, this corresponds to 
\begin{equation}
\lambda_s(k)= \frac{31}{70} \sigma_{v,k}^2, \quad\quad\quad \lambda_\nu(k)= \frac{78}{35} \sigma_{v,k}^2, \quad\quad\quad \kappa(k)=2. \label{eq:lambdaslambdanukappaOneloop}
\end{equation}

In our previous work \cite{viscousdm}, we had fixed the effective properties of a viscous fluid dynamic formulation of cosmological perturbations below the scale $k_m$ by matching
heuristically to the perturbative one-loop correction for the retarded propagator with loop momenta restricted to scales $\vert \bq\vert > k_m$.  As explained in section~\ref{sec4.2},
this procedure can be motivated by the RG flow of the retarded propagator, but it constrains only the sum $\gamma_s+\gamma_\nu$ of the viscous coefficients. 
In contrast, the projection prescription advocated here constrains   $\gamma_s$ and $\gamma_\nu$ independently and uniquely through two dynamical conditions.
It thus determines the truncated effective action \eqref{eq:truncationGamma} unambiguously.
The sum  $\gamma_s+\gamma_\nu$ obtained in this way agrees with the result from ref.\ \cite{viscousdm} within a few \%.

\subsection{Flow equations for effective fluid parameters and their solution}
\label{sec5.3}

We extend now the discussion of the projection prescription for the effective viscosity and sound velocity coefficients to $k$-dependent quantities $\gamma_{\nu,k}$ and $\gamma_{s,k}$. They are related to zero crossings $s_{j,k}(\bq,\eta)$ of  $\text{det} \, D^R_{ab,k}\left(\bq,\eta ; s\right)$ defined according to \eqref{eq:definitionsjqn}. Now they are also functions of the renormalization-group scale $k$. 

It follows directly from equation~\eqref{eq:projectionprescription} that the scale dependence of the effective viscous coefficients $\gamma_s$, $\gamma_\nu$ is determined by 
\begin{equation}
\begin{split}
\frac{\partial}{\partial k} \gamma_{\nu,k}(\eta) = & - \frac{\partial}{\partial k} \left( s_{1,k}^\prime(0,\eta) + s_{2,k}^\prime(0,\eta)  \right)\,  , \\
\frac{\partial}{\partial k} \gamma_{s,k}(\eta) = & - \frac{\partial}{\partial k} \left( \frac{3}{2} s_{1,k}^\prime(0,\eta) - s_{2,k}^\prime(0, \eta)  \right)\,  .
\label{flowvisc}
\end{split}
\end{equation}
To determine the $k$-derivative of $s_{j,k}(\bq, \eta)$, we use
\begin{equation}
\frac{d}{dk} \text{det} \, D^R_{ab,k}\left(\bq,\eta ; s\right)\Big\vert_{s=s_{j,k}(\bq, \eta)} = 0\, .
\label{callansymanzik}
\end{equation}
Equation~\eqref{callansymanzik} leads to
\begin{equation}
 \frac{\partial}{\partial k}s_{j,k}(\bq,\eta) = - \frac{\text{Tr} \left\{ \left[D^R_{k}\left(\bq,\eta ; s\right)\right]^{-1} \frac{\partial}{\partial k} D_k^R\left(\bq,\eta ; s\right)    \right\}}{ 
 \text{Tr} \left\{  \left[D^R_{k}\left(\bq,\eta ; s\right)\right]^{-1} \frac{\partial}{\partial s} D_k^R\left(\bq,\eta ; s\right)  \right\}} \Bigg\vert_{s=s_{j,k}(\bq, \eta)} \, .
 \label{callansymanzik2}
\end{equation}
In general,  the matrix $D_k^R$ is not known explicitly. However, for the evolution of $s_{j,k}^\prime(0,\eta)$, i.\ e.\ the first derivative with respect to momentum $\bq^2$ at $\bq=0$, it is sufficient to determine the $O(\bq^2)$ term on the right hand side of ~\eqref{callansymanzik2}. Because $\frac{\partial}{\partial k} D_k^R\left(\bq,\eta ; s\right) 
= \frac{\partial}{\partial k} \Sigma_k^R\left(\bq,\eta ; s\right)$ vanishes for $q=0$, the other matrices on the right hand side of eq.~\eqref{callansymanzik2} can be evaluated at $\bq=0$, where they assume their standard Einstein-de-Sitter form. In particular, one has there $ \textstyle\frac{\partial}{\partial s} D_{ab,k}^R\left(0,\eta ; s\right) = \delta_{ab} $ and we obtain
\begin{equation}
 \frac{\partial}{\partial k}s^\prime_{j,k}(0,\eta) = -  \text{Tr} \left\{ 
\left( \begin{array}{cc}  s+\textstyle\frac{1}{2} & ~~~1 \\  \frac{3}{2} & ~~~s \end{array} \right)
 \frac{\partial}{\partial \bq^2}\frac{\partial}{\partial k}  \Sigma_k^R\left(0,\eta ; s\right) \right\}   \Big /
 \text{Tr} \left\{  \left( \begin{array}{cc}  s+\textstyle\frac{1}{2} & ~~~1 \\  \frac{3}{2} & ~~~s \end{array} \right) \right\} 
  \bigg\vert_{s=s_{j,k}(0,\eta)}  \, .
 \label{callansymanzik3}
\end{equation}
Moreover, on the right hand side one can use $s_{1,k}(0,\eta)=1$ and $s_{2,k}(0,\eta)=-3/2$.

To obtain an explicit expression for ~\eqref{callansymanzik3}, one then needs to derive the flow of $D^R_k$ in equation \eqref{flowow}, with vertices 
and spectra defined by the effective theory. To this end, we start from the truncation \eqref{eq:truncationGamma}. Several complications that could arise in calculating the flow~\eqref{flowow} of $D^R_k$
for more general effective actions do not arise in this case.~\footnote{ For instance, in the effective viscous theory~\eqref{eq:truncationGamma}, the vertex functions do not get modified, 
and contributions with novel effective vertices (such as the four-leg vertex entering the third diagram in the first  row of fig.\ \ref{fig:FlowDR}) 
do not contribute. Also, one can see from fig. \ref{fig:FlowDR} that the scale-dependent spectrum $P_{ab,k}(\bq,\eta,\eta^\prime)$ does not feed into the loop expressions directly. 
This would be different for the flow equation of the function $H_{ab,k}(\bq,\eta,\eta^\prime)$, for which the corresponding diagrams are shown in the second row of fig.\ \ref{fig:FlowDR}.}
The perturbative calculation of eq.\ \eqref{eq4.13} thus generalizes rather directly to a flow equation for the full inverse propagator of the truncated effective theory,
with the loop expression
\begin{eqnarray}
	&&\partial_k \Sigma^R_{ab,k}(\bq,\eta,\eta') = 4 \int d{\bf r}\, 
G_{c'f,k}^R(- \br, \eta,0) \partial_k P_{fh,k}^0({\bf r}) \, G^R_{d'h,k}(\br, \eta',0)\nonumber \\
&& \qquad \qquad  \qquad \qquad \qquad  \times \gamma_{acc'}(\bq,{\bf r}+\bq,-{\bf r}) G_{cd,k}^R(\br +\bq,\eta,\eta') \gamma_{dd'b}({\bf r}+\bq,{\bf r},\bq)\, .
\label{eq:flowDRFull}
\end{eqnarray}
Here $G^R_{ab,k}(\bq,\eta,\eta^\prime)$ is the effective retarded, $k$-dependent propagator determined for the theory with sound velocity and viscosity coefficients as in \eqref{eq:truncgammasnupower}, while the vertices are unmodified.  To calculate ~\eqref{callansymanzik3}, one requires an explicit expression for the full dissipative retarded propagator $G^R_{ab,k}(\bq,\eta,\eta^\prime)$
 that enters \eqref{eq:flowDRFull}. Here, we first explain how this propagator can be obtained before presenting results for the flow equations. 

\subsubsection{The retarded propagator for viscous fluid dynamics}
For the equations of motion of viscous fluid dynamics that one obtains from varying the effective action \eqref{eq:truncationGamma}, we have given in
\cite{viscousdm} an exact solution in EdS for the density contrast 
\be
  \delta_{\bq}(\eta) = U_{\bq, \delta}(\eta)\cdot c\, ,
\ee
where $c=(c_1,c_2)$ parametrizes the initial conditions in terms of growing and decaying modes, and
\bea
U_{\bq, \delta}(\eta)&=&\Bigg( \exp(\eta)\,
_1 F_1\left(\frac{1}{\kappa}+ \frac{\lambda_s}{\kappa\lambda_\nu}, 1 + \frac{5}{2\kappa}, -X_{\bq}(\eta) \right),
\nonumber \\\
&& \exp(-3\eta/2)\,
_1 F_1 \left(-\frac{3}{2 \kappa}+ \frac{ \lambda_s}{\kappa \lambda_\nu}, 1 -\frac{5}{2 \kappa}, -X_{\bq}(\eta) \right) \Bigg)\, ,
\label{sollsHyp} 
\eea
with
\be
  X_{\bq}(\eta) \equiv \frac{\lambda_\nu\, \bq^2}{\kappa} \exp(\kappa \eta)\, .
\ee
Here, the effective sound velocity and viscosity are parametrized according to \eqref{eq:truncgammasnupower}. The viscous equations of motion relate $ \delta_{\bq}(\eta)$ to the 
 normalized velocity divergence $-\theta/{\cal H}$ by a time derivative
\be
  -\theta_{\bq}(\eta)/{\cal H} = U_{\bq, \theta}(\eta)\cdot c = dU_{\bq, \delta}/d\eta \cdot c\, .
\ee
One can then define the $2\times 2$-matrix
\be
  U_\bq(\eta) \equiv \left( U_{\bq, \delta}(\eta) \atop U_{\bq, \theta}(\eta) \right)\,,
\ee
in terms of which the retarded propagator is given explicitly by
\be
  G^R(\bq,\eta,\eta^\prime)  = \Theta(\eta-\eta')U_\bq(\eta) U_\bq^{-1}(\eta')\,.
  \label{eq:gvisc}
\ee
For algebraic manipulations, we find it convenient to expand the hypergeometric functions that enter \eqref{sollsHyp},
we have taken recourse to the expansion 
\be
  _1 F_1\left(b-a,b,-X\right) = e^{-X}\left(1+  \frac{a}{b}X + \frac{a(1+a)}{b(1+b)}\frac{X^2}{2} + \frac{a(1+a)(2+a)}{b(1+b)(2+b)}\frac{X^3}{6}+\dots \right)\, .
  \label{eq:expandHyp}
\ee
The viscous propagator \eqref{eq:gvisc} approaches the EdS propagator for $X\to 0$. In our case, $X_{\bq}(\eta) \propto \bq^2$, and the expansion
~\eqref{eq:expandHyp} turns out to be well-suited for identifying the long-wavelength $O(\bq^2)$ contributions to the Laplace transform 
of the self-energy $\Sigma^R_{ab,k}(\bq,\eta;s)$.

\subsubsection{Flow equations for the dissipative coefficients}
With the explicit form \eqref{eq:gvisc} for the retarded propagator of the effective theory, one can write an explicit algebraic form for the Laplace transform
of the flow~ \eqref{eq:flowDRFull} of the self-energy $\Sigma_{ab,k}^R$. With the help of a standard program for algebraic manipulations~\cite{mathematica},
we then use the series expansion~\eqref{eq:expandHyp} to write the Laplace transform of \eqref{eq:flowDRFull} as a sum of terms. Term by term,
we integrate out the angular components of the momentum variable ${\bf r}$ in \eqref{eq:flowDRFull}, and we determine the flow equations~\eqref{flowvisc}
of the effective viscosity and sound velocity by evaluating $\frac{\partial}{\partial k} s^\prime_{j,k}(0,\eta)$. We find
\begin{eqnarray}
 \partial_k\left(\lambda_\nu(k)\, e^{\kappa(k)\eta} \right) &=& \sum_\ell \left( b_1^\ell + b_2^\ell \right) \hat{Y}_\ell(\eta)\, , \nn\\
 \partial_k\left(\lambda_s(k)\, e^{\kappa(k)\eta} \right) &=& \sum_\ell \left( \tfrac{3}{2} b_1^\ell - b_2^\ell \right) \hat{Y}_\ell(\eta)\, ,
\label{eq:5.40}
\end{eqnarray}
where
\be
  \hat{Y}_\ell(\eta) \equiv \frac{4\pi}{3} e^{2\eta} P^0(k)\, (X_{k}(\eta))^\ell\, \exp\left(-2X_{k}(\eta)\right) .
  \label{yelltilde}
\ee
The sums in \eqref{eq:5.40} involve all orders in the series expansion of \eqref{eq:expandHyp}. 
To lowest order, the expansion coefficients $b_j^{(\ell)}$  read
\bea\label{eq:bi}
  b_1^{\ell=0} &=& - \frac{187}{175}\,,\nn\\
  b_2^{\ell=0} &=& - \frac{29}{25}\,,\nn\\
  b_1^{\ell=1} &=& + 2\frac{(2618 + 5973 \kappa + 2698 \kappa^2 + 180 \kappa^3)\lambda_s-(3927 + 4900 \kappa + 2274 \kappa^2 + 488 \kappa^3)\lambda_\nu}{175(1+\kappa)(5+2\kappa)(7+2\kappa)\lambda_\nu} \,,\nn\\
  b_2^{\ell=1} &=& - \frac{2(174 - 151 \kappa + 59 \kappa^2 + 90 \kappa^3)\lambda_s+(-522 - 155 \kappa + 346 \kappa^2 + 192 \kappa^3)\lambda_\nu}{25(1+\kappa)(-3+2\kappa)(5+2\kappa)\lambda_\nu}\,.
\eea
Coefficients for $\ell >1$ can be obtained easily by algebraic manipulation, but the expressions become rapidly more lengthy and will not be presented here.  

Any approximate solution of a functional renormalization group by truncation is based on assumptions about the functional dependence of the parameters entering the effective action.
Since the functional RG equation \eqref{eq:WetterichEqn} is an exact non-perturbative equation, this truncated ansatz cannot be expected to solve it exactly. In the present case, the ansatz for the 
time-dependence $\propto e^{\kappa(k) \eta}$ of the viscous coefficients \eqref{eq:truncgammasnupower} is motivated by perturbation theory but it cannot be expected to be exact. Indeed,
the set of equations \eqref{eq:5.40} derived from the functional renormalization group~\eqref{eq:WetterichEqn} allow on the right-hand side for a more general $\eta$-dependence that
does not necessarily match our ansatz. This indicates that we cannot
assume this $\eta$-dependence to hold exactly at all times $\eta$, but we may assume that it captures essential physics close to $\eta = 0$. We therefore 
write flow equations for $\lambda_\nu(k)$ and $\lambda_s(k)$ by evaluating \eqref{eq:5.40} for $\eta = 0$, and we derive a flow equation for $\kappa$ by a logarithmic derivative with respect to $\eta$
of either the first or second line of \eqref{eq:5.40}. Because the viscosity coefficient plays a numerically more important role, we use the first line. We obtain thus
\begin{equation}
\begin{split}
\partial_k \lambda_\nu(k) = & \sum_\ell \left( b_1^\ell + b_2^\ell \right) \hat{Y}_\ell(0)\, ,\\
\partial_k \lambda_s(k) = & \sum_\ell \left( \tfrac{3}{2} b_1^\ell - b_2^\ell \right) \hat{Y}_\ell(0)\, , \\
\partial_k \kappa(k) =&  \frac{1}{\lambda_\nu(k)}  \sum_l (b_1^l+b_2^l) \left[\frac{\partial}{\partial\eta} \hat Y_l(0) - \kappa \, \hat Y_l(0) \right]\, .
\end{split}
\label{eq:5.43}
\end{equation}
Given that our ansatz $\propto e^{\kappa(k) \eta}$ for the time-dependence of the viscous coefficients does not solve \eqref{eq:5.40} exactly, the question arises of whether one can test how
good this approximate solution is. According to the numerical results shown in the following section, $\kappa(k)$ stays throughout the RG-evolution close to the perturbative one-loop value
$\kappa = 2$, for which our ansatz is valid. This can be seen as an a posteriori indication that the ansatz is reasonable. In principle, one can try to improve the approximate solution by 
allowing for additional functional dependencies and checking whether the results for the viscous coefficients remain stable, We shall not explore this point further.
However, we do check in appendix \ref{appb} that the power-law ansatz for the time-dependence is a reasonable approximation, by evaluating the RG equations at non-zero values $\eta$, which
leads to consistent results.

To further analyze the flow equations~\eqref{eq:5.43}, we introduce the dimension-less quantities $\tilde \lambda_\nu(k) = k^2 \lambda_\nu(k)$, $\tilde \lambda_s(k) = k^2 \lambda_s(k)$. Restricting the sum over $l$ to the terms $l=0,1$ we find ($t=\ln k$)
\begin{equation}
\begin{split}
\partial_t \tilde \lambda_\nu = & 2\tilde \lambda_\nu + \frac{4\pi}{3} k^3 P^0(k) e^{-2\frac{\tilde\lambda_\nu}{\kappa}} {\Bigg [} -\frac{78}{35} 
-\tilde \lambda_\nu \frac{928 \kappa ^4+4084 \kappa ^3+4148 \kappa ^2-5719 \kappa -9828}{35 \kappa  (\kappa +1) (2 \kappa -3) (2
   \kappa +5) (2 \kappa +7)} \\
&    \quad\quad\quad\quad\quad\quad\quad\quad\quad\quad\quad\quad\quad
   -\tilde \lambda_s \frac{2 \left(180 \kappa ^4+76 \kappa ^3-615 \kappa ^2+1544 \kappa +3276\right)}{35 \kappa  (\kappa
   +1) (2 \kappa -3) (2 \kappa +5) (2 \kappa +7)}
   + {\cal O}\left(\tilde \lambda_{\nu,s}^2 \right)
{\Bigg ]} \, ,   \\
\partial_t \tilde \lambda_s = & 2\tilde \lambda_s + \frac{4\pi}{3} k^3 P^0(k) e^{-2\frac{\tilde\lambda_\nu}{\kappa}} {\Bigg [} -\frac{31}{70} 
 - \tilde \lambda_\nu \frac{ 48 \kappa ^4-1000 \kappa ^3-1170 \kappa ^2-1127 \kappa -1953 }{35 \kappa  (\kappa +1) (2 \kappa -3) (2 \kappa
   +5) (2 \kappa +7)} \\
&\quad\quad\quad\quad\quad\quad\quad\quad\quad\quad\quad\quad\quad+\tilde \lambda_s \frac{ 720 \kappa ^4+5008 \kappa ^3+2622 \kappa ^2-9595 \kappa -1302}{35 \kappa  (\kappa +1) (2
   \kappa -3) (2 \kappa +5) (2 \kappa +7)}   
   + {\cal O}\left(\tilde \lambda_{\nu,s}^2 \right)
{\Bigg ]} \, , \\
\partial_t \kappa = &  \frac{4\pi}{3} k^3 P^0(k) e^{-2\frac{\tilde\lambda_\nu}{\kappa}} {\Bigg [}  
\frac{78 (\kappa -2)}{35 \tilde\lambda_\nu}
 + \frac{ 2  \left(624 \kappa ^5+2504 \kappa^4-1432 \kappa ^3-12494 \kappa^2-2471\kappa
   +9828\right)}{35 \kappa  (\kappa +1) (2 \kappa -3) (2 \kappa +5) (2 \kappa +7)} \\
& \quad\quad\quad\quad\quad\quad\quad\quad\quad\quad\quad\quad\quad
-\frac{\tilde\lambda_s}{\tilde\lambda_\nu} \frac{4\left(180 \kappa ^4+76 \kappa ^3-615 \kappa ^2+1544 \kappa +3276\right)}{35 \kappa  (\kappa +1) (2 \kappa -3) (2 \kappa +5) (2 \kappa +7)}
+ {\cal O}\left(\tilde \lambda_{\nu,s}\right)
{\Bigg ]} \, .
\end{split}
\label{eq:flowlambdanulambdaskappa}
\end{equation}
Although these expressions contain contributions to all orders in $\tilde\lambda_\nu$ due to the exponential, they should only be trusted to the order indicated. Higher orders in the sum over $l$ contribute additional terms in principle (see appendix \ref{appb} for a discussion of the impact of these terms).

Before presenting numerical solutions of \eqref{eq:flowlambdanulambdaskappa} in the next section, we discuss the qualitative behaviour of the RG flow that we expect. The linear power spectrum $P^0(k)$ has a maximum at a scale $k_\text{max}$. For $k\gg k_\text{max}$, the dimensionless combination $k^3 P^0(k)$ is large and the RG flow of $\tilde\lambda_\nu$ and $\tilde\lambda_\nu$ in \eqref{eq:flowlambdanulambdaskappa} is dominated by the terms in square brackets. On the other hand, for $k\ll k_\text{max}$ one has $k^3 P^0(k)\ll 1$, and the flow is dominated by the first terms due to the scaling dimension. In that regime, the dimensionful quantities $\lambda_\nu =\tilde \lambda_\nu/k^2$,  $\lambda_s =\tilde \lambda_s/k^2$ should approach fixed-point values. The scale is set by $k_\text{max}$ and the coefficients are fixed by
the flow itself. 

A simple approximation to the flow equations is obtained by keeping only the lowest orders in $\tilde\lambda_\nu$ and $\tilde \lambda_s$, corresponding to the one-loop approximation,
\begin{equation}
\begin{split}
\partial_t \tilde \lambda_\nu = & 2\tilde \lambda_\nu - \frac{4\pi}{3} k^3 P^0(k) \frac{78}{35}, \\
\partial_t \tilde \lambda_s = & 2\tilde \lambda_s - \frac{4\pi}{3} k^3 P^0(k) \frac{31}{70}, \\
\partial_t \kappa = &  \frac{4\pi}{3} k^3 P^0(k) \frac{78 (\kappa -2)}{35 \tilde\lambda_\nu}.
\end{split}\label{eq:flowlambdanulambdaskappaOneLoop}
\end{equation}
The IR fixed points are in this case closely related to the one-loop values~\eqref{eq:lambdaslambdanukappaOneloop}
\begin{equation}
\lambda_\nu(0) = \lambda_\nu(\Lambda)+\frac{78}{35} \frac{4\pi}{3} \int_0^\Lambda dq P^0(q) , \quad\quad\quad
\lambda_s(0) =\lambda_s(\Lambda) + \frac{31}{70} \frac{4\pi}{3} \int_0^\Lambda dq P^0(q) ,\quad\quad\quad
\kappa(0) = 2.
\end{equation}
We denote here by $\Lambda$ the UV scale where the flow equations are initialized with some initial value. In principle, these initial values could represent a microscopic viscosity and sound velocity of dark matter. However, the typical situation is that the flow (and therefore the fixed point values $\lambda_\nu^*$, $\lambda_s^*$) is dominated by intermediate scales $k\approx k_\text{max}$ such that the initial values are not very important.
These expectations will be scrutinized and confirmed by a numerical analysis in the next section.

\section{Numerical results}
\label{sec6}

The approach discussed above suggests a concrete way to compute observables such as the two-point correlation function
in a two-step approach.
\begin{enumerate}
\item
  In a first step, the effective action of the system is computed by evolving the coarse-graining scale,
starting at an initial microscopic scale $\Lambda$, above the non-linear scale $k_{nl}\simeq 0.3\, h/$Mpc,
down to a scale $k_m<\Lambda$ that will be chosen around the BAO scale. This procedure
generates an effective action described, at first order in gradients, by Navier-Stokes equations with effective sound velocity and viscosity that depend on the matching scale $k_m$,  as described by the RG equations \eqref{eq:flowlambdanulambdaskappa}.
\item
In a second step, these effective equations are solved in order to obtain correlation functions for $k<k_m$.
\end{enumerate}
In the following we describe numerical results obtained for both steps, assuming a $\Lambda$CDM model.
Our discussion in the previous section was based on an EdS background. As discussed in \cite{viscousdm}, the $\Lambda$CDM 
model can be mapped to an EdS background to very good accuracy by redefining the time variable as  
$\eta\equiv \ln D_L$, where $D_L$ is the conventional linear growth factor. In addition, the field doublet takes the form $(\delta, -\theta/({\cal H}f))$
where $f=d \ln D_L/d\ln(1+z)$, and the effective viscosity and sound velocity parameters are 
\be
\gamma_\nu = \frac{4\eta/3+\zeta}{(\rho+p){\cal H}f}=\lambda_\nu \exp(\kappa \eta),\qquad
\gamma_s=\frac{c_s^2}{{\cal H}^2f^2}=\lambda_s \exp(\kappa \eta)\, .
\label{inter} 
\ee

\be
  \gamma
\ee
In the notation used in \cite{viscousdm}, $\gamma_\nu\equiv \tilde\alpha_\nu/k_m^2$ and $\gamma_s\equiv \tilde\alpha_s/k_m^2$.
All numerical results correspond to a $\Lambda$CDM 
model with parameters $\sigma_8=0.79$, $\Omega_m=0.26$, $\Omega_b=0.044$, $h=0.72$, $n_s=0.96$.

\subsection{RG evolution of effective sound velocity and viscosity}

\begin{figure}[htb]
\centering
$$
\includegraphics[width=0.32\textwidth]{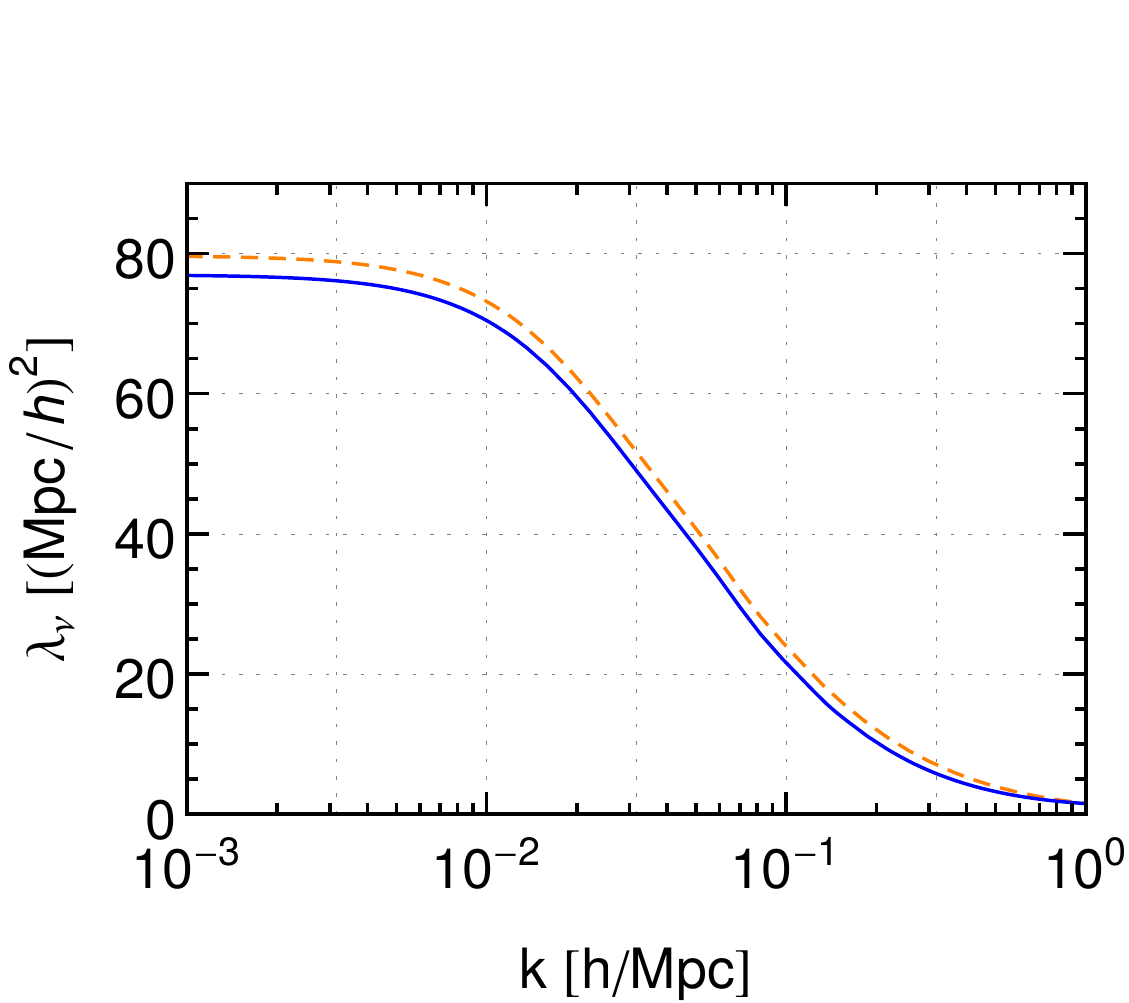}\hspace*{0.02\textwidth}
\includegraphics[width=0.32\textwidth]{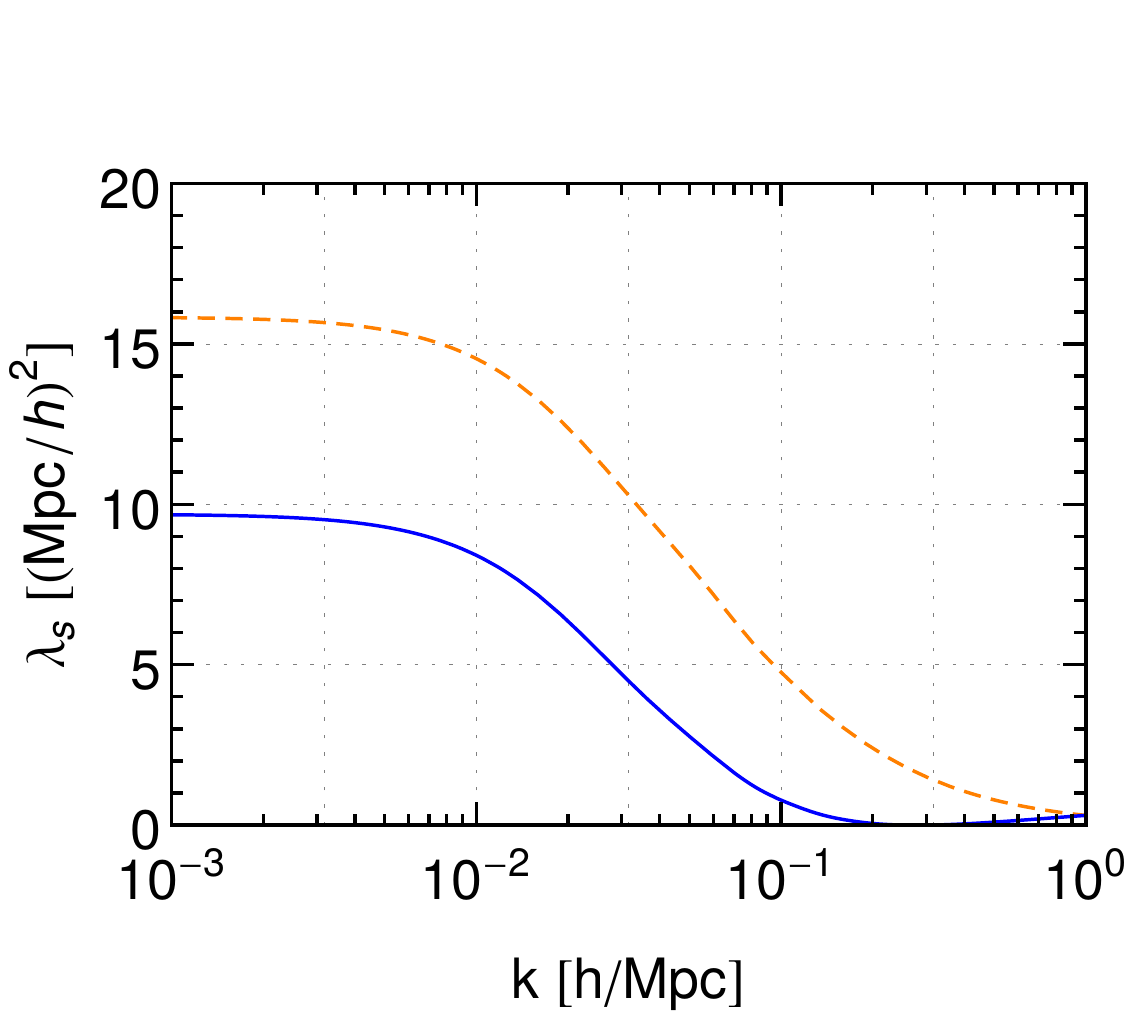}\hspace*{0.02\textwidth}
\includegraphics[width=0.32\textwidth]{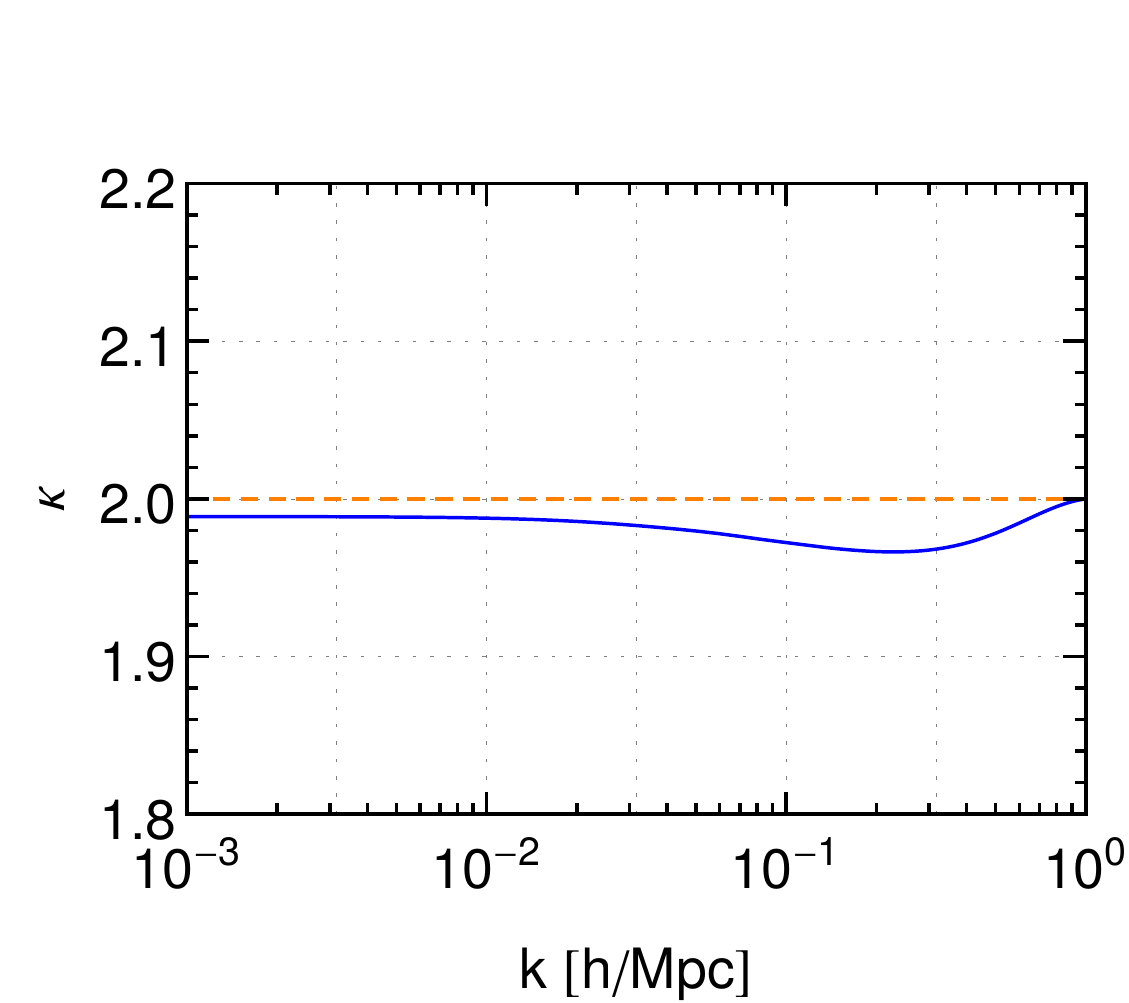}
$$
\caption{Renormalization group evolution of the coefficients $\lambda_\nu(k)$, $\lambda_s(k)$ and $\kappa(k)$. We have initialized the flow at $k=\Lambda=1\, h/$ Mpc with the one-loop values \eqref{eq:lambdaslambdanukappaOneloop}. The solid lines correspond to the solution of the full flow equations \eqref{eq:flowlambdanulambdaskappa} while the dashed lines correspond to the solution of the one-loop approximation \eqref{eq:flowlambdanulambdaskappaOneLoop}.
}
\label{fig:flow}
\end{figure}

As the first step, we determine the effective sound velocity and viscosity obtained from coarse graining of UV modes.
As described in section \,\ref{sec5}, we consider a truncation of the full effective action that is 
characterized by sound velocity and viscosity terms with an approximate power-law dependence on time.
Inserting this truncation in the functional RG equation yields the RG evolution equations \eqref{eq:flowlambdanulambdaskappa}
for $\lambda_i(k_m)$ and $\kappa(k_m)$. For the rest of this section we will omit the subscript of $k_m$ for brevity.

In fig.\ \ref{fig:flow} we show the solution of the flow equations \eqref{eq:flowlambdanulambdaskappa} (solid lines) and of the one-loop approximation \eqref{eq:flowlambdanulambdaskappaOneLoop} (dashed lines) for initial conditions $\lambda_\nu(\Lambda)$, $\lambda_s(\Lambda)$ and $\kappa(\Lambda)$ corresponding to the one-loop result \eqref{eq:lambdaslambdanukappaOneloop} at the scale $k=\Lambda=1$ h/Mpc. One observes that the viscosity coefficient $\lambda_\nu(k)$ is relatively well approximated by the one-loop equation, albeit the full solution leads to a slightly larger fixed point value. In contrast, the sound-velocity coefficient $\lambda_s(k)$ flows to a somewhat smaller value for the full solution. The exponent $\kappa(k)$, which governs the time dependence, does not differ from the one-loop result $\kappa=2$ very much. At macroscopic scales $k\to 0$ we find for the full solution $\lambda_\nu = 77.0 \,(\text{Mpc/h})^2$, $\lambda_s = 9.70 \, (\text{Mpc/h})^2$ and $\kappa=1.99$.
These values correspond today to sound velocity $c_s\simeq 750$ km/s and kinematic viscosity $\nu=\frac{\zeta+4\eta/3}{\rho}=77.0 \,\frac{100 \, \text{km} \,\text{Mpc}}{\text{s}\, h}\simeq 4 \times 10^{29}\, \text{m}^2/\text{s}$.
Note that the approximations performed to arrive at the RG equations \eqref{eq:flowlambdanulambdaskappa} require the dimensionless combinations $\lambda_i(k) k^2 $
to be smaller than order unity. Using the numerical solution one can check that this condition is indeed satisfied (see appendix \ref{appb} for further discussions of this point).

\begin{figure}[t]
\centering
$$
\includegraphics[width=0.5\textwidth]{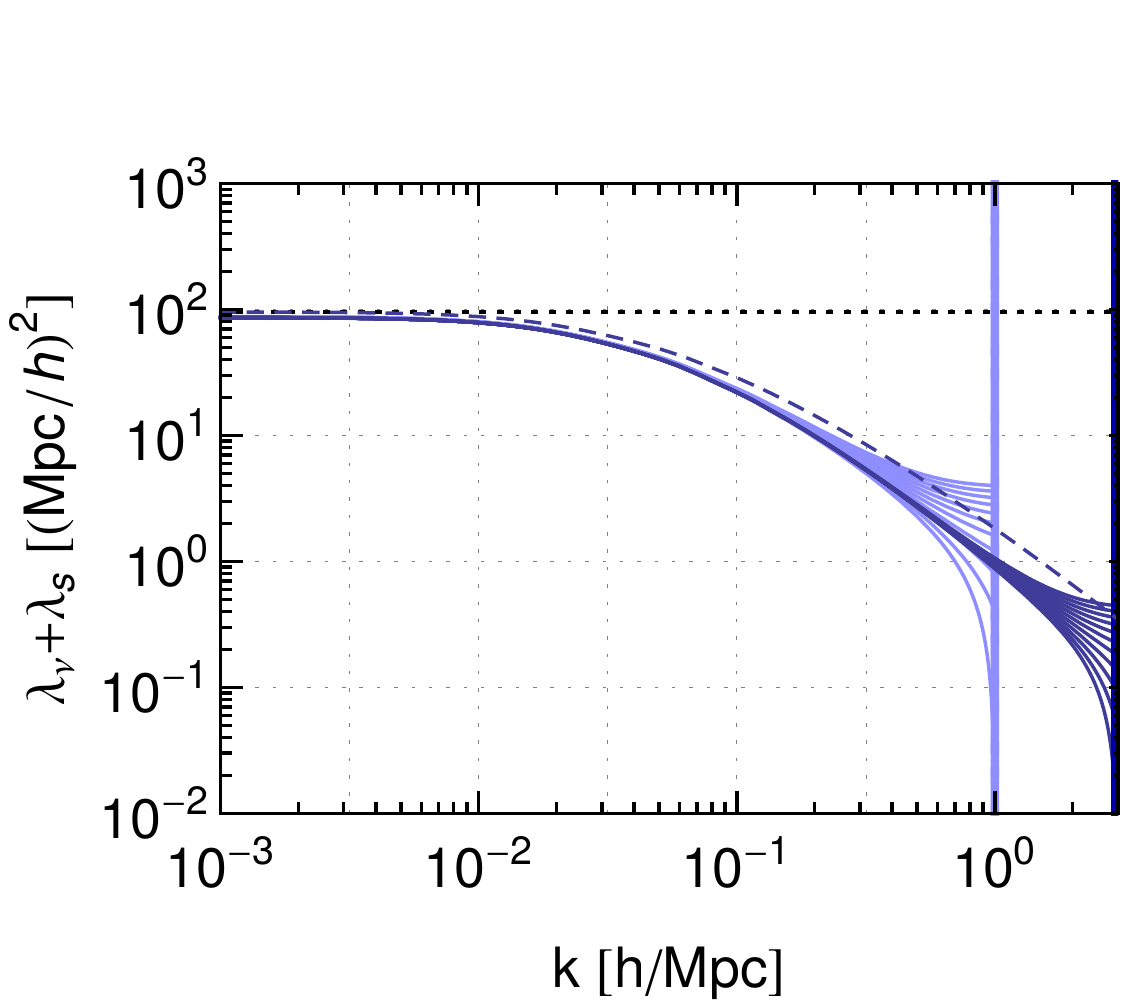} \qquad
$$
\caption{Renormalization group evolution of the sum $\lambda_\nu(k)+\lambda_s(k)$. The various lines correspond to the RG evolution
obtained when imposing initial values at $\Lambda=1\, h/$Mpc (light blue) or $\Lambda=3\, h/$Mpc (dark blue), respectively.
In both cases the various lines correspond to initial values within the interval $\lambda_i(\Lambda)\Lambda^2\in (0,2)$ for $i=\nu, s$. 
The dashed line shows the perturbative one-loop estimate \eqref{eq:lambdaslambdanukappaOneloop} for comparison, and
the dotted line corresponds to the IR fixed point.
}
\label{fig:RG}
\end{figure}

\begin{figure}[htb]
\centering
$$
\includegraphics[width=0.5\textwidth]{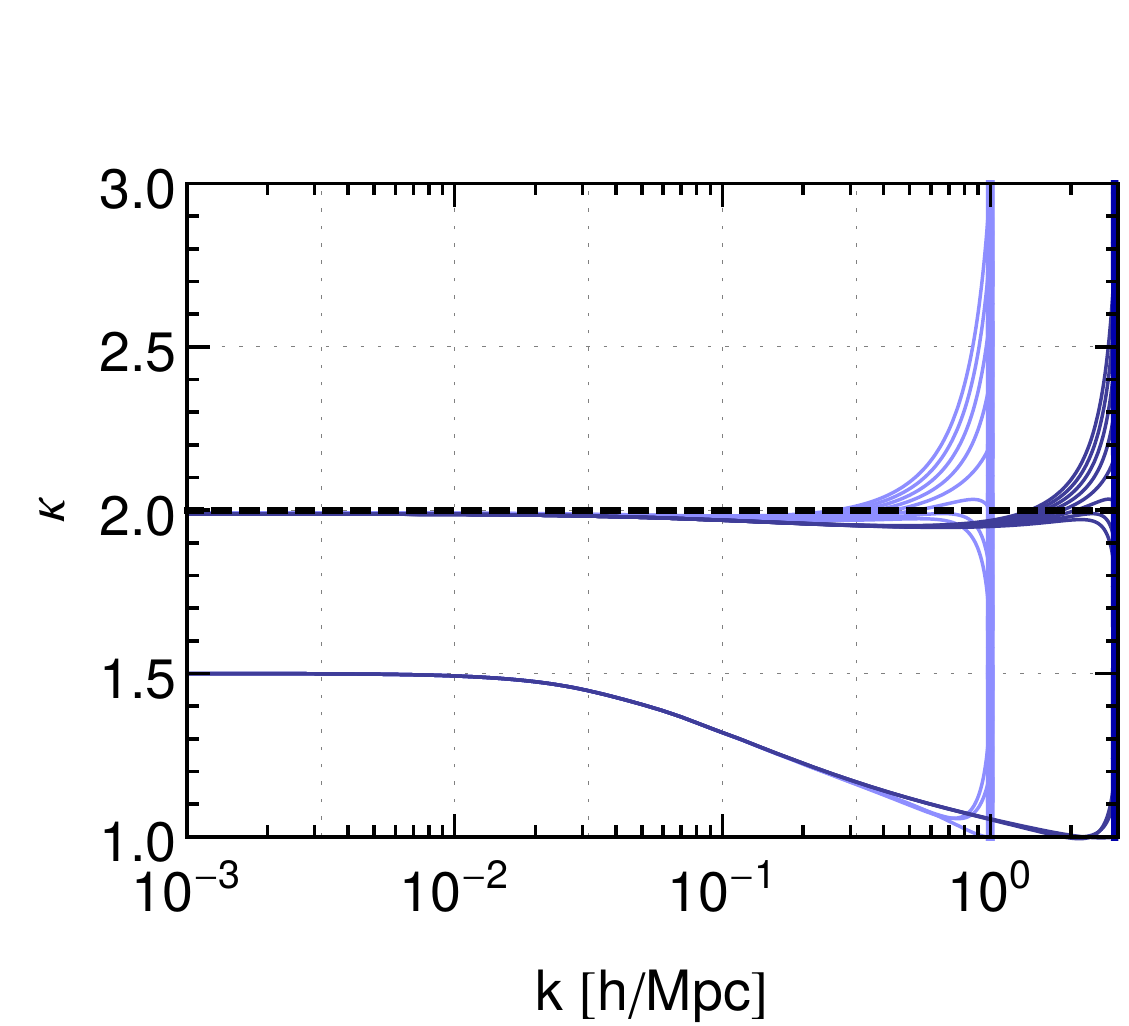} 
$$
\caption{Renormalization group evolution of the power law index $\kappa(k)$ characterizing the time-dependence
of the effective sound velocity and viscosity. The various blue lines show the RG evolution when initializing
the RG flow at $\Lambda=1\, h/$Mpc (light blue) or $\Lambda=3\, h/$Mpc (dark blue), respectively, with initial
values chosen in the range $\kappa(\Lambda)\in(0,3)$. The dashed line corresponds to the one-loop prediction
of the IR fixed point.
}
\label{fig:RGkappa}
\end{figure}

In Figs.\,\ref{fig:RG} and \ref{fig:RGkappa} we show the dependence of the RG flow on the initial conditions imposed at $k=\Lambda$, for the sum\footnote{The power spectrum is mainly sensitive to this linear combination \cite{viscousdm}.} $\lambda_\nu+\lambda_s$ as well as for $\kappa$. For illustration, we chose two
different values $\Lambda=1\, h/$Mpc (light blue lines) and $\Lambda=3\, h/$Mpc (dark blue lines). The various lines for each choice of $\Lambda$ show the dependence on the initial values $\lambda_i(\Lambda)$ and $\kappa(\Lambda)$ imposed at $k=\Lambda$. These initial values are chosen in the interval $0 \leq \lambda_i(\Lambda) \Lambda^2 \leq 2$
and $1\leq \kappa(\Lambda) \leq 3$, respectively. 
The RG evolution on scales
$k<\Lambda$ possesses an IR fixed-point behaviour in the limit $k\to 0$. This implies that the sensitivity to the initial value is drastically
reduced by virtue of the RG evolution. This behaviour can be clearly seen in fig.\,\ref{fig:RG}. Even for moderate
values $k\sim 0.1-0.3\, h/$Mpc the various RG trajectories have already converged to a small interval.
For a very large range of initial values of viscosity and sound velocity at the scale $\Lambda$, the RG evolution with decreasing $k$ drives their values  rapidly into a narrow regime.
A similar behaviour can be observed for the running of the power law coefficient $\kappa$ shown in fig.\,\ref{fig:RGkappa},
where the IR fixed point $\kappa \simeq 2.12$ is slightly offset from the one-loop value $\kappa=2$. This fixed point is approached as long
as $\kappa(\Lambda)>1$. For smaller initial values the RG evolution converges towards a different IR fixed point which is given by the unphysical pole at $\kappa= 1.5$ in the truncated
evolution equations~\eqref{eq:flowlambdanulambdaskappa}. 

\begin{figure}[htb]
\centering
$$
\includegraphics[width=0.5\textwidth]{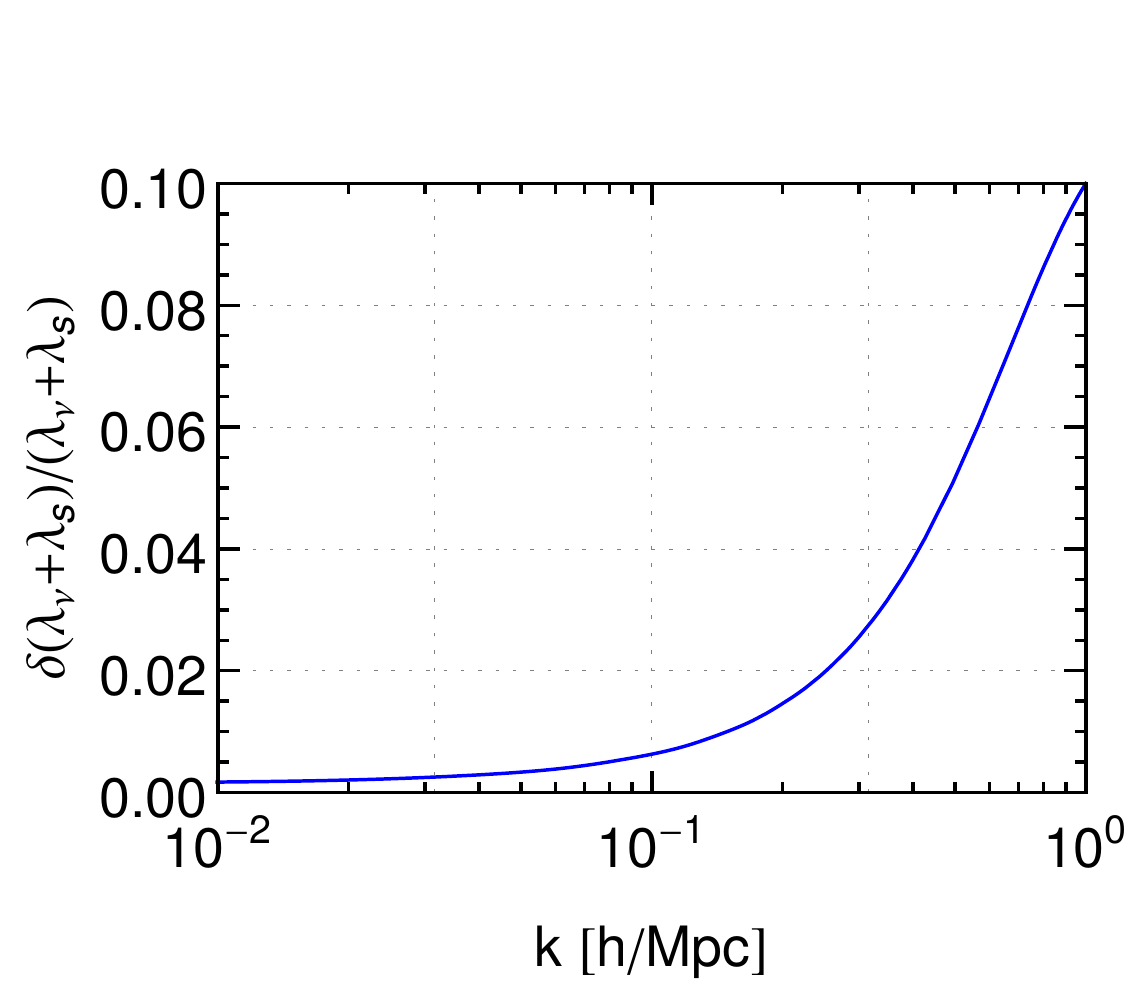} 
\includegraphics[width=0.5\textwidth]{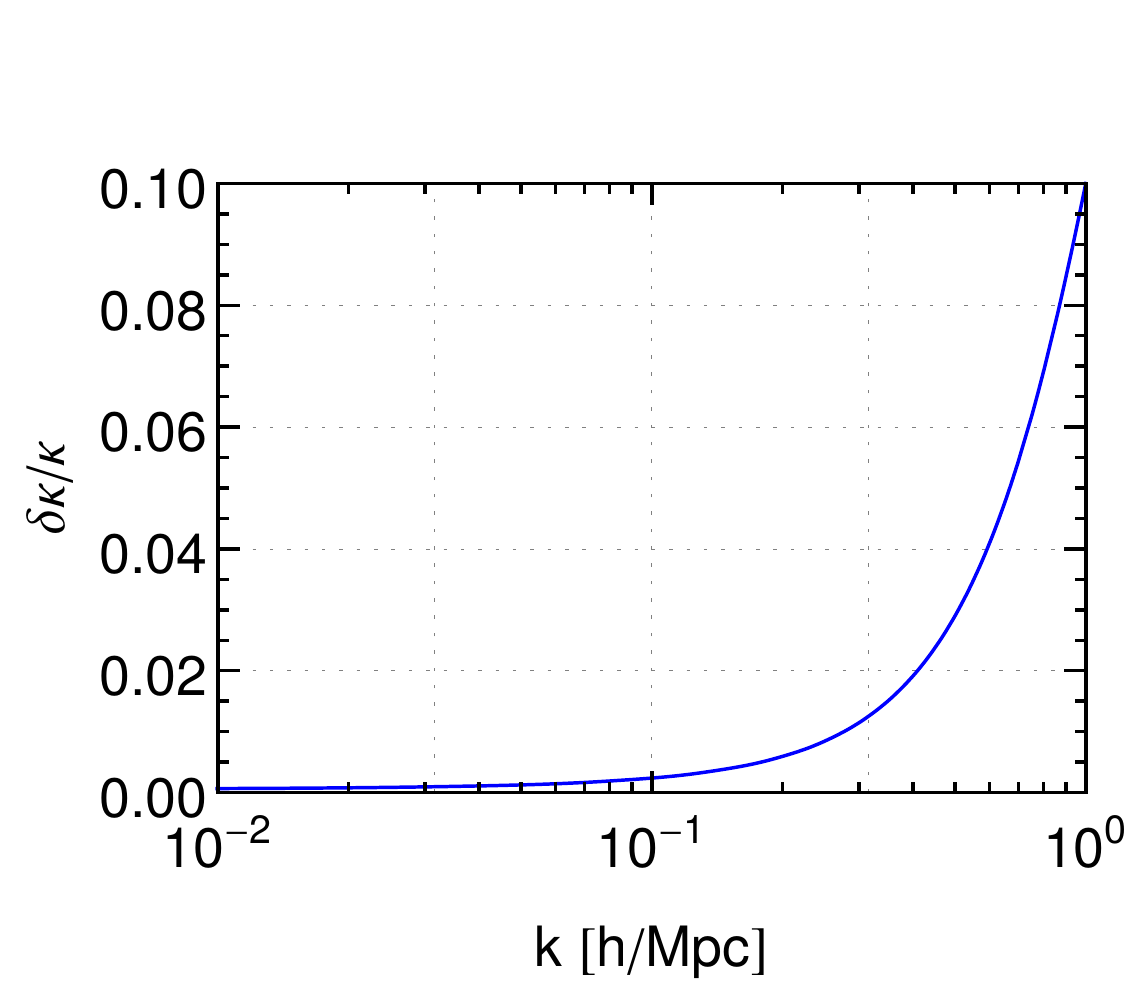} 
$$
\caption{Relative variation of $\lambda_\nu+\lambda_s$ (left) and $\kappa$ (right), obtained when varying the initial
condition at $\Lambda=1\, h/$Mpc by $10\%$. The IR fixed-point behaviour leads to a convergence of the RG trajectories such
that the relative uncertainty decreases for $k<\Lambda$. 
}
\label{fig:RGdelta}
\end{figure}

The sensitivity to the initial value can be characterized by the range of values $\lambda_i(k)$ for a given,
fixed value of $k$. This range quickly shrinks when going to smaller values of $k$, implying that the dominant contribution to effective
viscosity and sound velocity originates from coarse-graining the fluid modes in the interval between $k$ and $\Lambda$. This finding can be seen
as a justification of the matching prescription discussed in  \cite{viscousdm}.
It is in qualitative agreement with non-perturbative
arguments presented in \cite{Garny:2015oya}, as well as numerical studies based on  $N$-body response functions \cite{nishimichi}.
In order to quantify the (in-)sensitivity to the initial values, we vary them 
at $\Lambda=1\, h/$Mpc by
$\pm 5\%$ compared to  fig.\ \ref{fig:flow}. This yields solutions $\lambda_\pm(k)$ and $\kappa_\pm(k)$. In fig.\ref{fig:RGdelta} we
show the relative variation $\delta\lambda(k)/\lambda(k)=(\lambda_+(k)-\lambda_-(k))/\lambda(k)$, as well as $\delta\kappa(k)/\kappa(k)$.
This figure can be interpreted in the following way: if the effective sound velocity and viscosity are known to $10\%$ at $\Lambda=1\, h/$Mpc, then
the RG evolution towards smaller values $k<\Lambda$ reduces the relative uncertainty. For example, at the BAO scale the corresponding uncertainty
is at the sub-percent level.

Apart from the small sensitivity to the initial value, the overall value of $\lambda_\nu$ and $\lambda_s$ still evolves considerably
within the BAO range $k\sim 0.1-0.3\, h/$Mpc. However, this is actually expected, and is even required in order for observables such as
the power spectrum to be (approximately) independent of the value of the matching scale $k=k_m$, as we will discuss next.

 \newpage

\subsection{Power spectra}

The effective sound velocity and pressure obtained from the renormalization-group flow can be used as an input for
computing observables such as the power spectrum. 
For given effective sound velocity and viscosity at a renormalization scale denoted by $k_m$, the computation of observables such as the power spectrum for modes $k<k_m$ is organized as a perturbative
expansion in terms of the initial power spectrum, similarly to standard perturbation theory. However, based on the method described in \cite{viscousdm}, we fully take the effect of the time-dependent effective viscosity and
sound velocity on the non-linear propagation into account.

Since the choice of the matching scale $k_m$ is arbitrary, predictions for observable quantities must not depend on it. 
If we were able to solve exactly both the evolution equations for the effective action, and the resulting equations of motion, the $k_m$-dependence 
would drop out completely by definition. Therefore, any residual dependence is a crucial test of the
approximation scheme we employ, both for the determination of the effective sound velocity and viscosity, as well as the perturbative solution at $k<k_m$.

Motivated by this observation, we propose to use a variation in the scale $k_m$ as a quantitative indication for the theoretical error of our predictions.
Needless to say, control over the theoretical uncertainty is a crucial piece of information, and it enters directly e.g. in the determination of
confidence regions in parameter-estimation studies \cite{Audren:2012vy, Baldauf:2016sjb}. Nevertheless, it is important to keep in mind two limitations: (i) the range
in which we expect the perturbative solution of Navier-Stokes equations to converge implies an upper bound on $k_m$ not too far above the non-linear scale.
On the other hand, observables can only be computed for wavenumbers well below the matching scale, and we are interested primarily in the BAO range.
This prompts us to choose matching scales in the range $k_m\sim 0.4\, h/$Mpc $ - 1\, h/$Mpc. (ii) even if we were able to compute exactly the effective
viscosity and sound velocity generated by the RG evolution, and solve exactly the Navier-Stokes equations, we would still expect a residual dependence on $k_m$ coming
from the fact that we have to employ a truncation of the full effective action.

The first issue does not seem like a serious limitation in practice. Indeed, scale variation over a factor of two is a standard benchmark used
in many effective-theory calculations in high-energy physics. The second issue is potentially more limiting, since it implies an additional
theoretical uncertainty that is not captured well by the variation with $k_m$. Nevertheless, higher-gradient contributions to the equations of
motion at order $n>2$ lead to corrections to the power spectrum that scale with higher powers of the wavenumber ($\propto k^{2n}P_{lin}(k, z)$).
They are, therefore, parametrically suppressed on large scales compared to the viscosity and sound-velocity terms that contribute at $n=2$.
In addition, fluctuations in the hydrodynamic variables give rise to a stochastic force in the Euler equation that yields additional
contributions to the power spectrum known as noise terms, see e.g. \cite{Mercolli:2013bsa}. Within the renormalization-group approach these
are described by the contribution $\Phi$ in eq.\,(\ref{3.19}), and we will briefly comment on their impact below. In general, due to symmetry
arguments \cite{Goroff:1986ep, Mercolli:2013bsa}, their contribution can be shown to be suppressed by $k^4$ for small wavenumbers, and are
therefore sub-dominant compared to the corrections from effective viscosity and sound velocity. 
Keeping this limitation in mind, the theoretical error is expected to be represented faithfully by the $k_m$-variation for small enough wavenumbers.

\begin{figure}[htb]
\centering
$$
\includegraphics[width=0.5\textwidth]{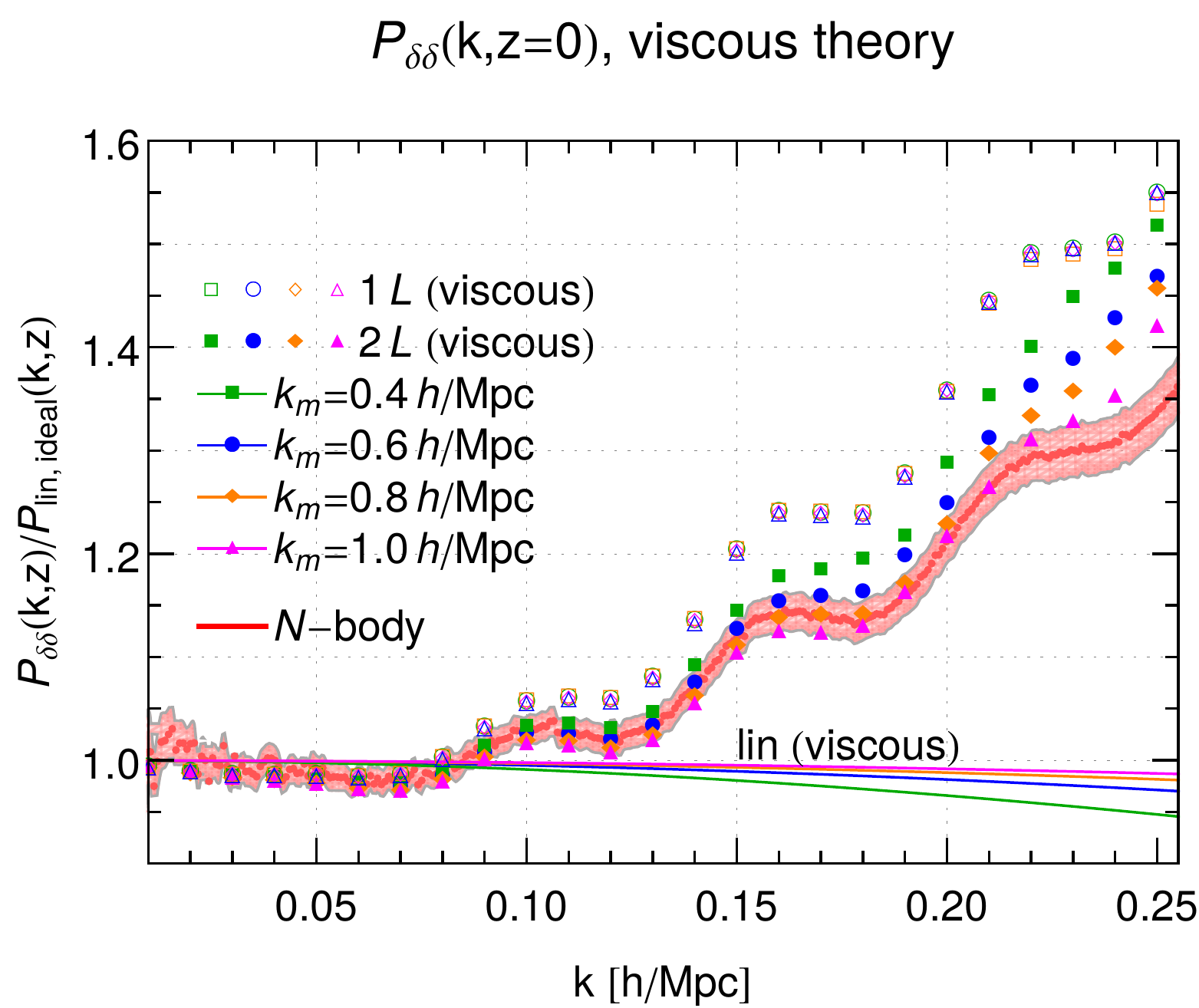} \qquad
\includegraphics[width=0.5\textwidth]{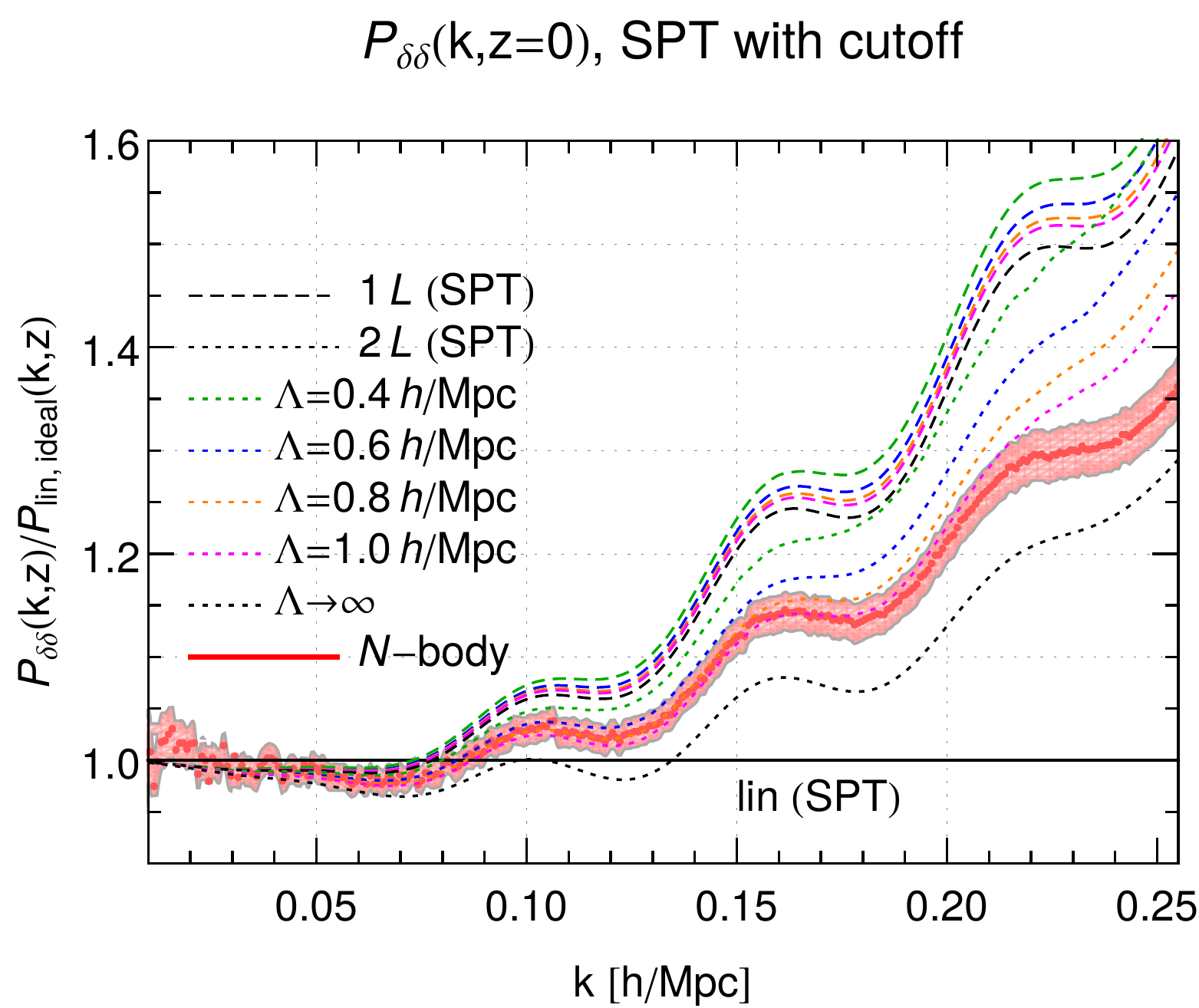}
$$
\caption{Density power spectrum for a $\Lambda${\rm CDM} model with $\sigma_8=0.79$, $\Omega_m=0.26$, $\Omega_b=0.044$, $h=0.72$, $n_s=0.96$.
The left figure shows results obtained in the coarse-grained theory with effective viscosity and sound velocity obtained from
the renormalization-group equations \eqref{eq:flowlambdanulambdaskappa},
and solving the Navier-Stoles equations at LO (linear), NLO (1-loop), NNLO (2-loop) in an expansion in the initial matter power spectrum,
for various values of the matching scale $k_m$. The right figure shows corresponding results in standard perturbation theory (pressureless
ideal fluid) with a cutoff $\Lambda$ imposed in the wavenumber integrals entering the 1- and 2-loop expressions. In both figures, the
thick red lines correspond to $N$-body results taken from \cite{Kim:2011ab}, and the shaded region
indicates the uncertainty in the $N$-body data. All curves are normalized to the conventional linear power spectrum for an ideal fluid, obtained from the CLASS code \cite{Blas:2011rf}.
}
\label{figkmdd}
\end{figure}

\begin{figure}[htb]
\centering
$$
\includegraphics[width=0.5\textwidth]{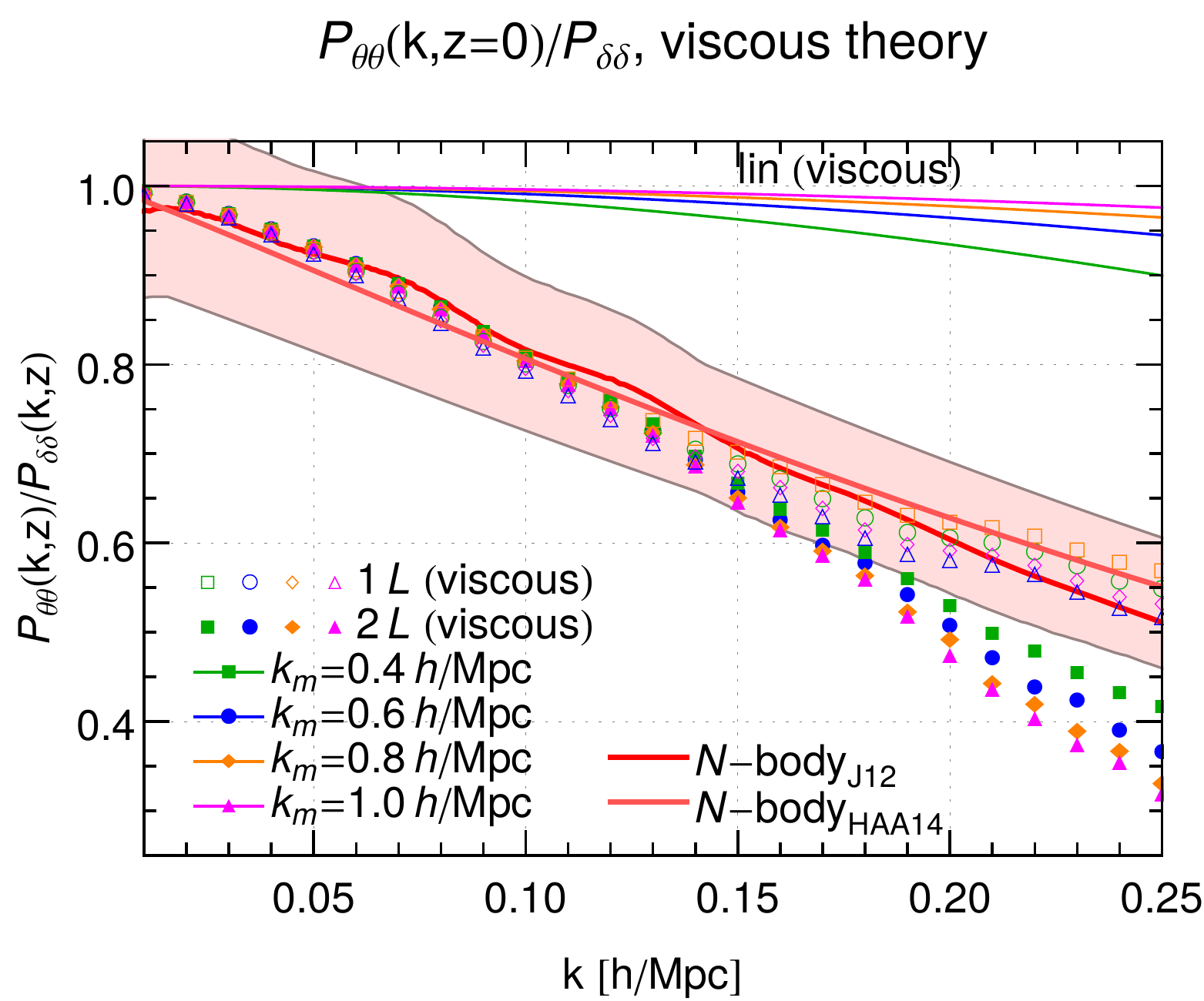} \qquad
\includegraphics[width=0.5\textwidth]{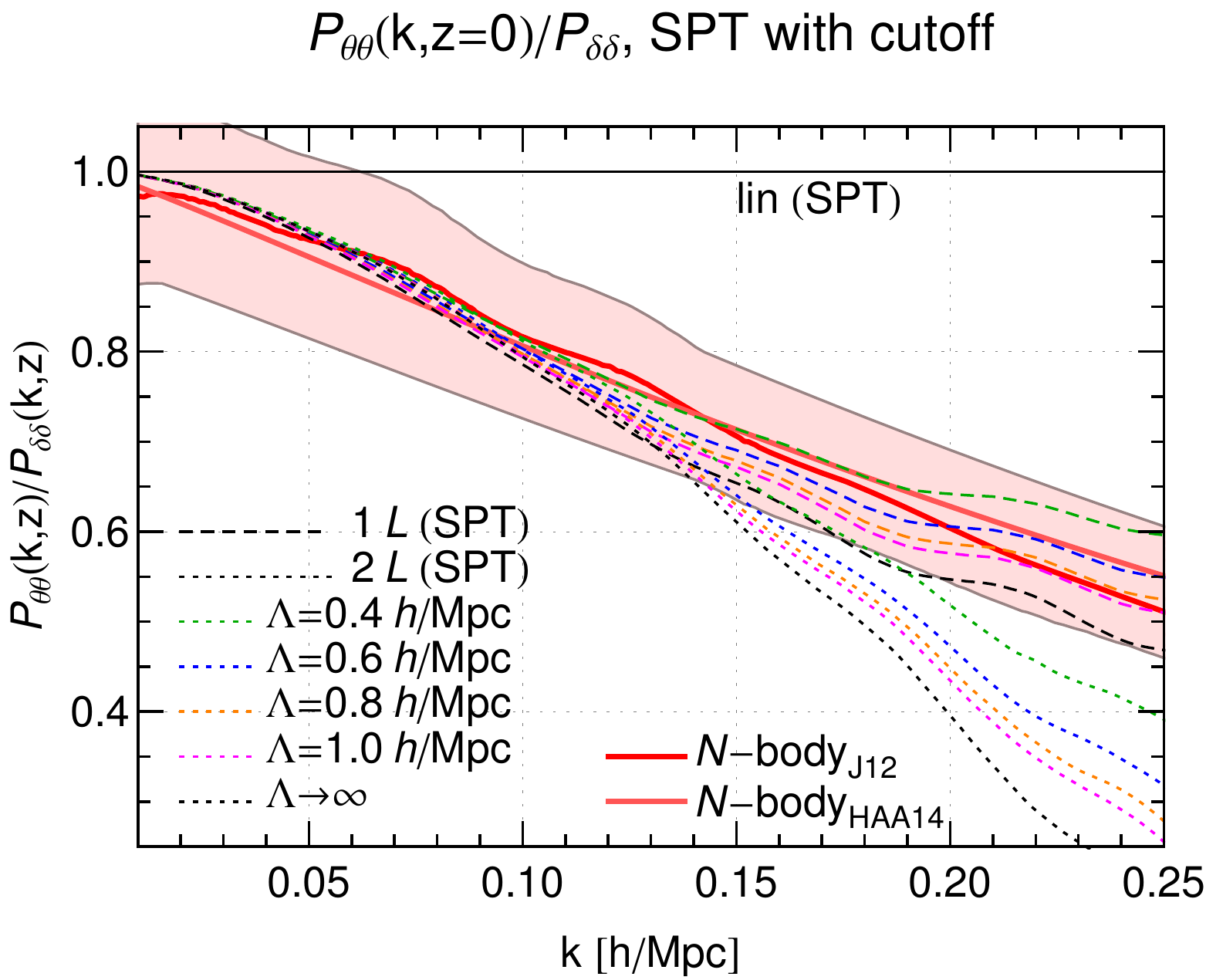}
$$
\caption{Ratio of the velocity divergence power spectrum to the density power spectrum, for the same $\Lambda${\rm CDM} model and the same
set of perturbative results as shown in fig.\,\ref{figkmdd}.
For comparison we show $N$-body results (thick red lines) for the ratio of velocity and density power spectra taken from \cite{Jennings:2012ej} (J12)
and \cite{Hahn:2014lca} (HAA14), and the shaded region indicates the uncertainty in the $N$-body data.
The velocity spectrum is normalized in such a way that $P_{\theta\theta}/P_{\delta\delta}\to 1$ for $k\to 0$.
}
\label{figkmtt}
\end{figure}

\begin{figure}[htb]
\centering
$$
\includegraphics[width=0.5\textwidth]{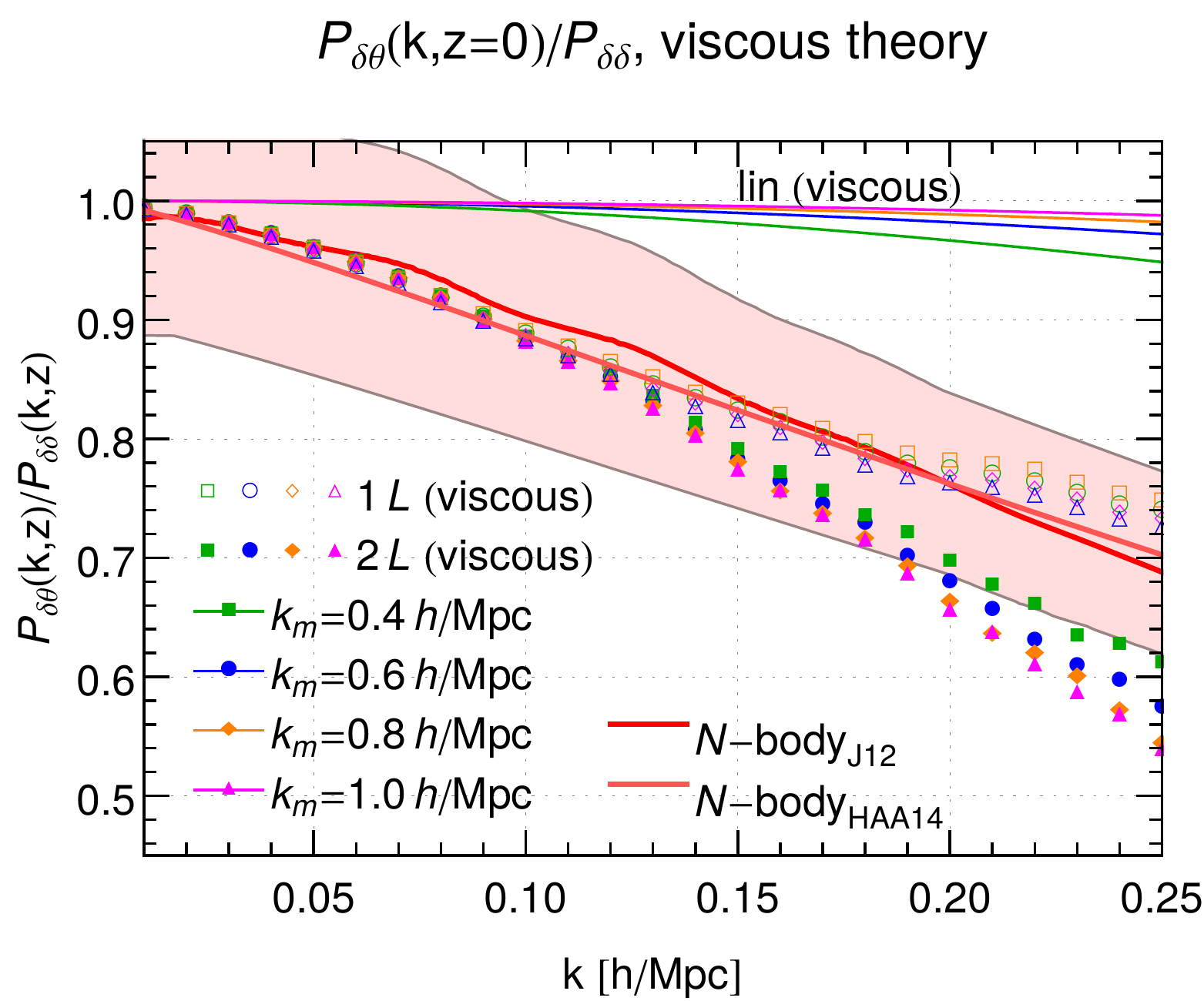} \qquad
\includegraphics[width=0.5\textwidth]{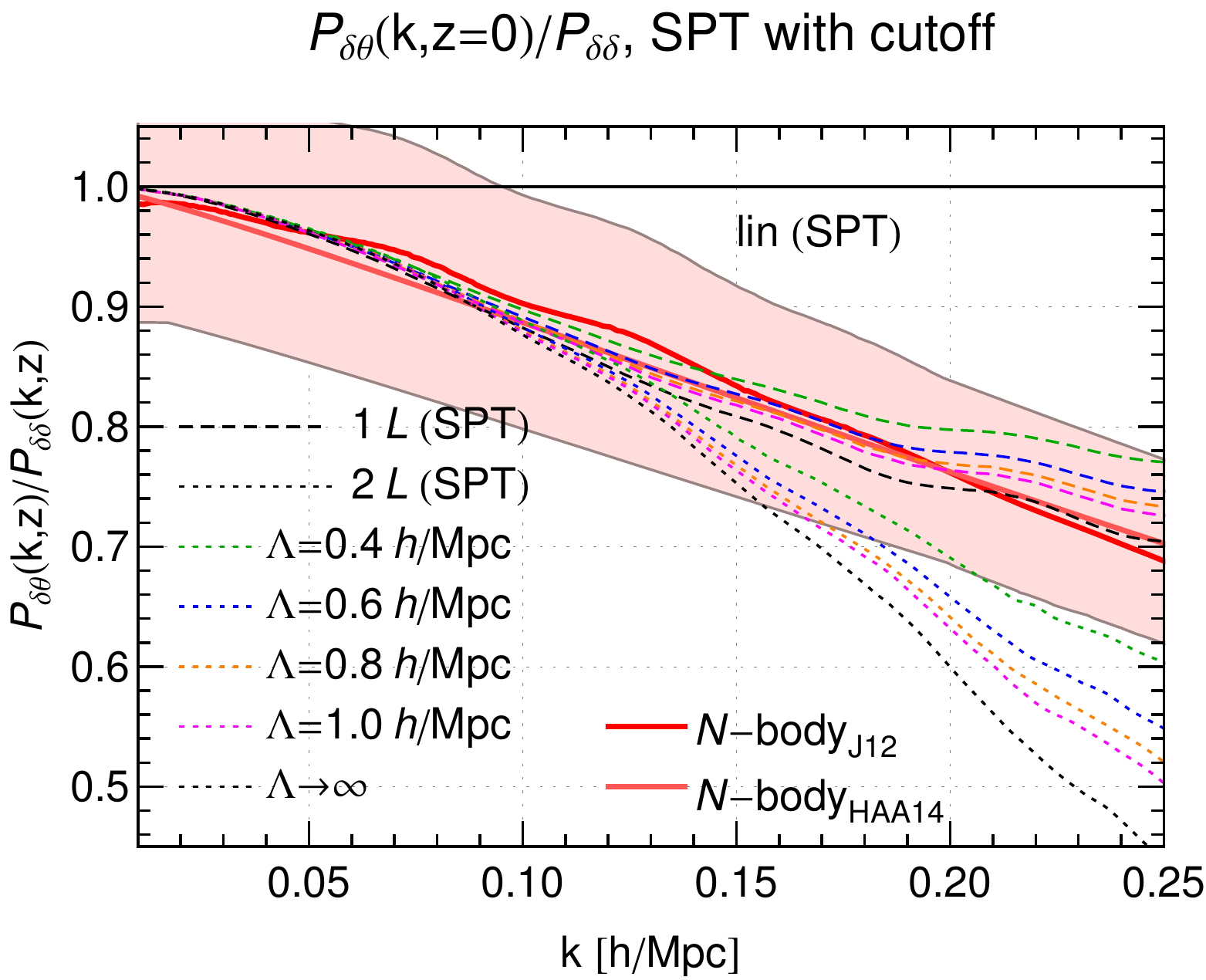}
$$
\caption{As fig.\,\ref{figkmtt}, but for the cross power spectrum of density and velocity divergence.
}
\label{figkmtd}
\end{figure}

In order to put these expectations to a test, we consider in the following the predictions obtained for various power spectra, and compare them
to results from numerical simulations. In fig.\,\ref{figkmdd} (left) we show the results for the density power spectrum at redshift $z=0$ within the viscous
theory.  Open symbols correspond to
solving the Navier-Stokes equations at NLO (1-loop), and filled symbols NNLO (2-loop). The thin coloured lines show the linear (LO) result.
For each level of approximation, we show the results obtained for four different values $k_m=0.4, 0.6, 0.8, 1.0\, h/$Mpc of the matching scale,
respectively. For comparison, the thick red line shows the correlator measured from a large-scale $N$-body simulation \cite{Kim:2011ab},
and the red-shaded region corresponds to our estimate of its uncertainty due to the finite boxsize $L=7.2\, h/$Gpc and resolution $d=L/N^{1/3}$
with $N=6000^3$. All curves are normalized to the linear power spectrum computed within the conventional ideal fluid paradigm.
The linear (LO) result for the viscous power spectrum is slightly suppressed compared to the ideal fluid case, and the amount of suppression
grows with the wavevector $k$, in accordance with expectations \cite{viscousdm}. The NLO (1-loop) results agree well with the $N$-body data, within
the error bars, up to about $k\lesssim 0.075\, h/$Mpc. For the NNLO (2-loop) results, on the other hand, the agreement is good up to about $k\lesssim 0.2\, h/$Mpc.
In addition, the residual variation for different choices of the matching scale $k_m$ brackets the remaining uncertainties, and represents a conservative
estimate for the theoretical error. For $k\lesssim 0.2\, h/$Mpc it is at the $\pm (2-3) \%$ level. This can be compared to the results obtained in
standard perturbation theory, which are shown in fig.\,\ref{figkmdd} (right). Here we computed the standard one- and two-loop contributions using a sharp
cutoff $\Lambda$ in $k$-space. As is well-known, the two-loop prediction depends on the choice of the cutoff, and varies at the level of $\pm 5\%$
for $0.4\, h/$Mpc$<\Lambda<1\, h/$Mpc, and even at the $\pm 10\%$ level when varying the cutoff in the interval $0.4\, h/$Mpc$<\Lambda\lesssim 5\, h/$Mpc. Thus, the viscous
description leads to a significant reduction in the uncertainty because of the treatment of UV modes as compared to standard perturbation theory.
The remaining uncertainty within the viscous framework is consistent with the expected impact of effects that are not captured by the fluid dynamical
description employed here, such as multistreaming and virialization, as well as generation of vorticity, on weakly non-linear scales \cite{Pueblas:2008uv}.

All solutions shown in fig.\,\ref{figkmdd} are based on the effective pressure and sound velocity obtained from the
flow equations \eqref{eq:flowlambdanulambdaskappa}, evaluated at the corresponding matching scale $k_m$. For concreteness, we
imposed the initial conditions at $3\, h/$Mpc using the perturbative estimate \eqref{eq:lambdaslambdanukappaOneloop} at that scale,
and then run the effective parameters down to the matching scale $k_m$. We checked that when initializing the renormalization
group evolution at the lower scale $1\, h/$Mpc instead, our results for the density power spectrum remain unchanged at the
percent level. This behaviour can be attributed to the attractor behaviour of the renormalization-group flow discussed in the previous
section, and can be taken as an indication for the robustness of the viscous description.

The effective action generated by the renormalization-group evolution in general encompasses also corrections that
cannot be captured by effective viscosity and sound-velocity terms. One example for such a contribution is $\Phi$ in eq.\,(\ref{3.19}). As has been mentioned above, contributions of this type can be
interpreted as noise terms arising from the fluctuations of the modes that are integrated out by the renormalization
group evolution. In order to obtain a rough estimate of their impact we use the one-loop results for $\Phi$
from eq.\,(\ref{Phi1L}), taking into account modes with $q>k_m$ in the one-loop integral. Its contribution to the power
spectrum can then be obtained using eq.\,(\ref{eq3.20}), which we evaluate using the leading-order expression for the
propagator. We find that the resulting correction to the density power spectrum at $z=0$ is below the percent
level for all values of $k_m$ considered here and for $k<0.2\, h/$Mpc.

In order to further scrutinize the renormalization-group approach, we consider also the power spectrum for the velocity divergence and the cross power spectrum.
Apart from being of theoretical interest, these power spectra are an important input for computing redshift-space
distortions \cite{Scoccimarro:2004tg} (\emph{cf.} also \cite{Mercolli:2013bsa, Carrasco:2013mua} for a discussion of velocity spectra in the EFTofLSS framework).
We show our results for the velocity spectrum at $z=0$ in fig.\,\ref{figkmtt} (left figure), and for the
cross spectrum in fig.\,\ref{figkmtd} (left figure). Both are normalized to the density power spectrum obtained under the same approximations.
The various symbols show again the dependence on the matching scale $k_m$, varied within the range discussed above.

For comparison, we also show results
obtained from extracting the velocity field from $N$-body simulations, either using a Delaunay tesselation \cite{Jennings:2012ej} or a
phase-space projection method \cite{Hahn:2014lca} (thick dark and light red lines, respectively). The shaded region is an estimate of the
error from reconstructing the velocity field quoted in \cite{Jennings:2012ej, Hahn:2014lca}. In addition, we show corresponding results
obtained in standard perturbation theory (right graphs in fig.\,\ref{figkmtt} and fig.\,\ref{figkmtd}, respectively), where we impose
a sharp cutoff $\Lambda$ in the loop-integrals over wavenumber space, that we vary in the same way as for $k_m$. It can be readily observed
that the scale-variation in the left figures is significantly smaller compared to the cutoff dependence obtained in SPT in the right plots.
Furthermore, the agreement with the $N$-body results is better in the former case. Taking the quoted uncertainty of $N$-body spectra at face value,
we find that the results obtained in the viscous theory are compatible with the $N$-body results up to $k\lesssim 0.2\, h/$Mpc. Furthermore, the theoretical
error estimated from the scale variation is consistent with the $N$-body results. Nevertheless, it is
apparent that the agreement starts to degrade quickly for $k\gtrsim 0.15\, h/$Mpc. Disregarding for a moment the error of the $N$-body results, one may
speculate about the mechanism that could lead to an increase in power in this range relative to our result. Using the expected scaling of higher-gradient
corrections, we note that it would require at least second-order gradient corrections. Otherwise, the good agreement at lower wavenumbers
would be upset. Nevertheless, taking into account the uncertainties, and the fact that we did not adjust any free parameters,
we find that the good agreement found for $k\lesssim 0.2\, h/$Mpc supports the effective viscous description.

\section{Conclusions and Outlook}

In summary, we have developed a renormalization-group approach that determines the effect of initial-state fluctuations on the effective action and equations of motion of the cosmological large-scale structure. The formalism is based on a functional renormalization-group equation for the one-particle irreducible effective action of a stochastic field theory. This formalism generates solutions of the classical field equations for stochastic initial conditions. Initial-state fluctuations are added gradually by lowering an infrared regulator scale. We discuss how RG flow equations for propagators and other correlation functions are obtained from the flow equation for the effective action and how they can be solved perturbatively or, within truncations, non-perturbatively. 

More concretely, we determine the RG trajectories of the parameters 
corresponding to the effective viscosity and sound velocity. 
These quantities receive a contribution from initial state fluctuations that is largely independent of their value at the microscopic UV scale. Correspondingly, the final value in the IR is also largely independent of the microscopic value as long as the latter is not too large.

Using this RG improved effective theory, we derive 
perturbative two-loop expressions for different two-point correlation functions of the large-scale theory, and study them numerically.
Our results for the density and velocity power spectra show reasonable agreement with results from $N$-body simulations for wavenumbers below about $0.2\,  h/$Mpc. For higher momentum scales, one expects that additional effects,
that have not been taken into account so far, become important. In particular, the initial values of effective viscosity and pressure for the renormalization-group flow will receive contributions coming from the higher moments of the distribution function at the microscopic scale. While the attractor behaviour of the RG flow that we have discussed in section~\ref{sec5} shows that the impact of the initial conditions is actually small, they are expected to become relevant
at a certain level of precision and for larger wavevectors. The impact of the stress tensor of deep UV modes on the BAO range has been quantitatively
analyzed in Ref.\,\cite{Pueblas:2008uv} based on $N$-body data. For the density power spectrum at $z=0$ its impact was found to be at the percent level for modes with wavenumber $k\simeq 0.2\, h/$Mpc, and growing rapidly for larger values of $k$. A contribution of this size is consistent
with the level of accuracy that we obtain for the density power spectrum. On the one hand, this result is reassuring: the effective RG
treatment improves over standard perturbation theory, and yields an error estimate that is consistent with the expected size of
physical effects that have been neglected. On the other hand, it is straightforward to include these effects in future extensions
of the RG framework (see below). 

It is also straightforward to extend the RG framework to theories for which dark matter has non-trivial material properties on the microscopic level \cite{Kunz:2016yqy, Kopp:2016mhm}.
In particular, a possible viscosity or sound velocity due to fundamental (self-)interactions can be taken into account in a straightforward way
in the form of additional contributions to the initial condition for the RG flow at the microscopic scale.

We conclude with a few more general remarks:
\begin{enumerate}
\item The exact functional renormalization-group equation \eqref{eq:WetterichEqn} for the effective action $\Gamma_k$ specifies how the effective theory of structure formation changes with the coarse-graining scale $k$. In principle, for a given microscopic action one can follow the RG evolution and determine the RG trajectories of the parameters that describe the effective dynamics on macroscopic scales.
The analysis of the RG flow provides a way to quantify the sensitivity of the macroscopic evolution of the LSS to properties in the UV.
 For instance, as illustrated in section~\ref{sec6}, the RG-flow \eqref{eq:WetterichEqn} can converge so rapidly to a (partial) IR fixed point that certain parameters of the effective theory are practically insensitive to the precise initial value at the UV scale. Other macroscopic parameters could be more sensitive to the microscopic physics. 

\item For the present paper, we have specialized from the onset to a simplified description with a relatively small set of fields. However, the derivation of the flow equation for $\Gamma_k$ given in section~\ref{sec3} generalizes to more complete dynamical descriptions. For instance,  for calculations of LSS at scales $> 0.2 \, \text{h/Mpc}$,
where virialization becomes gradually more important, it is interesting to account also for additional effects, such as those induced by fluid vorticity or by higher moments of the distribution function. To this end, one can formulate, for a more comprehensive field content, an action that  encodes the initial state fluctuations and that obeys an exact RG flow of the form \eqref{eq:WetterichEqn}.

\item The RG flow equation~\eqref{eq:WetterichEqn} provides also a tool for testing the self-consistency of any proposed effective dynamics. More precisely, any specific effective action \eqref{eq:operatorExp}
is a truncation of the full dynamics at the coarse-graining scale $k$ to a finite-dimensional subspace. The exact RG flow \eqref{eq:WetterichEqn} will generate also terms that lie outside this 
finite dimensional subspace. A truncation is a good approximation as long as these terms are unimportant for the further evolution. We believe that it is interesting and possible to
develop techniques that employ these properties of the exact equation  \eqref{eq:WetterichEqn} in order to specify the range of applicability of a specific effective action, and to assign 
theoretical uncertainties to the scale-dependence of the effective parameters. 

\end{enumerate}

\subsubsection*{Acknowledgments}

MG acknowledges partial support by the Munich Institute for Astro- and Particle
Physics (MIAPP) of the DFG cluster of excellence ``Origin and Structure of the
Universe".

\appendix
\section{Appendix A}
\label{appa}
In this appendix, we derive an explicit expression for the bi-spectrum in terms of 3rd functional derivatives of the effective action $\Gamma$.
Our discussion will also provide further information about some of the results mentioned in the main text. We start from the observation that
the second functional derivatives of the generating functional for connected Green's functions, ${\bf W}^{(2)}$ and of the effective action 
${\bf \Gamma}^{(2)}$ are matrices that are inverse of each other,
\begin{equation}
	{\bf W}^{(2)}\cdot {\bf \Gamma}^{(2)}  = \mathbbm{1} =  {\bf \Gamma}^{(2)} \cdot {\bf W}^{(2)}\, ,
	\label{eqa1}
\end{equation}
where
\begin{equation}
	W_{ij}^{(2)}(a,\bk,\eta;b,\bk',\eta') =  \left(
	\begin{array}{cc}
	\frac{\delta^2 W}{\delta J_a(\bk,\eta)\, \delta J_b(\bk',\eta^\prime)}  & \frac{\delta^2 W}{\delta J_a(\bk,\eta)\, \delta K_b(\bk',\eta^\prime)}\\
         \frac{\delta^2 W}{\delta K_a(\bk,\eta)\, \delta J_b(\bk',\eta^\prime)}  & \frac{\delta^2 W}{\delta K_a(\bk,\eta)\, \delta K_b(\bk',\eta^\prime)}
  \end{array}
 \right)\, ,
	\label{eqa2}
\end{equation}
\begin{equation}
	\Gamma_{ij}^{(2)}(a,\bk,\eta;b,\bk',\eta') =  \left(
	\begin{array}{cc}
	\frac{\delta^2 \Gamma}{\delta \phi_a(\bk,\eta)\, \delta \phi_b(\bk',\eta^\prime)}  & \frac{\delta^2 W}{\delta \phi_a(\bk,\eta)\, \delta \chi_b(\bk',\eta^\prime)}\\
         \frac{\delta^2 W}{\delta \chi_a(\bk,\eta)\, \delta \phi_b(\bk',\eta^\prime)}  & \frac{\delta^2 W}{\delta \chi_a(\bk,\eta)\, \delta \chi_b(\bk',\eta^\prime)}
  \end{array}
 \right)\, .
	\label{eqa3}
\end{equation}
In the short-hand notation of this appendix, the unit matrix $\mathbbm{1}$ denotes a $\delta$-function in all discrete and continuous variables.
Matrix multiplication involves a summation over the internal index $i,j = 1,2$, as well as summation (integration) over
all internal discrete (continuous) variables. We sometimes avoid writing all external and internal variables explicitly, but they are always present. For instance,
\begin{eqnarray}
	W^{(2)}_{ij} \Gamma^{(2)}_{jl} &\equiv& \sum_{j,\bar{b}} \int d\bar\bk d\bar\eta\, W^{(2)}_{ij}(a,\bk,\eta;\bar{b},\bar\bk,\bar\eta)\,  \Gamma^{(2)}_{jl}(\bar{b},\bar\bk,\bar\eta;c,\bk',\eta')
			\nonumber \\
				&=& \delta_{il}\, \delta_{ac}\, \delta(\bk - \bk')\, \delta(\eta - \eta')\, .
	\label{eqa4}
\end{eqnarray}
Equations (\ref{eqa1})-(\ref{eqa4}) hold for arbitrary source fields $J_a$, $K_b$. We are mainly interested in expressions for vanishing source fields, when function derivatives
of $W$ and $\Gamma$ are evaluated for the solutions of the unsourced field equations. In this case, $W_{22}^{(2)} \big{|}_{J_a=K_b=0}= 0 = \Gamma_{11}^{(2)} \big{|}_{J_a=K_b=0}$.
It then follows from the component $\left({\bf W}^{(2)} \cdot {\bf \Gamma}^{(2)} \right)_{11}$ of equation (\ref{eqa1})  and from the corresponding (22) component that
\begin{equation}
	\Gamma_{12}^{(2)}.W^{(2)}_{21}\big{|}_{J_a=K_b=0} = \mathbbm{ 1}_{11}  = \Gamma_{21}^{(2)} \cdot W^{(12)}_{21}\big{|}_{J_a=K_b=0}\, .
         \label{eqa5}
\end{equation}
This equation (\ref{eqa5}) is a rederivation of equation (\ref{3.15}). 
Another interesting expression is obtained from the off-diagonal component of eq.~(\ref{eqa1}),
\begin{equation}
	\left( {\bf W}^{(2)} \cdot {\bf \Gamma}^{(2)} \right)_{21} = \Gamma_{21}^{(2)} \cdot W^{(2)}_{11}  +    \Gamma_{22}^{(2)} \cdot W^{(2)}_{21} = 0\, .
        \label{eqa6}
\end{equation}
Acting on this expression with $W^{(2)}_{12}$ from the left, requiring vanishing source terms and using (\ref{eqa5}), one finds
\begin{equation}
	W_{11}^{(2)}\big{|}_{J_a=K_b=0} = - W_{12}^{(2)} \cdot \Gamma_{22}^{(2)}�\cdot W_{21}^{(2)}\big{|}_{J_a=K_b=0}\, .
	\label{eqa7}
\end{equation}
In the short-hand matrix notation of this appendix, (\ref{eqa7}) is just the rederivation of the equation (\ref{eq3.20}) for the spectrum in terms of second functional derivatives of the effective action. 
This matrix notation is also useful to derive relations between higher-point functions. Of particular physical interest is the bi-spectrum
\begin{eqnarray}
	\delta\left(\bk+\bk^\prime+\bk^{\prime\prime} \right)\, B(\bk,\bk^\prime,\eta,\eta^\prime,\eta^{\prime\prime})
	 &=& \frac{\delta^3 W}{\delta J_a(\bk,\eta)\, \delta J_b(\bk^\prime,\eta^\prime)\, \delta J_c(\bk^{\prime\prime},\eta^{\prime\prime})} \Big{|}_{J_a=K_b=0} \nonumber \\	  
	    &=& \frac{\delta W_{11}^{(2)}}{\delta J_c} \Big{|}_{J_a=K_b=0}\, ,
	\label{eqa8}
\end{eqnarray}
which, as indicated, can be obtained from the functional derivative of one particular component of $W^{(2)}$. To express this bispectrum in terms of functional derivatives of the
effective action, one just notes that the perturbation of the inverse of a general matrix ${\bf M}$ is $\delta\left({\bf M}^{-1}\right) = -  {\bf M}^{-1}.\left(\delta {\bf M}\right).{\bf M}^{-1}$
and hence
\begin{equation}
	\delta {\bf W}^{(2)} = \delta \left( \left({\bf \Gamma}^{(2)}\right)^{-1}\right) = - {\bf W}^{(2)} \cdot \delta{\bf \Gamma}^{(2)} \cdot {\bf W}^{(2)}\, .
	\label{eqa9}
\end{equation} 
Using the $(1,1)$-component of this matrix equation and specifying that the variation is with respect to $J_c$, we find for the bi-spectrum
\begin{eqnarray}
	\frac{\delta W_{11}^{(2)}}{\delta J_c} \Big{|}_{J_a=K_b=0} &=& - 
		\left[ \left( {\bf W}^{(2)} \cdot \frac{\delta{\bf \Gamma}^{(2)}}{\delta \phi_{\bar c}} \cdot {\bf W}^{(2)} \right)_{11} W_{11}^{(2)}(\bar c,c)\right] \Big{|}_{J_a=K_b=0} \nonumber \\
	&& - \left[ \left( {\bf W}^{(2)} \cdot \frac{\delta{\bf \Gamma}^{(2)}}{\delta \chi_{\bar c}} \cdot {\bf W}^{(2)} \right)_{11} W_{21}^{(2)}(\bar c,c)\right] \Big{|}_{J_a=K_b=0}\, .
	\label{eqa10}
\end{eqnarray}
Here, to write the functional derivative of $\Gamma^{(2)}$ with respect to $J_c$, we have used the chain rule and the identities
$W_{11}^{(2)}(\bar c,c) = \frac{\delta \phi_{\bar c}}{\delta J_c}$, $W_{21}^{(2)}(\bar c,c) = \frac{\delta \chi_{\bar c}}{\delta J_c}$. Evidently, not only the discrete index $\bar c$ is
summed over in (\ref{eqa10}), but also all other discrete and continuous arguments of $\phi_{\bar c}$ and $\chi_{\bar c}$.

\section{Appendix B}
\label{appb}

In this appendix we discuss the robustness of the renormalization-group equations for
the effective sound velocity and viscosity (\ref{eq:flowlambdanulambdaskappa}), as well
as the dependence on redshift.

\begin{figure}[htb]
\centering
$$
\includegraphics[width=0.5\textwidth]{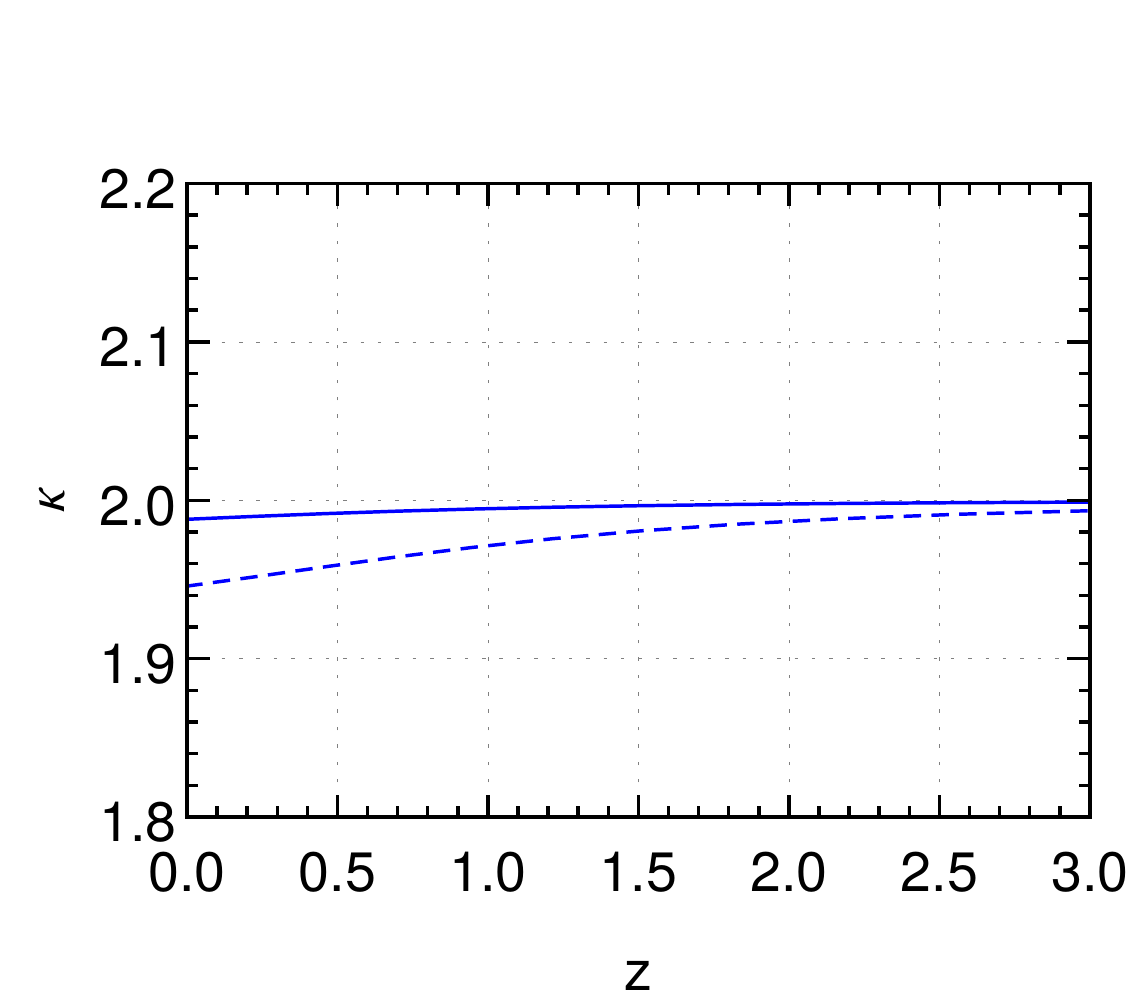} \qquad
\includegraphics[width=0.5\textwidth]{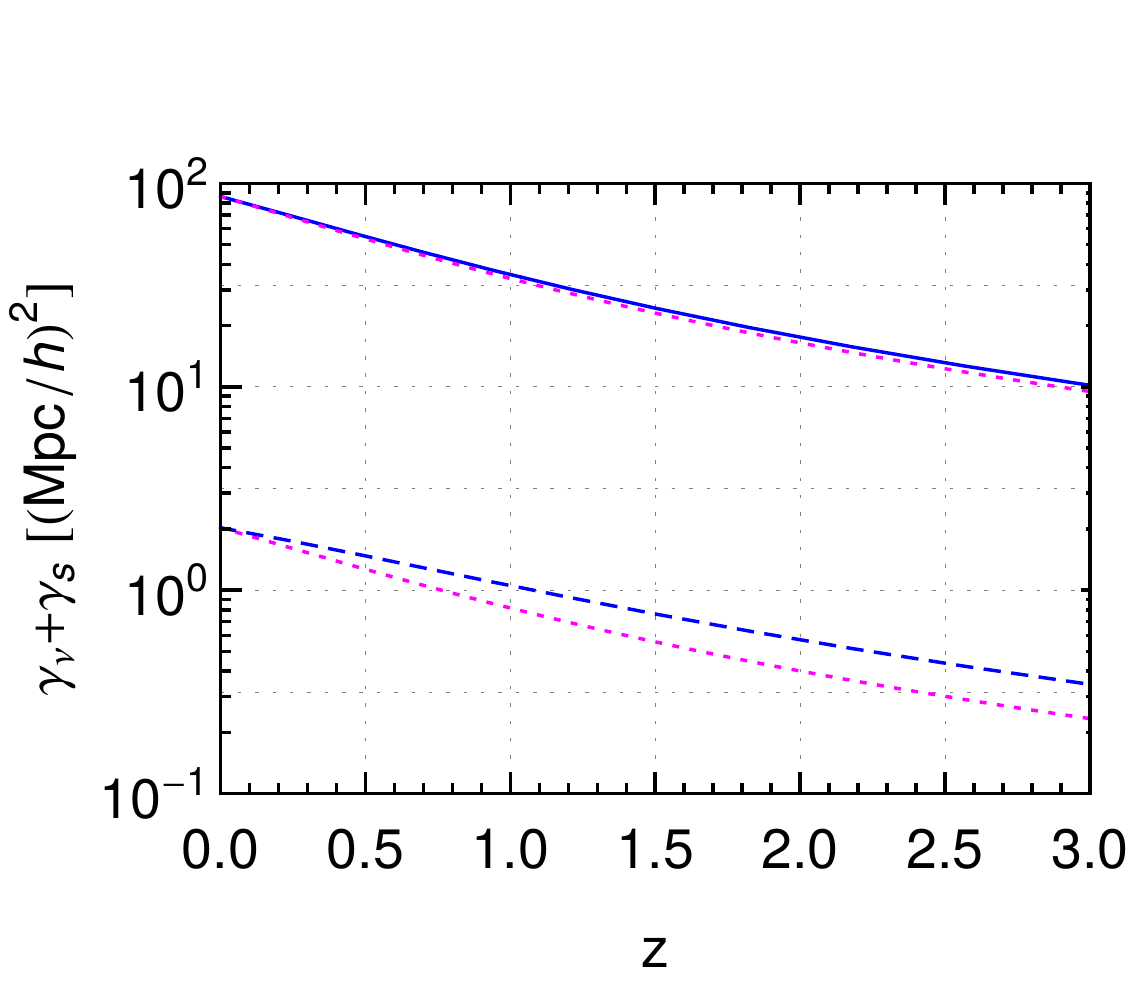}
$$
\caption{Running of the power law coefficient $\kappa$ (left) as well as the sum of effective
viscosity and sound velocity $\gamma_\nu+\gamma_s$ (right) with redshift $z$. The blue solid line shows
the values in the IR limit $k\to 0$, and the blue dashed for renormalization scale $k=0.6\, h/$Mpc.
The dotted lines show the result obtained from the RG flow evaluated at $\eta_0=0$ for comparison.
}
\label{fig:redshift}
\end{figure}

The RG flow equations (\ref{eq:flowlambdanulambdaskappa}) have been obtained by
evaluating (\ref{eq:5.43}) as well as its first derivative with respect to $\eta=\ln(D_L(z))$ at $\eta=0$,
i.e. at redshift $z=0$. It is straightforward to extend the derivation of the
RG flow equations to non-zero redshift. For the power-law ansatz $\gamma_\nu=\lambda_\nu e^{\kappa\eta}$
and $\gamma_s=\lambda_s e^{\kappa\eta}$ to be consistent, it is necessary that the value obtained for
$\kappa$ from the RG flow equation evaluated at $\eta\not=0$ varies sufficiently slowly with time,
$d\kappa/d\eta \ll \kappa$. In fig.\,\ref{fig:redshift} (left) we show the dependence of $\kappa$ on
redshift for two values of the renormalization-group scale. One can check that the adiabaticity
condition $d\kappa/d\eta \ll \kappa$ is safely satisfied. 

In a next step, one can use the slow running
of $\kappa$ to obtain improved results for the time-dependence of $\gamma_\nu(\eta)$ and $\gamma_s(\eta)$.
To see this, we denote by $\lambda_i=\lambda_i(k;\eta_0)$ ($i=\nu, s$) and $\kappa=\kappa(k;\eta_0)$ the solutions
of the RG flow obtained when evaluating the RG equations at time $\eta_0$. Then the dependence of
$\kappa(k;\eta_0)$ on $\eta_0$ corresponds to the
time dependence discussed above, i.e. the one shown in fig.\,\ref{fig:redshift} (left, translated to redshift using $\eta_0=\ln(D_L(z))$).
The effective viscosity and pressure are therefore given by $\gamma_i(k,\eta;\eta_0)=\lambda_i(k;\eta_0)\,e^{\kappa(k;\eta_0)\eta}$.
We checked that $\lambda_i(k;\eta_0)$ depend also only midly on $\eta_0$. Nevertheless, since the RG flow equation evaluated at $\eta_0$
is expected to be most accurate for time $\eta\approx\eta_0$, one can obtain an improved result for the redshift dependence by evaluating
$\gamma_i$ at $\eta_0=\eta$. The corresponding result is shown by the blue lines in fig.\,\ref{fig:redshift} (right), again for
two choices of the RG scale $k$. For comparison, also the result using a fixed $\eta_0=0$ is shown by the dotted lines. As expected,
the differences are very small, particularly at $z\lesssim 1$. Therefore it is well-justified to use the solutions obtained for
$\eta_0=0$ as an input when computing predictions for the power spectra at $z=0$ within the viscous fluid description. On the other hand,
if one is interested in power spectra at high redshift, the improved results with $\eta_0=\eta$ can be used.

\begin{figure}[htb]
\centering
$$
\includegraphics[width=0.5\textwidth]{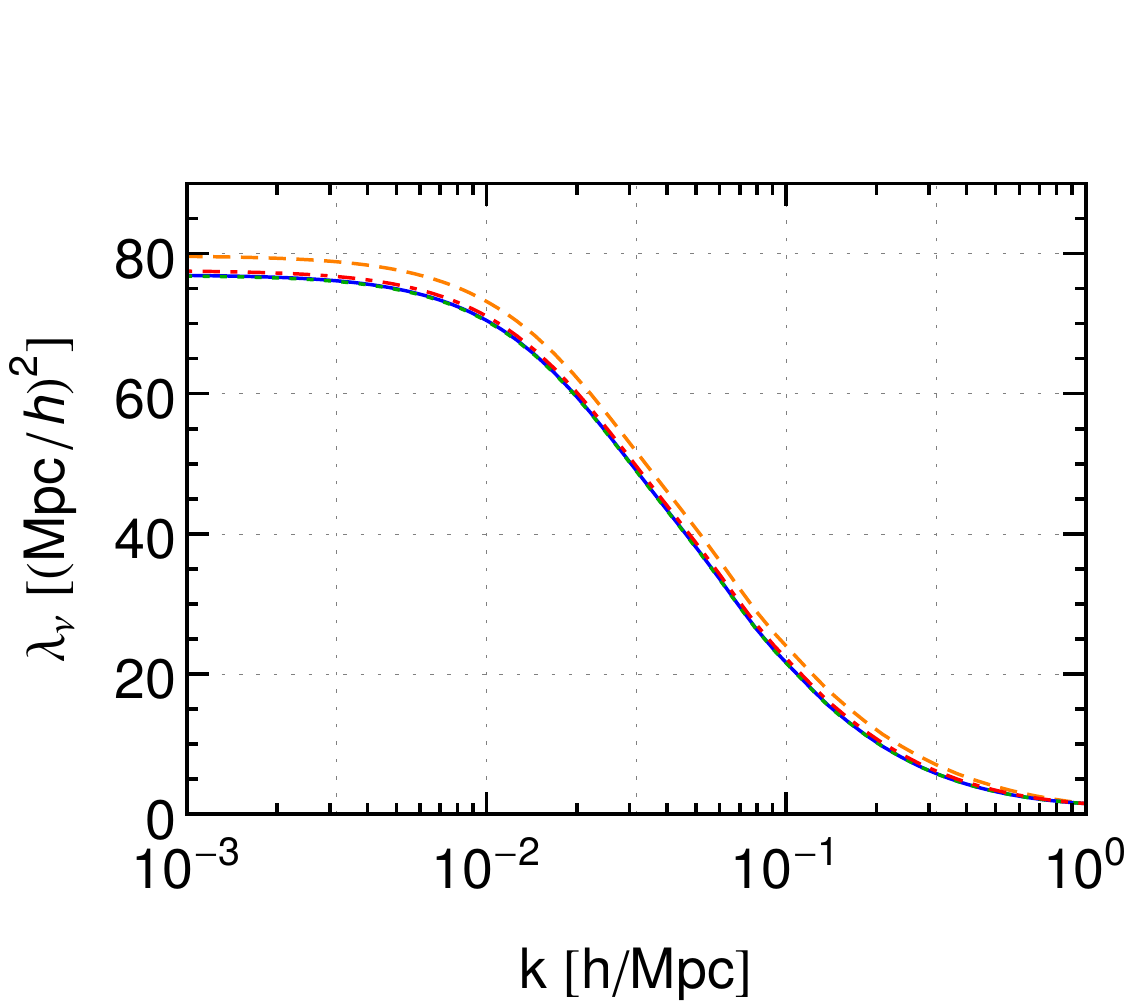} \qquad
\includegraphics[width=0.5\textwidth]{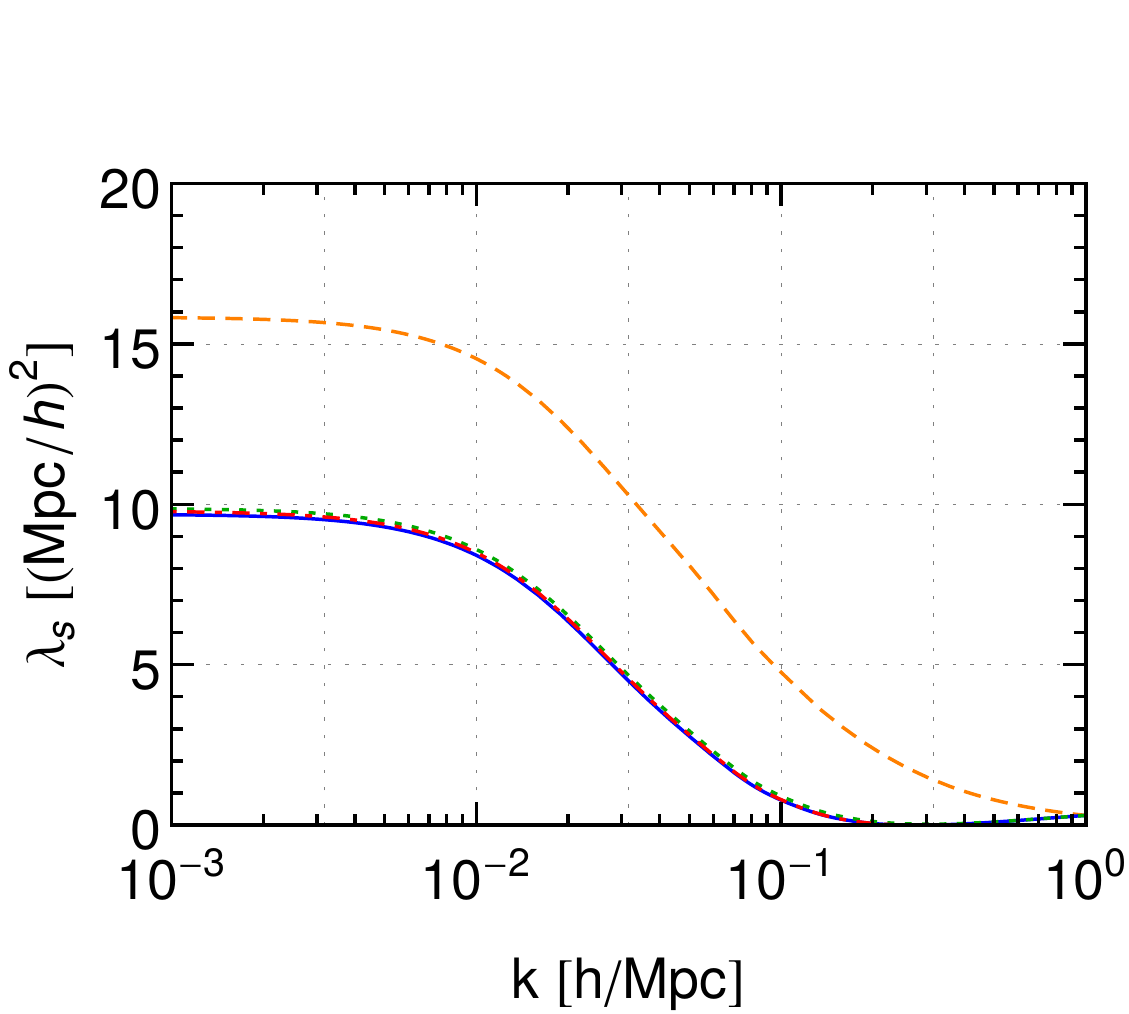}
$$
\caption{As fig.\,\ref{fig:flow}, but including two additional solutions of the RG flow. The green dotted line corresponds to the solution
of (\ref{eq:flowlambdanulambdaskappa2}) with $\kappa=2$ held fixed. The red dot-dashed line corresponds to the result when taking the full
viscous propagator into account.
}
\label{fig:flowFull}
\end{figure}

\begin{figure}[htb]
\centering
$$
\includegraphics[width=0.5\textwidth]{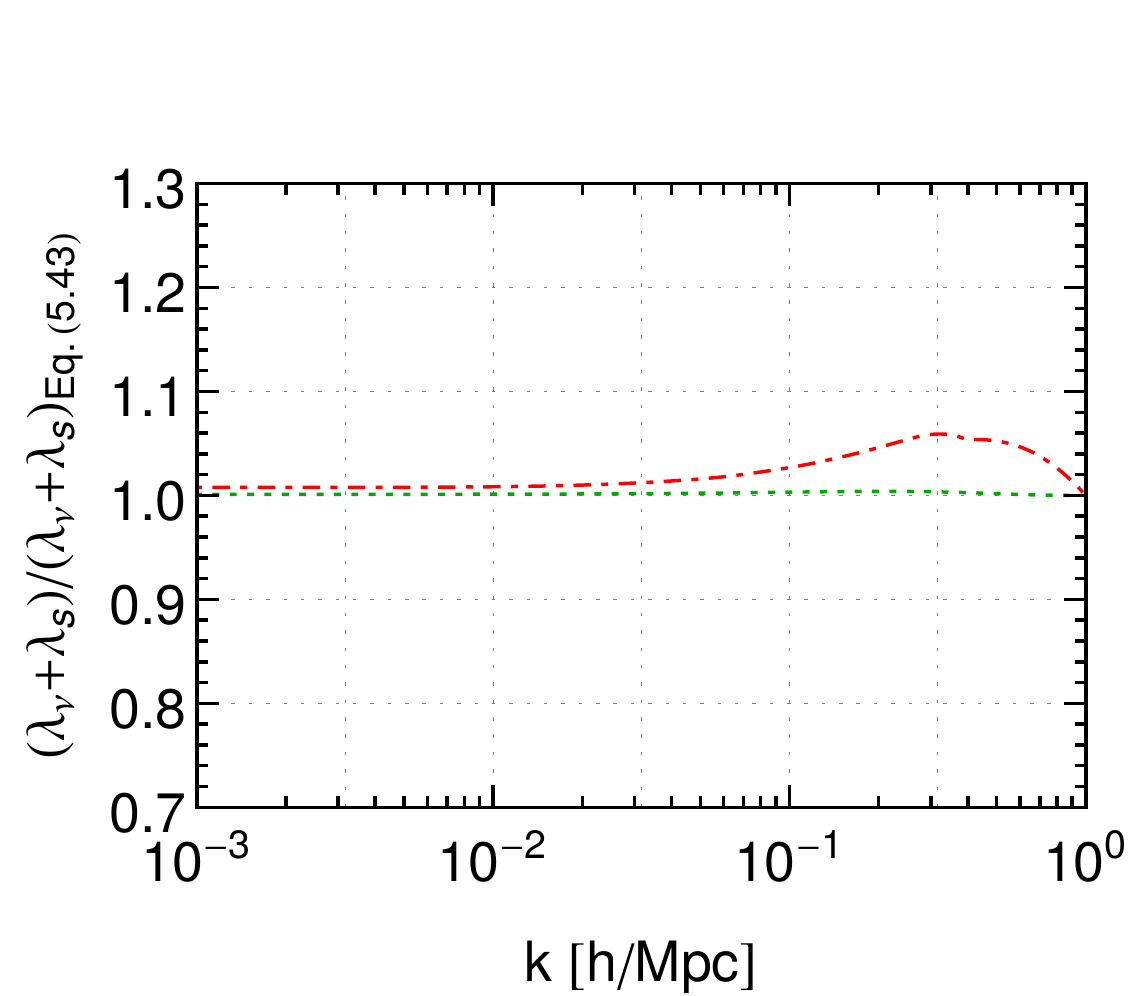} 
$$
\caption{Ratio of the sum of effective viscosity and sound velocity compared to the solution obtained from (\ref{eq:flowlambdanulambdaskappa}).
As in fig.\,{fig:flowFull}, the green dotted line corresponds to the solution
of (\ref{eq:flowlambdanulambdaskappa2}) and the red dot-dashed line corresponds to the result when taking the full
viscous propagator into account.
}
\label{fig:flowComparison}
\end{figure}

Instead of including an RG evolution for the power law coefficient $\kappa$, one
can impose also a fixed value for it. In this case the RG equations for $\lambda_\nu$ and
$\lambda_s$ remain unchanged, while the RG equation for $\kappa$ can be omitted. For the particularly
interesting case $\kappa=2$, which corresponds to the time-dependence obtained at one-loop,
the RG flow equations then simplify to
\begin{equation}
\begin{split}
\partial_t \tilde \lambda_\nu = & 2\tilde \lambda_\nu + \frac{4\pi}{3} k^3 P^0(k) e^{-2\frac{\tilde\lambda_\nu}{\kappa}} {\Bigg [} -\frac{78}{35} 
-\tilde \lambda_\nu \frac{7141}{3465} 
   -\tilde \lambda_s\frac{32}{45}
   + {\cal O}\left(\tilde \lambda_{\nu,s}^2 \right)
{\Bigg ]} \, ,   \\
\partial_t \tilde \lambda_s = & 2\tilde \lambda_s + \frac{4\pi}{3} k^3 P^0(k) e^{-2\frac{\tilde\lambda_\nu}{\kappa}} {\Bigg [} -\frac{31}{70} 
 + \tilde \lambda_\nu \frac{597}{770} +2\tilde \lambda_s 
   + {\cal O}\left(\tilde \lambda_{\nu,s}^2 \right)
{\Bigg ]} \, . \\
\end{split}
\label{eq:flowlambdanulambdaskappa2}
\end{equation}
The result for the RG flow is shown in fig.\,\ref{fig:flowFull} for $\lambda_\nu$ and $\lambda_s$, respectively (green dotted lines).
We also reproduce the solutions shown in fig.\,\ref{fig:flow} for comparison. For both $\lambda_\nu$ and $\lambda_s$ the difference compared to the
flow equation (\ref{eq:flowlambdanulambdaskappa}) is very small at all scales (blue line). The relative difference for $\lambda_\nu+\lambda_s$
is shown in fig.\,\ref{fig:flowComparison} (green dotted line).

\begin{figure}[htb]
\centering
$$
\includegraphics[width=0.5\textwidth]{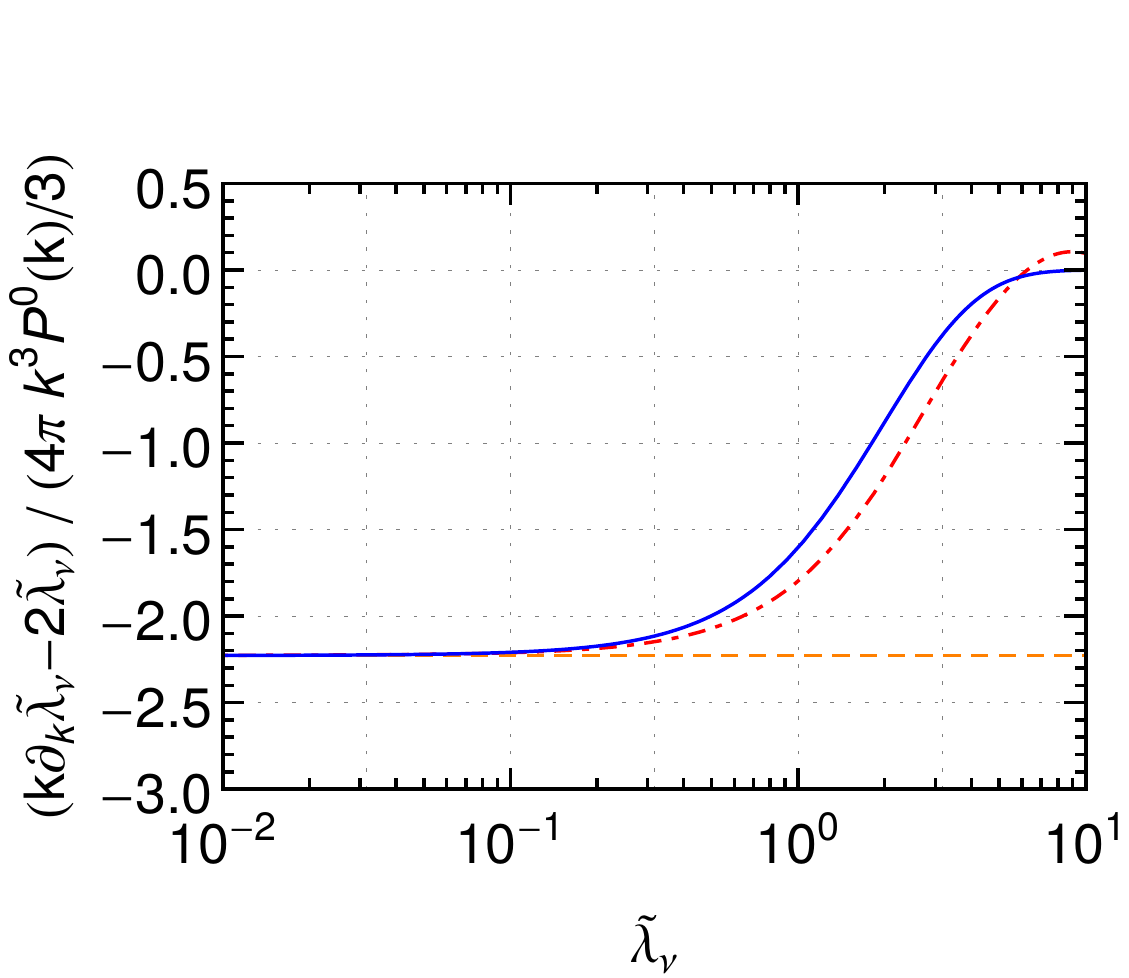} \qquad
\includegraphics[width=0.5\textwidth]{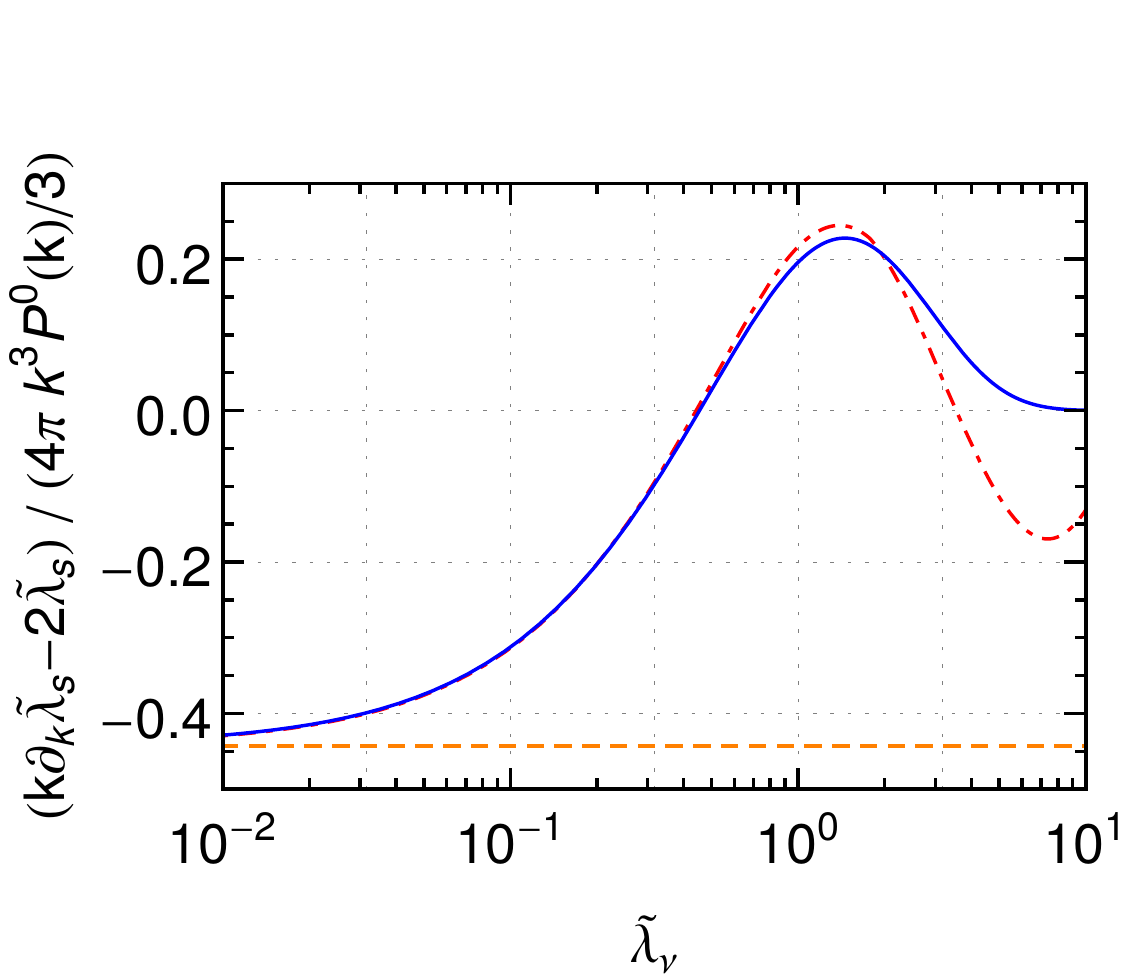}
$$
\caption{Right-hand side of the RG flow equation (`beta function') evaluated for different approximations, as a function of
the dimensionless viscosity coefficient $\tilde\lambda_\nu=\lambda_\nu k^2$. The orange dashed
line corresponds to the one loop approximation (\ref{eq:flowlambdanulambdaskappaOneLoop}), the blue line to (\ref{eq:flowlambdanulambdaskappa}), and the
red dot-dashed line corresponds to the result when taking the full viscous propagator into account. In these figures
we fixed the ratio $\lambda_s/\lambda_\nu=0.1$ as well as $\kappa=2$.
}
\label{fig:betafctn}
\end{figure}

Finally, one may also wonder about the convergence of the expansion of the viscous propagator that lead to the expansion in
powers of $\tilde\lambda_{\nu, s}$ on the right-hand side of the RG flow equations (\ref{eq:flowlambdanulambdaskappa}). Strictly
speaking this expansion is justified for $\tilde\lambda_{\nu, s}=\lambda_{\nu, s}k^2\ll 1$. One can check that this condition is
satisfied very well for $k\lesssim 0.5\, h/$Mpc. For larger values, the dimensionless coefficients can become of order one.
Therefore, in order to check the dependence of the RG flow on this approximation, we computed the Laplace transform of the self-energy
entering (\ref{callansymanzik3}) numerically using the full viscous propagator (\ref{eq:gvisc}) in (\ref{eq:flowDRFull}). Via eq.\,(\ref{flowvisc}) this yields RG evolution equations that
correspond to a resummation over $\ell$ in (\ref{eq:5.40}). 
In order to assess the impact of these terms we consider the rescaled `beta functions'
\be
 \beta_\nu(\tilde\lambda_\nu,\tilde\lambda_s,\kappa) \equiv \frac{k\partial_k\tilde\lambda_\nu - 2\tilde\lambda_\nu}{\frac{4\pi}{3} k^3 P^0(k)}, \qquad
 \beta_s(\tilde\lambda_\nu,\tilde\lambda_s,\kappa) \equiv \frac{k\partial_k\tilde\lambda_s - 2\tilde\lambda_s}{\frac{4\pi}{3} k^3 P^0(k)}\,.
\ee
In the perturbative one-loop approximation (\ref{eq:flowlambdanulambdaskappaOneLoop}), and assuming an EdS background, they take the constant values $\beta_\nu^{1L}=-78/35$ and $\beta_s^{1L}=-31/70$.
In fig.\,\ref{fig:betafctn} we compare the numerical result for the beta functions based on the full viscous propagator (red dotdashed lines) to the analytical result for the beta
functions given in Eq.\,(\ref{eq:flowlambdanulambdaskappa}) (blue lines) for $\beta_\nu$ (left figure) and for $\beta_s$ (right figure).
Both approximations agree well for $\tilde\lambda_{\nu, s}\lesssim 1$.
We also show the constant one-loop values (orange dashed lines) that are approached for $\tilde\lambda_{\nu, s}\to 0$.

The impact on the solution of the RG equations is shown in fig.\,\ref{fig:flowFull} (red dotdashed lines).
Due to the mild running of $\kappa$, we performed this analysis
for fixed $\kappa=2$. 
The differences compared to (\ref{eq:flowlambdanulambdaskappa}) shown as blue lines are again very small for both $\lambda_\nu$ and $\lambda_s$. 
The relative difference is shown in fig.\,\ref{fig:flowComparison}.
Therefore, we conclude that the RG flow equations
expanded up to ${\cal O}({\tilde\lambda_{\nu,s}})$ capture the dominant behaviour of the RG flow very accurately. We also checked that the attractor behaviour and the insensitivity
to the initial condition are exhibited by the resummed RG flow equations as well.


\end{document}